\documentclass[twocolumn,linenumbers]{aastex631}
\usepackage[T1]{fontenc}
\usepackage{savesym}
\savesymbol{tablenum}
\usepackage[section]{placeins}
\usepackage{amsmath} 
\usepackage{siunitx} 
\restoresymbol{SIX}{tablenum}
\usepackage{amssymb}
\usepackage{natbib}
\usepackage{changepage}
\usepackage{graphicx}
\usepackage{ragged2e}
\justifying
\usepackage{url}
\usepackage{xcolor}
\usepackage[flushleft]{threeparttable}

\shorttitle{Star Formation Histories from LEGA-C}
\shortauthors{Kaushal et al.}
\graphicspath{{./}{figures/}}

\begin{document}
\nolinenumbers

\title{A census of star formation histories of massive galaxies at 0.6 < z < 1 from spectro-photometric modeling using Bagpipes and Prospector}

\author[0000-0003-4382-4467]{Yasha Kaushal}
\affiliation{Department of Physics and Astronomy and PITT PACC,
University of Pittsburgh, 3941 O'Hara St, Pittsburgh, PA 15213, USA}

\author[0000-0001-6843-409X]{Angelos Nersesian}
\affiliation{Sterrenkundig Observatorium, Universiteit Gent,
Krijgslaan 281 S9, B-9000 Gent, Belgium}

\author[0000-0001-5063-8254]{Rachel Bezanson}
\affiliation{Department of Physics and Astronomy and PITT PACC,
University of Pittsburgh, 3941 O'Hara St, Pittsburgh, PA 15213, USA}

\author[0000-0002-5027-0135]{Arjen van der Wel}
\affiliation{Sterrenkundig Observatorium, Universiteit Gent,
Krijgslaan 281 S9, B-9000 Gent, Belgium}

\author[0000-0001-6755-1315]{Joel Leja}
\affiliation{Department of Astronomy \& Astrophysics, The Pennsylvania State University, University Park, PA 16802, USA}
\affiliation{Institute for Computational \& Data Sciences, The Pennsylvania State University, University Park, PA 16802, USA}
\affiliation{Institute for Gravitation and the Cosmos, The Pennsylvania State University, University Park, PA 16802, USA}

\author[0000-0002-1482-5818]{Adam Carnall}
\affiliation{Institute for Astronomy, School of Physics and Astronomy, University of
Edinburgh, Royal Observatory, Edinburgh, EH9 3HJ, UK}

\author[0000-0002-9656-1800]{Anna Gallazzi}
\affiliation{INAF-Osservatorio Astrofisico di Arcetri, Largo Enrico Fermi 5, I-50125 Firenze, Italy}

\author[0000-0003-1734-8356]{Stefano Zibetti}
\affiliation{INAF-Osservatorio Astrofisico di Arcetri, Largo Enrico Fermi 5, I-50125 Firenze, Italy}

\author[0000-0002-3475-7648]{Gourav Khullar}
\affiliation{Department of Physics and Astronomy and PITT PACC,
University of Pittsburgh, 3941 O'Hara St, Pittsburgh, PA 15213, USA}

\author[0000-0002-8871-3026]{Marijn Franx}
\affiliation{Leiden Observatory, Leiden University, P.O.Box 9513, NL-2300 AA Leiden, The Netherlands}

\author[0000-0002-9330-9108]{Adam Muzzin}
\affiliation{Department of Physics and Astronomy, York University, 4700 Keele St., Toronto, Ontario, M3J 1P3, Canada}


\author[0000-0002-2380-9801]{Anna de~Graaff}
\affiliation{Leiden Observatory, Leiden University, P.O.Box 9513, NL-2300 AA Leiden, The Netherlands}
\affiliation{Max-Planck-Institut f\"ur Astronomie, K\"onigstuhl 17, D-69117, Heidelberg, Germany}

\author[0000-0003-4196-0617]{Camilla Pacifici}
\affiliation{Space Telescope Science Institute, 3700 San Martin Drive, Baltimore, MD 21218, USA}

\author[0000-0001-7160-3632]{Katherine E. Whitaker}
\affiliation{Department of Astronomy, University of Massachusetts, Amherst, MA 01003, USA}
\affiliation{Cosmic Dawn Center (DAWN), Niels Bohr Institute, University of Copenhagen, Jagtvej 128, K{\o}benhavn N, DK-2200, Denmark} 

\author[0000-0002-5564-9873]{Eric F. Bell}
\affiliation{Department of Astronomy, University of Michigan, Ann Arbor, MI 48109, USA}

\author[0000-0003-2373-0404]{Marco Martorano}
\affiliation{Sterrenkundig Observatorium, Universiteit Gent,
Krijgslaan 281 S9, B-9000 Gent, Belgium}




\begin{abstract}

We present \textit{individual} star-formation histories of $\sim3000$ massive galaxies (log($\mathrm{M_*/M_{\odot}}$) > 10.5) from the Large Early Galaxy Astrophysics Census (LEGA-C) spectroscopic survey at a lookback time of $\sim$7 billion years and quantify the population trends leveraging 20hr-deep integrated spectra of these $\sim$ 1800 star-forming and $\sim$ 1200 quiescent galaxies at 0.6 < $z$ < 1.0. Essentially all galaxies at this epoch contain stars of age < 3 Gyr, in contrast with older massive galaxies today, facilitating better recovery of previous generations of star formation at cosmic noon and earlier. We conduct spectro-photometric analysis using parametric and non-parametric Bayesian SPS modeling tools - \texttt{Bagpipes} and \texttt{Prospector} to constrain the median star-formation histories of this mass-complete sample and characterize population trends. A consistent picture arises for the late-time stellar mass growth when quantified as $t_{50}$ and $t_{90}$, corresponding to the age of the universe when galaxies formed 50\% and 90\% of their total stellar mass, although the two methods disagree at the earliest formation times (e.g. $t_{10}$). Our results reveal trends in both stellar mass and stellar velocity dispersion as in the local universe - low-mass galaxies with shallower potential wells grow their stellar masses later in cosmic history compared to high-mass galaxies. Unlike local quiescent galaxies, the median duration of late-time star-formation ($\tau_{SF,late}$ = $t_{90}$ - $t_{50}$) does not consistently depend on the stellar mass. This census sets a benchmark for future deep spectro-photometric studies of the more distant universe.

\end{abstract}

\keywords{galaxies: evolution, galaxies: formation, galaxies: stellar content, galaxies: star formation, galaxies: statistics}

\section{Introduction} \label{section1}

Galaxies are complex systems of multiple stellar populations. Observational constraints on the timescales of star-formation and hence stellar mass growth and assembly enable us to understand the role of various physical mechanisms and their environment in guiding the formation and cosmic evolution of galaxies \citep{panter:07,leitner:12}. Understanding these processes on various physical scales (radial and spatially resolved) and temporal scales also helps shed light on some of the long-standing puzzles, such as when and how galaxies stop forming stars (quenching) \citep{schawinski:14,wild:16,schreiber:16,maltby:18,carnall:18} and the transition from star-forming to quiescent that leaves a bimodal population. \citep{bell:04,muzzin:13,vulcani:14,smethurst:15,taylor:15}.

Our current knowledge of the formation and cosmic evolution of galaxies is fundamentally limited by trade-offs between the quality and quantity of observations. Building a coherent picture of galaxy evolution demands (a) looking back in time for galaxies in the younger universe without compromising the quality of observations, and (b) a statistically large representative sample size to connect the dots between similar populations of galaxies at multiple epochs. This is difficult owing to technological and observational challenges as light from galaxies at increasing distances dims and shifts to less accessible infrared regime.

As stellar sources predominantly emit in UV/optical to NIR wavelengths, broadband multi-wavelength spectral energy distribution (SED) of a galaxy contains information about its total stellar mass and dust reddening. Stellar Population Synthesis (SPS) modeling is a widely used technique to derive physical properties (e.g., age, mass, dust, metallicity) of a composite stellar population (often a galaxy). This modeling is often limited to broadband photometric data due to its relative availability and the speed and simplicity of modeling a handful of data points compared to modeling a higher resolution spectrum. However, broadband photometry itself may not be sufficient to completely break the dust-age-metallicity degeneracy or to constrain higher order moments of star formation histories (SFHs) (\citet{leja:17,leja:19:a,tacchella:22}, A. Nersesian et al. submitted). The numerous, old low mass stars have long lifetimes and slow spectral evolution, yet leave weak imprints on observed data owing to their faint intrinsic luminosities. Therefore, inferring SFHs from photometry only is strongly susceptible to even small perturbations in data, leading to different recovery of SFHs (see e.g. \citet{ocvirk:06}). The recovered SFHs from mocks show biases up to 0.2 dex in the mass-weighted formation times. This is even worse with real-world broadband photometric data having more systematic uncertainties involved \citep{leja:19:a,carnall:19:a}. Photometry alone is not always sufficient to differentiate amongst modeling assumptions \citep{belli:18,carnall:19:a,tacchella:22}, thus properties measured from SED-only fits will always suffer from systematic uncertainties e.g. flux calibrations and different physical models obtained from stellar templates that vary with uncertainties on the data.

To robustly recover and analyze the timescales of a galaxy's stellar mass growth, we need to decipher spectral signatures containing strong imprints of stellar evolution. Continuum spectroscopy in the rest-frame optical regime contains information about the nature of past star-formation (bursty, uniform, rising, declining etc.) and metal enrichment within a galaxy in features like G4300, Fe4383, Fe4531, $Mg_{2}$, Balmer lines and 4000-$\si{\angstrom}$ break. These signatures evolve most rapidly in young stellar populations, since more massive galaxies today are generally older by several Gyr, studies of their SFHs should be more accurate and precise at earlier times. To obtain better signal-to-noise ratios, many analyses of high-redshift galaxy surveys involve mass matched stacking of spectra, vanishing any information about any individual galaxy's evolution \citep{schiavon:06,choi:14,siudek:17,cullen:19}. High signal-to-noise continuum spectroscopic data when combined with deep broadband photometric data has the potential to produce much stronger constraints on SFHs of galaxies \citep{gallazzi:08,pacifici:12,thomas:17,carnall:19:b,iyer:19,tacchella:22,webb:20} and provide clues on the number of major star formation episodes and the timescales of rejuvenation, starbursts, and quiescence. Spectroscopy alone also suffers from systematics like instrumental noise and outlier pixels and emphasizes the importance of spectro-photometric modeling.

\begin{figure*}[!t]
    \includegraphics[width=\textwidth]{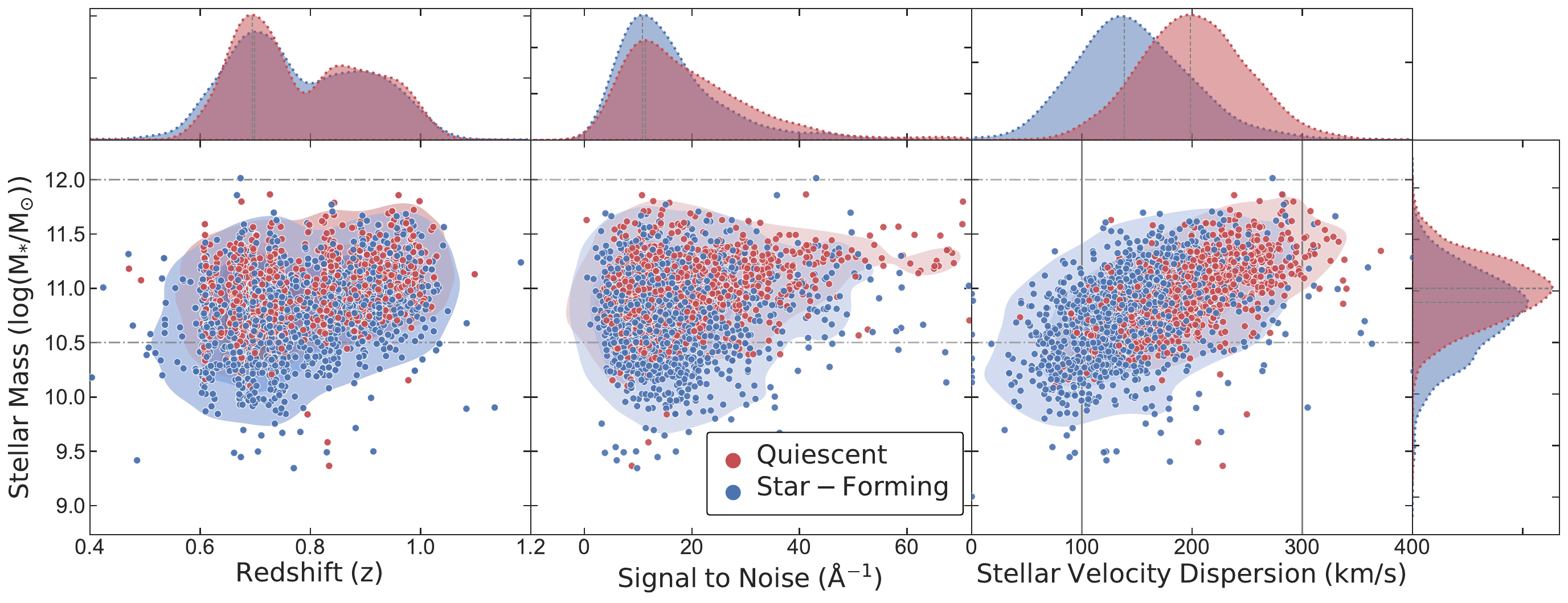}
    \caption{Stellar mass vs redshift (left), median signal to noise of LEGA-C spectra (center) and stellar velocity dispersion (right) for all 3005 primary LEGA-C Galaxies from the DR3 catalog with density distributions of star-forming (blue) and quiescent (red) populations based on rest-frame UVJ colors. Our effective sample selections are shown with dashdot lines (mass-selected) and solid lines (velocity dispersion-selected).}
    \label{fig:sample_distribution}
\end{figure*}

It is worth noting that even spectro-photometric modeling can be insufficiently sensitive to more slowly evolving old stellar populations, more so at later epochs. However, it has been argued that photometry can only probe the last $\sim$1 Gyr of a SFH, whereas adding spectroscopic data can help to probe the SFH further back in time \citep{chaves-montero:19}. Many studies have been conducted to recover SFHs of local galaxies \citep{thomas:05,cis-fernandes:07,panter:07,tojeiro:09,mcDermid:15,citro:16,ibarra-medel:16}. However, this approach of using fossil records at $z$ = 0 may not lead to reliable results, especially for more massive galaxies for which the major episode of their star-formation happened very early in the universe and these galaxies are now left with predominantly old stellar populations that suffer from strong outshining effects. This underscores the importance deep spectroscopic studies with high signal-to-noise at large lookback times. 

Recent advances in computational and sampling techniques \citep{10.1214/06-BA127,feroz:08,feroz:13,feroz:19} have led to the development of a variety of tools that provide fast full-Bayesian fitting of models to spectro-photometric data, like MCSED \citep{bowman:20}, BEAGLE \citep{chevallard:16}, \texttt{Prospector} \citep{Johnson:21,johnson:17}, and \texttt{Bagpipes} \citep{carnall:18}. These Bayesian methods give a better handle on the priors assumed and are robust to classical problem of over-fitting the data with complex models e.g. \citep{leja:19:a}. However, different SSP libraries and modeling assumptions in these tools can influence the derived star-formation histories \citep{martins:20,pacifici:22}. 

Though real SFHs of galaxies are complex, a simplified way to model them is by parametrizing the SFHs using a functional form. The most commonly used forms are exponentially declining \citep{mortlock:17,mclure:18,wu:18}, delayed exponentially declining \citep{ciesla:17,chevallard:19}, log-normal \citep{gladders:13,abramson:15,diemer:17,cohn:18} and double-power law \citep{ciesla:17,carnall:18}. These analytic prescriptions have been shown to match well with many SFHs from simulations \citep{simha:14,diemer:17} and are widely used as they minimize computational requirements. Increasing complexity by adding bursts of star-formation can bring parametric SFHs even closer to realistic scenarios. However, capturing events like rejuvenation and sudden quenching can still be challenging for parametric models. A more flexible method is non-parametric SFHs - which adopts a series of periods of constant star-formation in fixed or flexible time bins \citep{chauke:18,cappellari:17,leja:17,iyer:17,ocvirk:06,cid-fernandes:05}. Non-parametric approaches have higher flexibility with a wider range of possible solutions and therefore can allow a broader range of priors on SFHs and decrease biases on recovered results with more realistic episodes of star-formation in predefined time bins \citep{iyer:17, leja:19:a, suess:22a, suess:22b}. 

Until recently, SPS modeling of statistically representative populations has been limited to the analysis of photometry-only data sets. Numerous efforts have been put toward recovering the SFHs of both star-forming and quiescent populations from deep broadband photometric surveys spanning wide redshift ranges \citep{dye:08,wuyts:09,pforr:12,pacifici:16:a,pacifici:16:b,iyer:17,iyer:19,aufort:20,olsen:21} due to greater data availability and lower computational demands. In contrast, fewer studies have tested modeling spectro-photometric data for high redshift galaxies \citep{carnall:19:b,estrada-carpenter:20,forrest:20,Johnson:21,tacchella:22,khullar:22,hamadouche:23} owing to the dearth of high signal-to-noise continuum spectroscopy at significant lookback times. Large spectroscopic surveys like DEEP2 \citep{newman:12}, MOSDEF \citep{kriek:15} and KBSS \citep{rudie:12,steidel:14,strom:17} have been primarily sufficient to characterize emission line properties of thousands of star-forming galaxies. On the other hand, smaller spectroscopic studies of hundreds of quiescent galaxies have provided windows into the stellar populations of quiescent systems out to z$\sim$2, but with a significant bias towards the brightest, most massive subset. 

The Large Early Galaxy Astrophysics Census (LEGA-C) \citep{vanderwel:21,straatman:18,vanderwel:16} provides a novel opportunity to characterize the full population of Milky Way-mass and larger progenitors at significant lookback time. This deep spectroscopic survey of $\sim$3000 star-forming and quiescent galaxies looking 6-8 billion years back in time (0.6 $<$ z $<$ 1) includes deep imaging available for each galaxy from the UltraVISTA survey \citep{muzzin:13:a}. At this redshift, most stars in LEGA-C survey galaxies are $<$3 billion years in age, enabling more robust characterization of their SFHs. 

The primary goal of this paper is to measure the SFHs of the full LEGA-C DR3 sample by applying two commonly used Bayesian SPS modeling techniques (with modeling choices optimized for each) on spectro-photometric data and investigate the evolution of massive galaxies before z$\sim$0.8. We quantify these SFHs in two widely used metrics, namely $t_{50}$ and $t_{90}$, corresponding to the times when a galaxy formed 50\% and 90\% of its total stellar mass respectively and study the population trends of these formation times with stellar mass and stellar velocity dispersion.

The structure of the paper is as follows. Section \ref{section2} describes the LEGA-C dataset and our approach to spectro-photometric modeling using \texttt{Bagpipes} and \texttt{Prospector}, and some example demonstration of modeling results. Section \ref{section3} describes the SFHs of star-forming and quiescent galaxies and of the full population as recovered from spectro-photometric fits. We show cumulative median mass growth trends in stellar mass bins and further quantify population trends of $t_{50}$ and $t_{90}$ with stellar mass and stellar velocity dispersion. In Section \ref{section4} we expand on the interpretation of our results with respect to previous low and high redshift studies and how this impacts our understanding of the formation of both star-formation and quiescent systems. In Section \ref{section5} we conclude our study and highlight some major takeaways from this analysis. We also speculate on potential future works that could be a successor of this study to help us better constrain the evolution of massive galaxies and answer more broader questions. Throughout this paper we assume $\Omega_{M}$ = 0.3, $\Omega_{\Lambda}$ = 0.7, $H_{0}$ = 70 km s$^{-1}$ Mpc$^{-1}$ and all magnitudes are in quoted in the AB system.

\section{Data and Modeling Methods} \label{section2}

\subsection{Data and Sample} \label{data-and-sample}
The Large Early Galaxies Astrophysics Census (LEGA-C) \citep{vanderwel:21,straatman:18,vanderwel:16} is a 130-night public spectroscopic survey of $\sim$3000 Ks-band selected galaxies targeting redshift range 0.6 < $z$ < 1.0 in the COSMOS field, looking 6-8 billion years back in time. Each spectrum has an approximate observed spectral coverage of 6300$\si{\angstrom}$ - 8800$\si{\angstrom}$ corresponding to 3600$\si{\angstrom}$ - 5200$\si{\angstrom}$ rest-frame optical regime. The survey was conducted on Very Large Telescope (VLT) using ViMOS (VIsible Multi Object Spectrograph) \citep{lefevre:03} and completed its third and final data release (DR3) in August 2021 \citep{vanderwel:21}. 

The full spectroscopic sample consists of 4081 galaxies - 3029 primary targets and 1052 fillers. Targets were Ks-band selected from the UltraVISTA catalog \citep{muzzin:13:a} to include massive galaxies above $\sim$ $10^{10} M_{\odot}$ and to have Ks-band magnitudes brighter than a redshift dependent limit of [20.7 - 7.5log((1+z)/1.8)].  This selection criterion is independent of any derived quantities from the spectra. The full primary sample can be re-weighted using sampling and volume corrections following \citet{vanderwel:21} to be mass representative above log($\mathrm{M_*/M_{\odot}}$) > 10.5, which we adopt as the mass threshold for this study. Each galaxy was observed for $\sim$ 20 hours yielding S/N $\approx$ 20 $\si{\angstrom}^{-1}$ continuum with high fidelity absorption and emission line features for both dusty, blue as well as faint, red galaxies. 

As the LEGA-C survey targets only a subset of the full photometric sample,  we account for missing galaxies with appropriate weighting to individual galaxies to ensure spectroscopic completeness above the previously specified magnitude limit in a full census of the properties of massive galaxies at that epoch. Hence throughout this paper we apply a multiplicative factor \textit{Tcor} - corresponding to volume and sample correction - from the DR3 catalog \citep{vanderwel:21} to individual object counts to make it representative of that redshift. The photometric information is taken from UltraVISTA catalog \citep{muzzin:13:a}. We use \textit{BvrizYJ} bands for our primary spectro-photometric analysis. 

The stellar mass (log($\mathrm{M_*/M_{\odot}}$)), spectroscopic redshift (z), signal-to-noise (S/N) and integrated stellar velocity dispersion ($\sigma_*$) density distributions for the primary sample are shown in Fig.~\ref{fig:sample_distribution}. All values are taken from the LEGA-C DR3 catalog with (total) stellar mass estimates from \texttt{Prospector} photometry-only stellar populations fits (LOGM\_MEDIAN), spectroscopic redshift from the LEGA-C spectra (Z\_SPEC), the median S/N per pixel of the LEGA-C spectrum (S/N) and stellar velocity dispersion (SIGMA\_STARS) estimated from Gaussian broadening of theoretical single stellar population models as described in \citet{bezanson:18}. We split the sample into star-forming and quiescent populations using \citet{muzzin+13:b} rest-frame U-V and V-J color-color cuts, identifying quiescent galaxies with U-V > (V-J) $\times$ 0.88 + 0.69. Note that throughout this study, we adopt the LEGA-C DR3 catalog value of $\sigma_*$ and $M_*/M_{\odot}$ to maintain a single set of labels for each individual galaxy and to facilitate consistent comparisons with other studies. There are 1774 star-forming (medians: log($\mathrm{M_*/M_{\odot}}$) $\sim$ 10.8 and $\sigma_*$ $\sim$ 140 km s$^{-1}$) and 1231 quiescent galaxies (median: log($\mathrm{M_*/M_{\odot}}$) $\sim$ 11.2, $\sigma_*$ $\sim$ 200 km s$^{-1}$) in the primary sample of 3005 objects shown in red and blue colors respectively (obtained using PRIMARY flag in LEGA-C DR3 catalog, more details in \citet{vanderwel:21}). Throughout this paper, we focus on objects above the approximate mass completeness limit of the LEGA-C survey (10.5 < log($\mathrm{M_*/M_{\odot}}$) $\leq12$ and 100 < $\sigma_*$ $\leq300$ km s$^{-1}$). For mass-limited sample we have an effective sample size of 2703 unique galaxies (1459 star-forming and 1244 quiescents) and the velocity-dispersion sample includes 2823 unique galaxies (1575 star-forming and 1248 quiescents). This sample selection is shown with dashed and solid lines in Fig \ref{fig:sample_distribution}.

\begin{deluxetable*}{ccccccc}
\tablecaption{{\tt \texttt{Bagpipes}} modeling parameters and prior distribution functions 
\label{tab:bagpipes}}
\tablehead{
\colhead{No} & \colhead{Property} & \colhead{Parameters} & \colhead{Symbol/Unit} & \colhead{Prior} & \colhead{Range} & \colhead{Remark}
}
\startdata
1 & SFH & alpha & $\alpha$ &  logarithmic   &  (0.01,1000)  &  falling slope       \\
2 & (double-power law) & beta & $\beta$ &  logarithmic   &  (0.01,1000)  &  rising slope      \\
3 &  & tau & $\tau$/Gyr &  uniform   &  (0.1,15)  & peak of star-formation      \\
4 &  & stellar metallicity & $Z_{*}/Z_{\odot}$ &  uniform   &  (0.02,2.5)  &  From grid interpolation      \\
5 &  & stellar mass & $M_{*}/M_{\odot}$ &  logarithmic   &  (0,13)  &  priors from \citet{carnall:19:b}      \\
\hline
6 & Burst-1 & age & Gyr &  uniform   &  (0,13)  &  delta function burst      \\
7 &  & stellar mass & $M_{*}/M_{\odot}$ &  logarithmic   &  (0,10)  & with 3 free parameters    \\
8 &  & stellar metallicity & $Z_{*}/Z_{\odot}$ &  uniform   &  (0,2.5)     \\
\hline
9 & Burst-2 & age & Gyr &  uniform   &  (0,13)  &  same as above      \\
10 &  & stellar mass & $M_{*}/M_{\odot}$ &  logarithmic   &  (0,10)   \\
11 &  & stellar metallicity & $Z_{*}/Z_{\odot}$ &  uniform   &  (0,2.5)  &      \\
\hline
12 & Dust & V-band attenuation & Av / mag &  uniform  &  (0.,2.0)  &  \citet{Charlot:00}     \\
13 & & slope of attenuation & n &  Gaussian  &  (0.3,2.5)  &  $\mu$ = 0.7, $\sigma$ = 0.3     \\
\hline
14 & Stellar Velocity Dispersion & sigma & $\sigma_{\star}$/(km/s) &  logarithmic  &  (40,400)  & free parameter     \\
\hline
15 & Redshift & z & $z_{spec}$ & LEGA-C  &    & fixed parameter     \\
\hline
16 & Spectral White Noise & scaling factor & a &  logarithmic  &  (0.1,10)  & uncorrelated spectroscopic noise    \\
\hline
17 & Calibration & 0th order & c0 &  Gaussian  &  (0.9,1.1)  &  $\mu = 1$, $\sigma = 0.05$    \\
18 & Calibration & 1st order & c1 &  Gaussian  &  (-0.5,0.5)  &  $\mu = 0$, $\sigma = 0.1$   \\
19 & Calibration & 2nd order & c2 &  Gaussian  &  (-0.5,0.5)  &  $\mu = 0$, $\sigma = 0.1$
\enddata
\end{deluxetable*}


\begin{deluxetable*}{cccccccccc}[t]
\tablecaption{{\tt \texttt{Bagpipes} } Spectro-photometric modeling results
\label{tab:Bagpipes-results}}
\tablehead{\colhead{ID} & \colhead{Mask} & \colhead{LEGAC\_ID} & \colhead{$z_{spec}$} & \colhead{$\sigma_{*,cat}$} & \colhead{$log(M_{*,cat}/M_{\odot})$} & \colhead{$log(M_{*,fit}/M_{\odot})$} & \colhead{SFR} & \colhead{t50} & \colhead{t90} \\ {} & {} & {} & {} & {(km/s)} & {} & {} & {($M_{\odot}/yr$)} & {(Gyr)} & {(Gyr)}}
\startdata
103041 & 17 & 1159 & 0.82 & 75.9 & 10.52 & 10.63$\pm_{0.07}^{0.1}$ & 5.91$\pm_{1.14}^{2.02}$    & 5.07$\pm_{0.18}^{0.24}$ & 6.01$\pm_{0.14}^{0.13}$ \\
103061 & 16 & 1160 & 0.72 & 193.1 & 11.03  & 11.25$\pm_{0.03}^{0.03}$ & 0.00$\pm_{0.00}^{0.00}$ & 3.54$\pm_{0.03}^{0.24}$ & 5.72$\pm_{0.06}^{0.03}$ \\
103155 & 19 & 1161 & 0.64 & 103.1 & 11.07 & 11.36$\pm_{0.06}^{0.07}$ & 17.26$\pm_{2.54}^{3.67}$ & 2.79$\pm_{0.03}^{0.03}$ & 6.29$\pm_{0.01}^{0.00}$ \\
103179 & 16 & 1162 & 0.62 & 108.1 & 10.75  & 11.02$\pm_{0.04}^{0.04}$ & 0.00$\pm_{0.00}^{0.00}$ & 3.22$\pm_{0.29}^{0.41}$ & 5.18$\pm_{0.15}^{0.20}$ \\
103274 & 14 & 1163 & 0.92 & 175.9 & 11.06 & 11.43$\pm_{0.04}^{0.04}$ & 6.12$\pm_{0.47}^{0.74}$  & 1.02$\pm_{0.12}^{0.30}$ & 3.56$\pm_{0.26}^{0.52}$
\enddata
\tablenotetext{*}{We display a table snippet for formatting; the full data table is available to download in the MRT format.}
\end{deluxetable*}

\subsection{Spectro-photometric Modeling} \label{spec-phot-modeling}

In this work, we use two state-of-the-art Bayesian SPS modeling tools, \texttt{Bagpipes} \citep{carnall:18} and \texttt{Prospector}  \citep{johnson:17,Johnson:21}, to perform a full spectro-photometric fitting of the primary LEGA-C galaxies. Details of the two methods used in this work are discussed in sections \ref{bagpipes} and \ref{prospector}. We use \textit{BvrizYJ} bands in our primary spectro-photometric analysis following the conclusions in \citet{vanderwel:21}'s Appendix B. That study finds that using photometry longward of rest-frame $\sim$0.8 $\mu$m disagrees with popular SPS models and leads to systematic errors in fits to the SED of $\sim$20\% of their sample, which in turn propagated into the stellar mass, star formation rate, dust attenuation, and other population parameters. We check the robustness of our results by performing comparisons with two more parametric set of fits altering the photometric data - (1) spectra +\textit{vriz} bands, and  (2) spectra + 20 bands (B to 24 micron, namely - \textit{B, g, IA484, IA527, V, IA624, r, IA679, IA738, i, z, y, J, H, Ks, ch1, ch2, ch3, ch4, mips24)} covering 15 Subaru bands (B to Ks), 4 IRAC bands (ch1 to ch4) and 1 Spitzer MIPS24 band, ranging from UV to far-IR wavelengths. Results of this comparison are shown in Figure \ref{fig:appendix1} in our Appendix A. 

\subsubsection{Bagpipes} \label{bagpipes}
\texttt{Bagpipes} (Bayesian Analysis of Galaxies for Physical Inference and Parameter EStimation) is a SPS modeling package built on the updated BC03 \citep{bruzual:03} spectral library\footnote{\url{www.bruzual.org/~gbruzual/bc03/Updated_version_2016}} with the 2016 version of the MILES library of empirical spectra that includes 2.5 $\si{\angstrom}$ resolution in 3525$\si{\angstrom}$ - 7500$\si{\angstrom}$ wavelength range \citep{falcon-barroso:11}. It is built on a \citet{kroupa:01} IMF assumption and utilizes a Multi-Nest nested sampling algorithm \citep{feroz:19} to produce posterior distributions of physical parameters. We perform parametric full spectral SPS modeling of LEGA-C spectra with \textit{BvrizYJ} UltraVISTA broadband photometry using a double-power law SFH \citep{ciesla:17,carnall:18}, parameterized as: 

\begin{equation}
\text{SFH(t)} \propto \left[\left(\frac{t}{\tau}\right)^{\alpha} + \left(\frac{t}{\tau}\right)^{-\beta}\right]^{-1}
\label{eq1}
\end{equation}


\begin{deluxetable*}{cccc}[t]
\tablecaption{{\tt \texttt{Prospector}} modeling parameters and prior distribution functions  \label{tab:Prospector}}
\tablehead{\colhead{No} & \colhead{Parameter} & \colhead{Description} & \colhead{Prior}}
\startdata
1 & $\sigma_*$ & Velocity dispersion & fixed to LEGA-C values \\
2 & $\log\left(M/\mathrm{M}_\odot\right)$ & total stellar mass formed & uniform in log space: min=7, max=12\\
3 & $\log\left(Z/\mathrm{Z}_\odot\right)$ & stellar metallicity & uniform in log space: min=-1.98, max=0.4\\
4 & SFR ratios & Ratios of adjacent SFRs & Student's t-distribution  ($\sigma=0.3$, $\nu=2$)\\
5 & $z_{spec}$ & redshift & prior: LEGA-C spectroscopic redshift $\pm 0.005$\\
\hline
6 & $\hat{\tau}_{\lambda, 2}$ & diffuse dust optical depth & clipped normal: min=0, max=4, mean=0.3, $\sigma=1$\\
7 & $\hat{\tau}_{\lambda, 1}$ & birth-cloud dust optical depth & clipped normal in ($\hat{\tau}_{\lambda, 1}/\hat{\tau}_{\lambda, 2}$): \\
 & & & min=0, max=4, mean=0.3, $\sigma=1$\\
8 & $n$ & \citet{Kriek_2013ApJ...775L..16K} dust law slope & uniform: min=-1, max=0.4\\
\hline 
9 & $\log\left(Z_\mathrm{gas}/\mathrm{Z}_\odot\right)$ & Gas-phase metallicity & uniform: min=-2, max=0.5\\
10 & $\log U$ & Ionization parameter & uniform: min=-4, max=-1\\
11 & $\sigma_\mathrm{eline}$ & Emission line amplitude & uniform: min=30, max=300\\
\hline
12 & $\nu$ & Spectral white noise & uniform: min=1, max=3\\
13 & $cp$ & Photometric calibration & uniform: min=$10^{-5}$, max=0.5\\
14 & $cs$ & Spectroscopic calibration & uniform: min=$10^{-5}$, max$=0.5$
\enddata
\end{deluxetable*}


\begin{deluxetable*}{cccccccccc}[!t]
\tablecaption{{\tt \texttt{Prospector} } Spectro-photometric modeling results
\label{tab:Prospector-results}}
\tablehead{\colhead{ID} & \colhead{Mask} & \colhead{LEGAC\_ID} & \colhead{$z_{spec}$} & \colhead{$\sigma_{*,cat}$} & \colhead{$log(M_{*,cat}/M_{\odot})$} & \colhead{$log(M_{*,fit}/M_{\odot})$} & \colhead{SFR} & \colhead{t50} & \colhead{t90} \\ {} & {} & {} & {} & {(km/s)} & {} & {} & {($M_{\odot}/yr$)} & {(Gyr)} & {(Gyr)}}
\startdata
103041 & 17 & 1159 & 0.82 & 75.9 & 10.52 &  10.64$\pm_{0.06}^{0.05}$ & 1.27$\pm_{0.47}^{8.80}$    & 3.72$\pm_{0.09}^{0.11}$ & 5.94$\pm_{0.43}^{0.03}$ \\
103061 & 16 & 1160 & 0.72 & 193.1 & 11.03  & 11.37$\pm_{0.02}^{0.02}$ & 0.15$\pm_{0.12}^{2.50}$ & 2.40$\pm_{0.01}^{0.01}$ & 4.66$\pm_{0.19}^{0.36}$ \\
103155 & 19 & 1161 & 0.64 & 103.1 & 11.07 & 11.19$\pm_{0.02}^{0.03}$ & 14.34$\pm_{2.01}^{15.61}$ & 4.32$\pm_{0.66}^{0.52}$ & 6.70$\pm_{0.02}^{0.34}$ \\
103179 & 16 & 1162 & 0.62 & 108.1 & 10.75  & 11.04$\pm_{0.02}^{0.02}$ & 0.19$\pm_{0.13}^{2.05}$ & 2.60$\pm_{0.17}^{0.11}$ & 5.27$\pm_{0.26}^{0.27}$ \\
103274 & 14 & 1163 & 0.92 & 175.9 & 11.06 & 11.35$\pm_{0.04}^{0.04}$ & 3.82$\pm_{1.80}^{7.41}$  & 2.36$\pm_{0.11}^{0.01}$ & 4.83$\pm_{0.36}^{0.04}$
\enddata

\tablenotetext{*}{We display a table snippet for formatting; the full data table is available to download in the MRT format.}
\end{deluxetable*}

This model has three free parameters describing the rising ($\beta$), falling ($\alpha$) and peak ($\tau$) of star formation, whereas other widely used options (e.g. exponential, delayed-tau and log-normal) have two or less free parameters. Hence, by construction, the double-power law SFH has more flexibility. Tau models are shown to fail to recover mock SFHs \citep{carnall:19:a} and simulation SFHs \citep{pacifici:12}. A double-power law SFH is chosen owing to its ability to recover the redshift evolution of cosmic SFRD \citep{behroozi:13, gladders:13, madau:14} and owing to the agreement with simulation results \citep{pacifici:16:a, diemer:17}. The rising and falling slopes essentially do not change beyond the prior ranges chosen (become either flat or vertical) owing to the analytical functional form (Eq.\ref{eq1}) and cover full variations of possible SFHs within this limit. The stellar metallicity is assumed to be the same for all stars born. This value is linearly interpolated on a grid of SSP models and is allowed as a free parameter varying from (0.02,2.5)$Z_{\odot}$ uniformly in linear space. The $Z_{\odot}$ value is assumed to be 0.02 in BC03 models. We test the impact of choosing linear and logarithmic priors on the stellar metallicity on the derived posterior values in our spectro-photometric fits for a subset of galaxies and find no strong deviations. The last free SFH parameter is the stellar mass formed in the entire lifetime of a galaxy until the point of observation (without mass return to the ISM); we allow $\mathrm{log(M_{*}/M_{\odot})}$ to vary logarithmically from (0, 13). Two bursts on top of a double power law are included to account for any abrupt variation in star formation activity. Each burst is given flexibility of age, stellar mass and stellar metallicity and hence three free parameters. We use \citet{Charlot:00} dust model with two free parameters - V-band attenuation and the slope of attenuation. We adopt a second order spectral calibration and white uncorrelated noise for spectral pixels, for which a detailed description can be found in \citet{carnall:19:b}. Dust emission models from \citet{Draine:07} are implemented with fixed $Q_{pah}$ = 2, $U_{min}$ = 1 and $\gamma_{e}$ = 0.01. While testing multiple parameter options and analyzing the full posterior distributions of the output stellar population properties, we found that nebular emission line modeling in Bagpipes biased the stellar metallicities of star-forming galaxies to high values ($\mathrm{log(Z_{*}/Z_{\odot})}$ > 0.35). Additionally, we were concerned that the limiting ionizing radiation of young stars could inappropriately describe emission lines produced by either evolved stars or AGN (and bias the SFR estimates), especially in such a diverse and massive sample of galaxies \citep{carnall:19:b}. Thus, we choose to mask the emission lines from the fit. Stellar velocity dispersion is another free parameter modeled with a variable Gaussian kernel in velocity space. Although the DR3 LEGA-C spectra used in this study are flux calibrated using the UltraVISTA photometry \citep{vanderwel:21}, we include an additional polynomial function of wavelength to address any higher order spectro-photometric calibration uncertainties. We use a second order Chebyshev polynomial function with Gaussian priors - 0th order centered around unity and 1st and 2nd orders centered around zero (see \S4.3.1 in \citet{carnall:19:b} for more details). This modeling requires an average of $\sim$70 CPU hours per galaxy. Further description of all modeling parameters and prior distributions is shown in Table \ref{tab:bagpipes}. Modeling results are included in Table \ref{tab:Bagpipes-results}. We notice a subset (total 126 in the full sample / 4.2\% of the full population) of quiescent population that prefers very similar best-fitting SFHs with ages $t_{mw}\sim3$Gyr and 3.160 $<$ $t_{lw}$ $<$ 3.163. We investigated these objects and found them well constrained in parameter space, spanning a range in empirical properties like spectral indices, UV-VJ colors and redshifts.


\subsubsection{Prospector}  \label{prospector}

\texttt{Prospector} is a Bayesian SPS modeling tool that allows a non-parametric modeling of the SFH of a galaxy in piece-wise constant SFR time bins. \citep{Johnson:21}. It deploys the Flexible Stellar Population Synthesis (FSPS) package with the MILES stellar library and MIST isochrones to model stellar properties \citep{conroy:09}. We use a \citet{chabrier:09} IMF and the \citet{Kriek_2013ApJ...775L..16K} dust law with nebular continuum and line emissions modeled with CLOUDY \citep{ferland:13}. Note that this choice differs from the \texttt{Bagpipes} modeling; we found that modeling nebular emission with Cloudy grids within \texttt{Prospector} produced well behaved posterior distributions. We also tested the impact of including/excluding physical modeling of emission lines in Prospector on a a subset of total 300 galaxies (150 quiescent and 150 star-forming) with significantly-detected emission lines and high signal-to-noise spectra (OII EW > 4 and SN > 12). In general, the recovered  SFHs agree for each galaxy within uncertainties. In a small subset of quiescent galaxies (N=11), non-physical modeling of the emission lines results in maximally old stellar populations that are formed in dramatic, but uncertain, bursts of star formation in the earliest time bin. Given the overall agreement between the two sets of models and slightly more extended SFHs in the aforementioned subset, we include the physical line ratio modeling in our \texttt{Prospector} fitting. We use dynesty \citep{speagle:20} nested sampling option for posterior sampling, similar to \texttt{Bagpipes}. This non-parametric approach is capable of recovering complex SFHs and capturing abrupt star formation processes like sudden quenching and rejuvenation events. On the flip side, fitting both the galaxy spectra and SED is highly computationally expensive and requires about $\sim$100 CPU hours per galaxy to fit an 8 fixed bin SFH model with 14-free parameters.  

In this work, we adopt a continuity prior piece-wise constant SFH with Student's-t distribution that fits the change in log(SFR(t)) in adjacent time bins while weighing against abrupt changes in SFR(t). This prior has also been shown to robustly reproduce mock and more importantly, simulated SFHs \citep{lower:20}. The pioneering work of \citet{ocvirk:06} show that a maximum of 8 episodes of SFH can be independently recovered from an optical spectra of resolution R=10,000, S/N=100 and wavelength coverage $\lambda$ = 4000$\si{\angstrom}$ - 6800$\si{\angstrom}$, with the distinguishability of simple stellar populations proportional to the separation in logarithmic time. Hence we use an 8 time bin SFH model (5 logarithmically spaced) in our analysis. 
The 8 bins of constant SFRs are distributed as follows (in lookback time) -
\begin{center}
0 < t < 30 Myr \\
30 Myr < t < 100 Myr \\
100 Myr < $t_{log1,2,3,4}$ < 0.85 $t_{univ}$ (5 log bins) \\
0.85 $t_{univ}$ < t < $t_{univ}$ \\
\end{center}
The two most recent fixed bins capture signatures of any recent abrupt star formation, one earliest fixed bin corresponding to oldest stellar populations spans the first 15\% of cosmic time, with 5 logarithmically-spaced bins in between. Redshift is set to the LEGA-C spectroscopic redshift with allowed $\pm$ 0.005 variation and stellar velocity dispersion is fixed to the LEGA-C DR3 catalog values. A full description of the free and fixed parameters of the Prospector model and their adopted priors are included in Table \ref{tab:Prospector} and modeling results are reported in Table \ref{tab:Prospector-results}.

\begin{figure}[t]
    \includegraphics[width=0.45\textwidth]{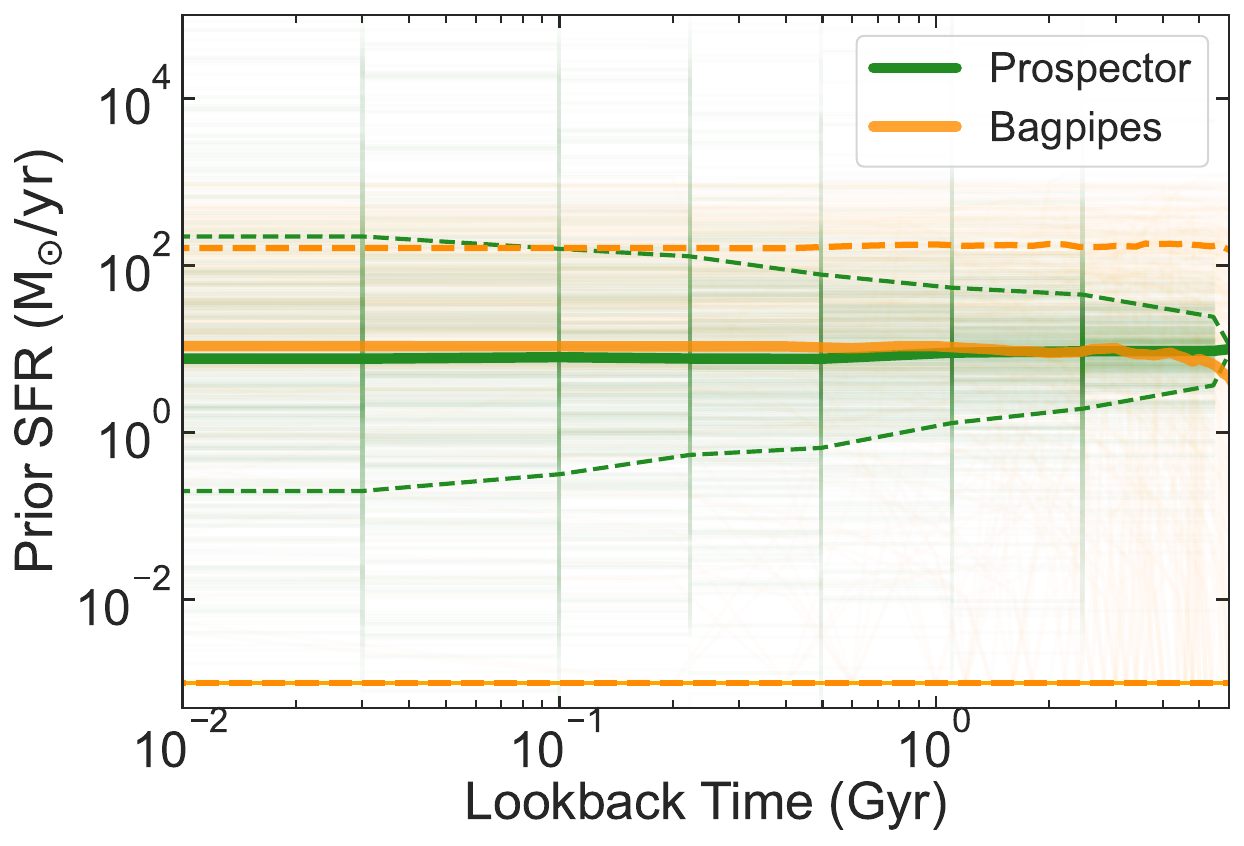}
    \caption{Prior density distribution of SFH obtained by drawing 1000 samples from parametrization described in Table \ref{tab:bagpipes} (\texttt{Bagpipes}) and Table \ref{tab:Prospector} (\texttt{Prospector}). Thick lines show the medians (solid) and dashed 1-$\sigma$ (16th - 84th percentile) scatter. The median values follow closely for both the codes, except at the earliest times the analytic function requires SFR(t=0) = 0 and non-parametric models assign non-zero star-formation in the earliest bin.}
    \label{fig:sfh-prior}
  \hfill
\end{figure}

One might be concerned that priors on the SFHs could drive differences in the inferred SFHs that are derived from the two software packages. We test this by drawing from the prior distributions in each set of models. Figure \ref{fig:sfh-prior} depicts the prior probability density from \texttt{Bagpipes} (orange) and \texttt{Prospector} (green) when 1000 random samples are drawn from the SFH parametrization described in Table \ref{tab:bagpipes} and Table \ref{tab:Prospector} (more details in \S\ref{bagpipes} and \S\ref{prospector}). For ease of comparison, we assign a floor SFR value of 0.001 $M_\odot\;yr^{-1}$ to arbitrarily small SFR values. The median distributions of prior SFHs are shown in solid lines whereas 16th and 84th percentiles are shown with dotted lines. The median values follow closely for both the codes, except at the earliest times the analytic function requires SFR(t=0) = 0 whereas non-parametric models assign non-zero star-formation in the earliest bin. This figure suggests no strong biases in SFRs with lookback time from the priors adopted in the two codes.

As shown in this section, we find the expected differences in the SFHs for individual galaxies as derived by the two modeling methods. We explore the impact of these choices on the full population of massive galaxies in the remainder of the paper.







\begin{figure*}[!h]
\centering
    \includegraphics[width=0.48\textwidth]{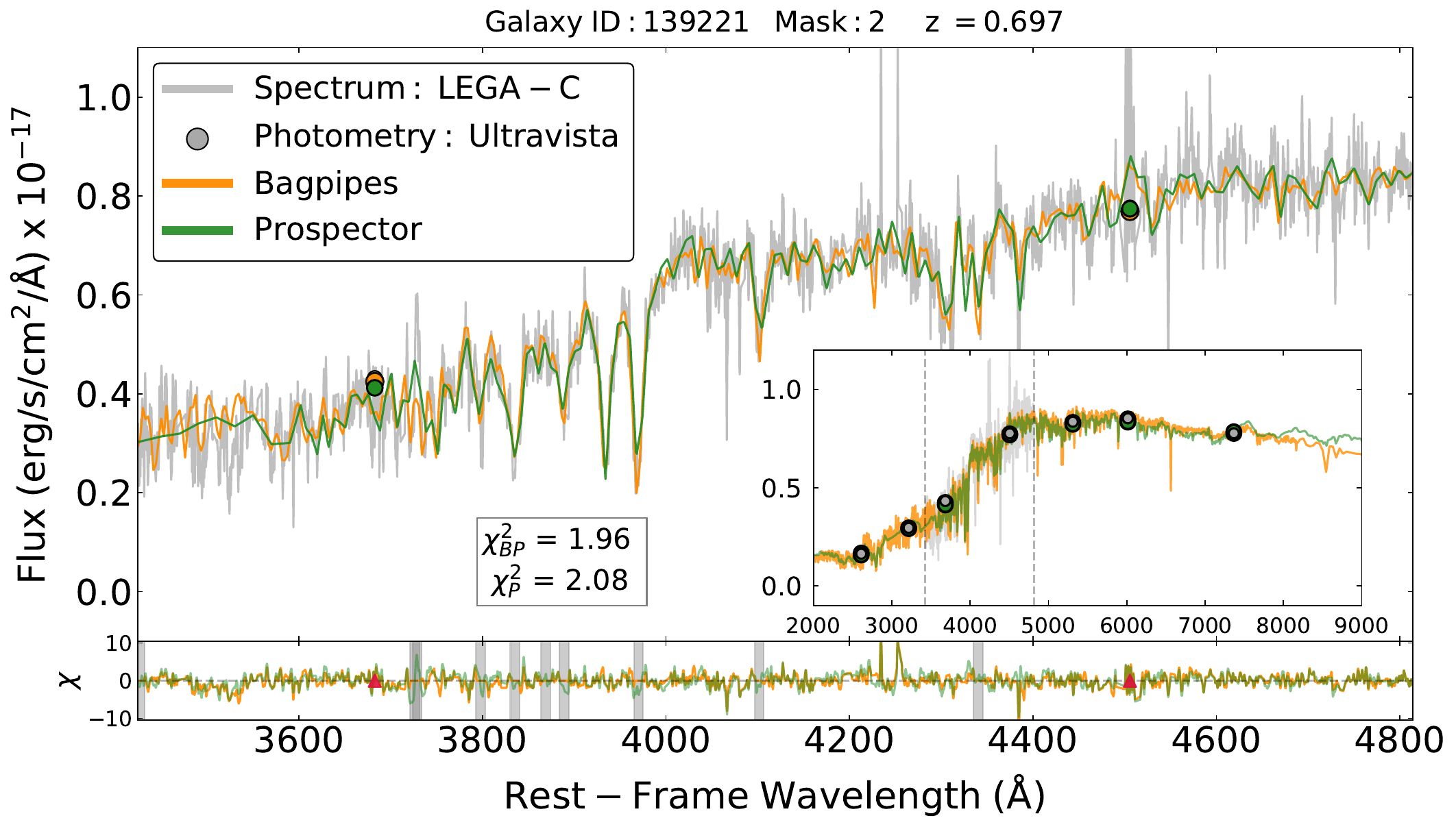}
    \includegraphics[width=0.48\textwidth,height=135pt]{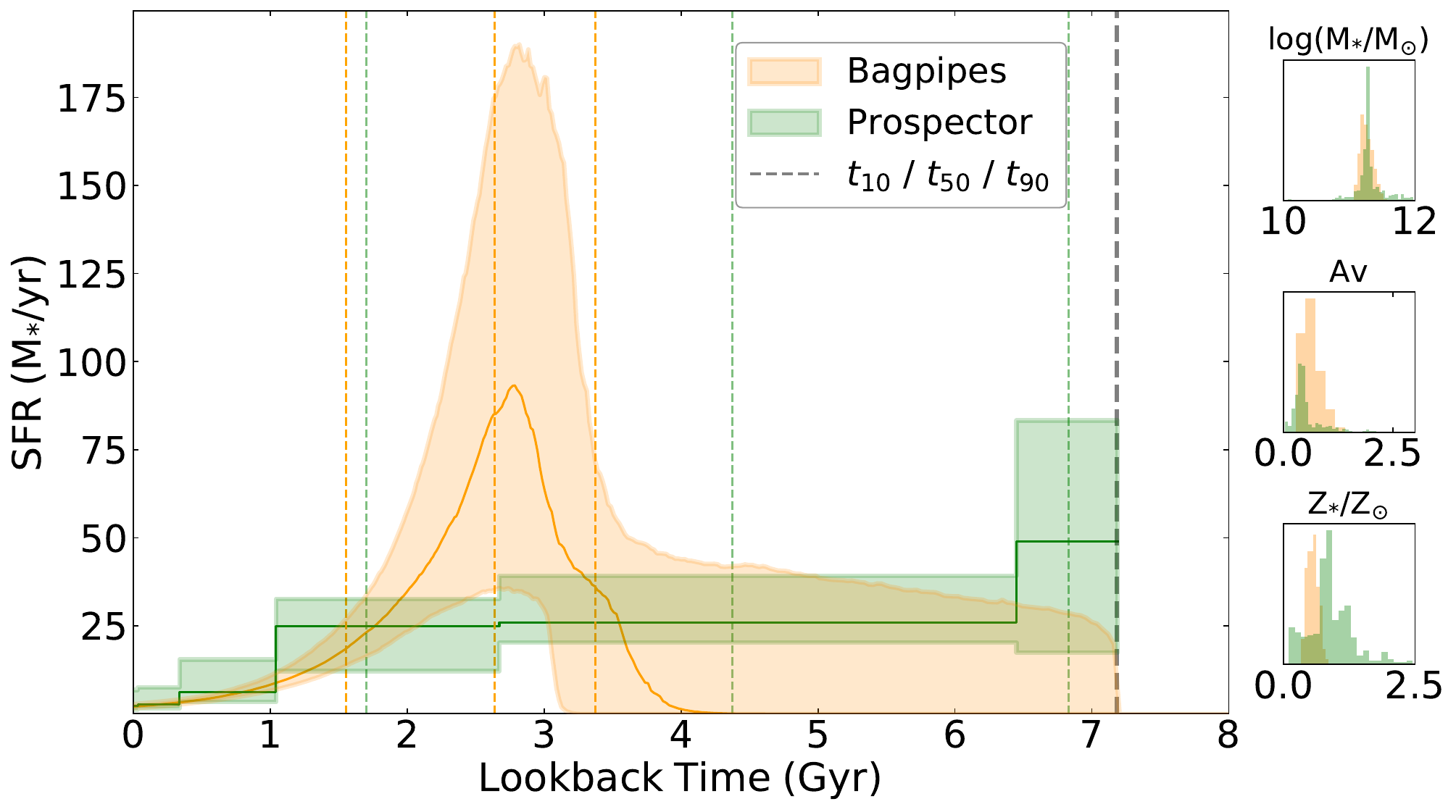}

    \includegraphics[width=0.48\textwidth]{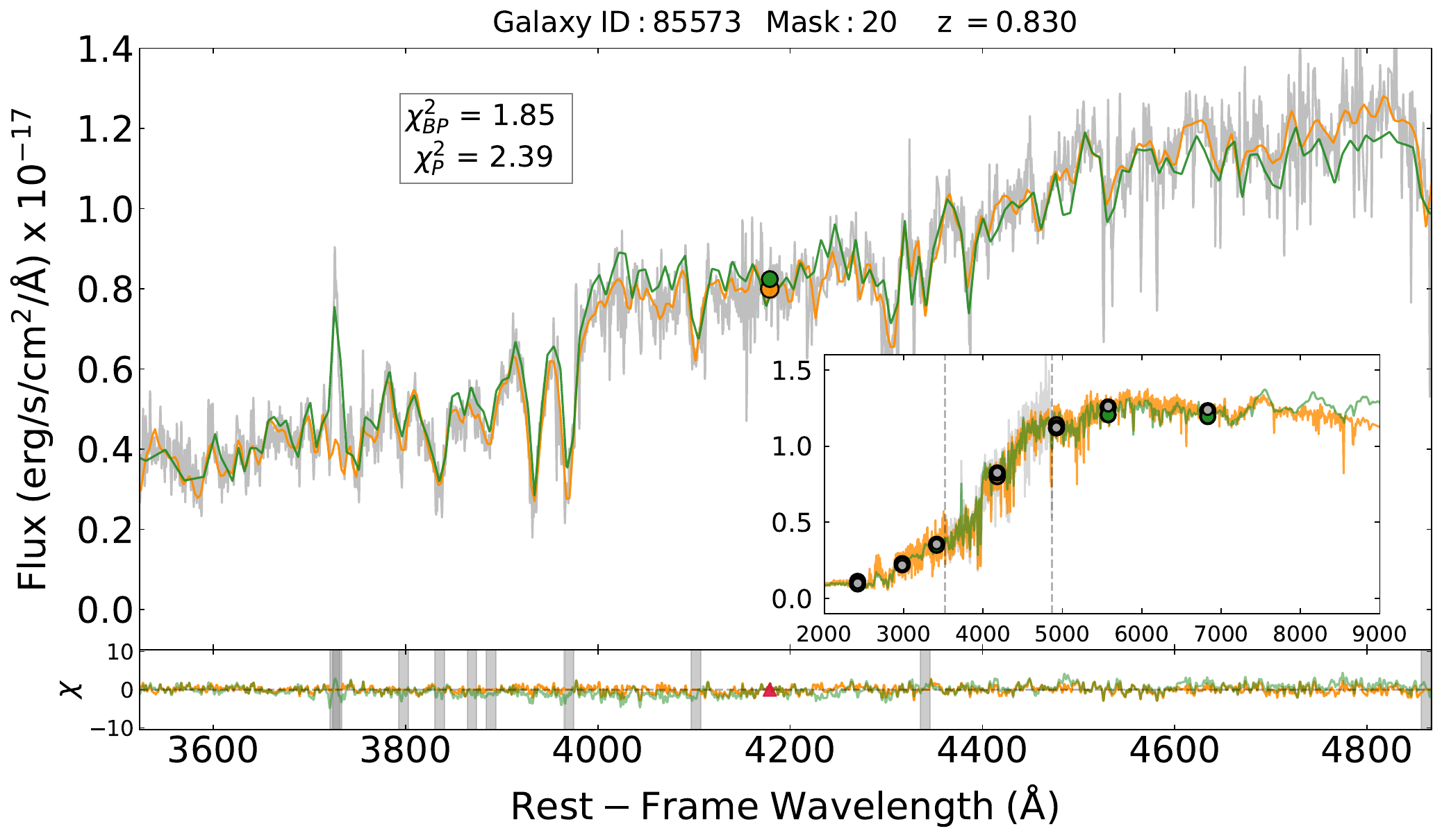}
    \includegraphics[width=0.48\textwidth,height=135pt]{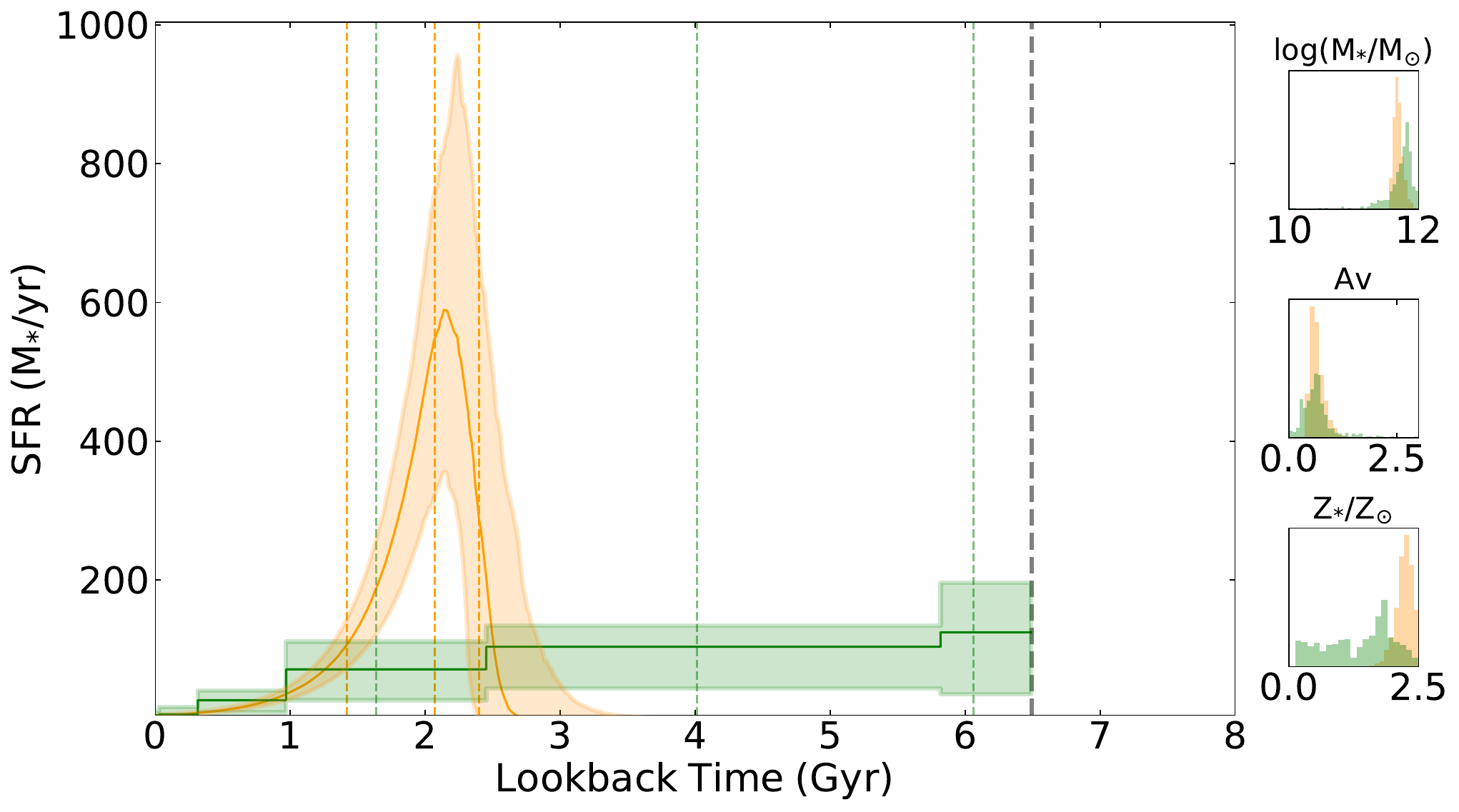}

    \includegraphics[width=0.48\textwidth]{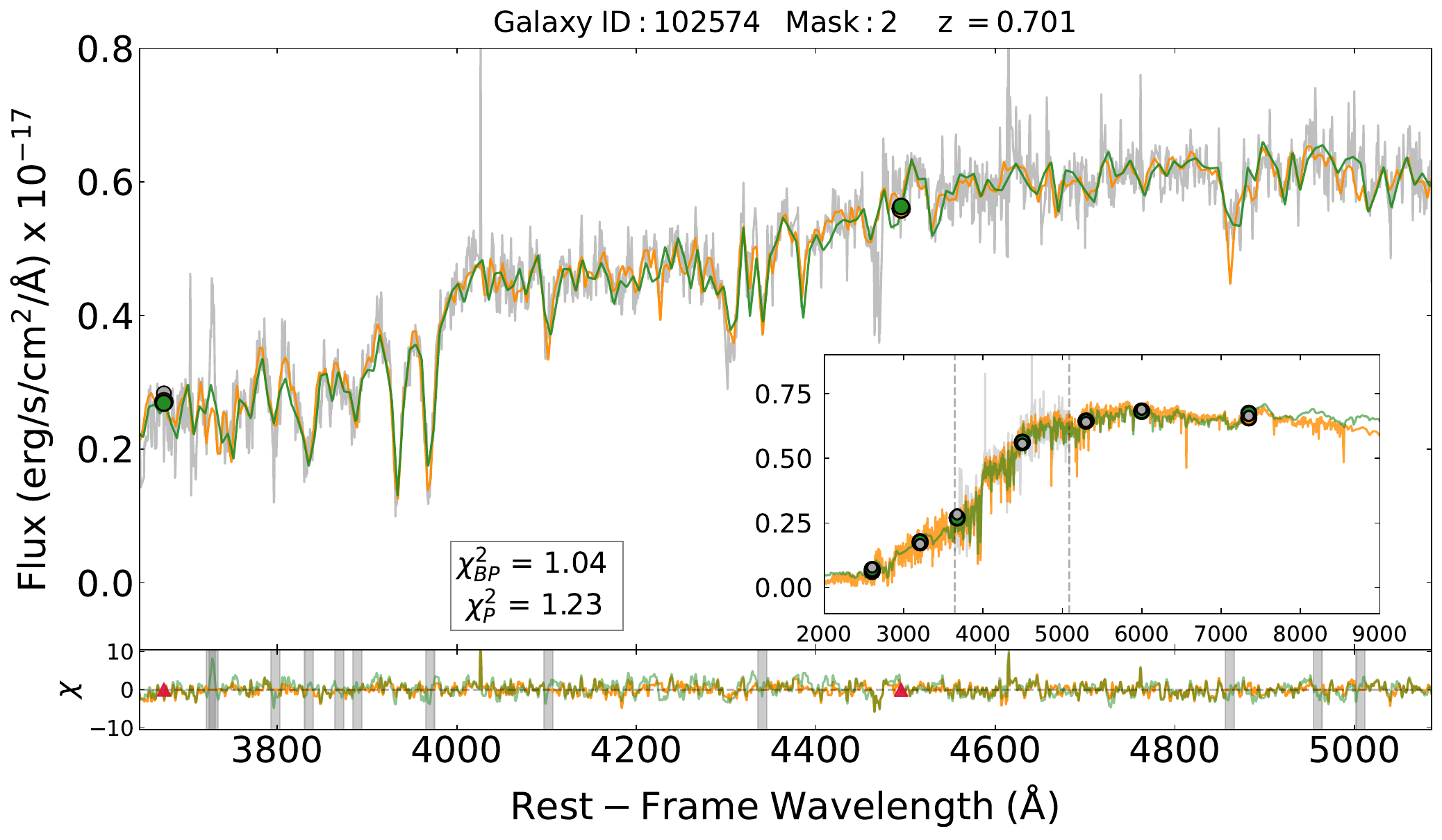}
    \includegraphics[width=0.48\textwidth,height=135pt]{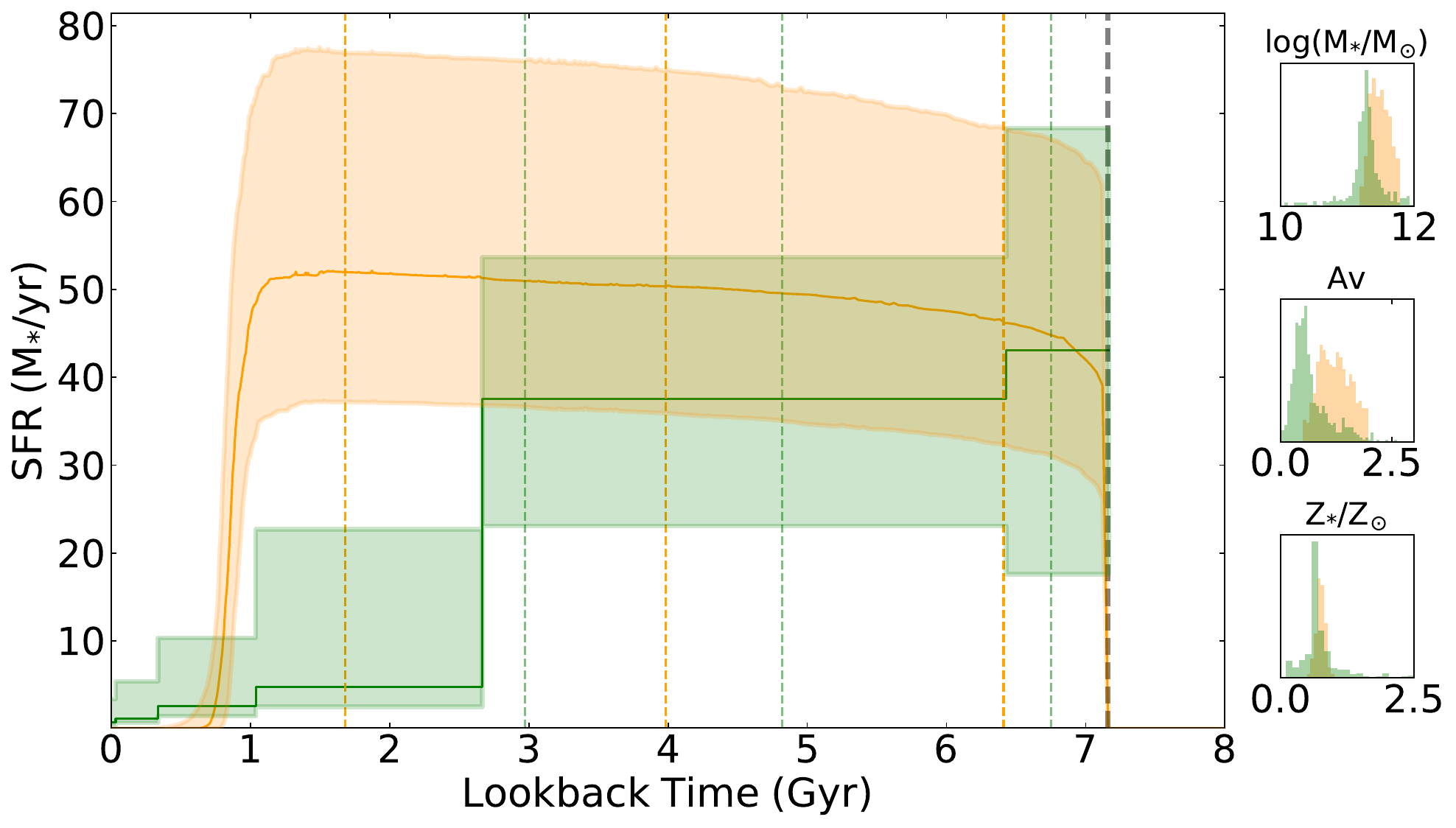}

    \includegraphics[width=0.48\textwidth]{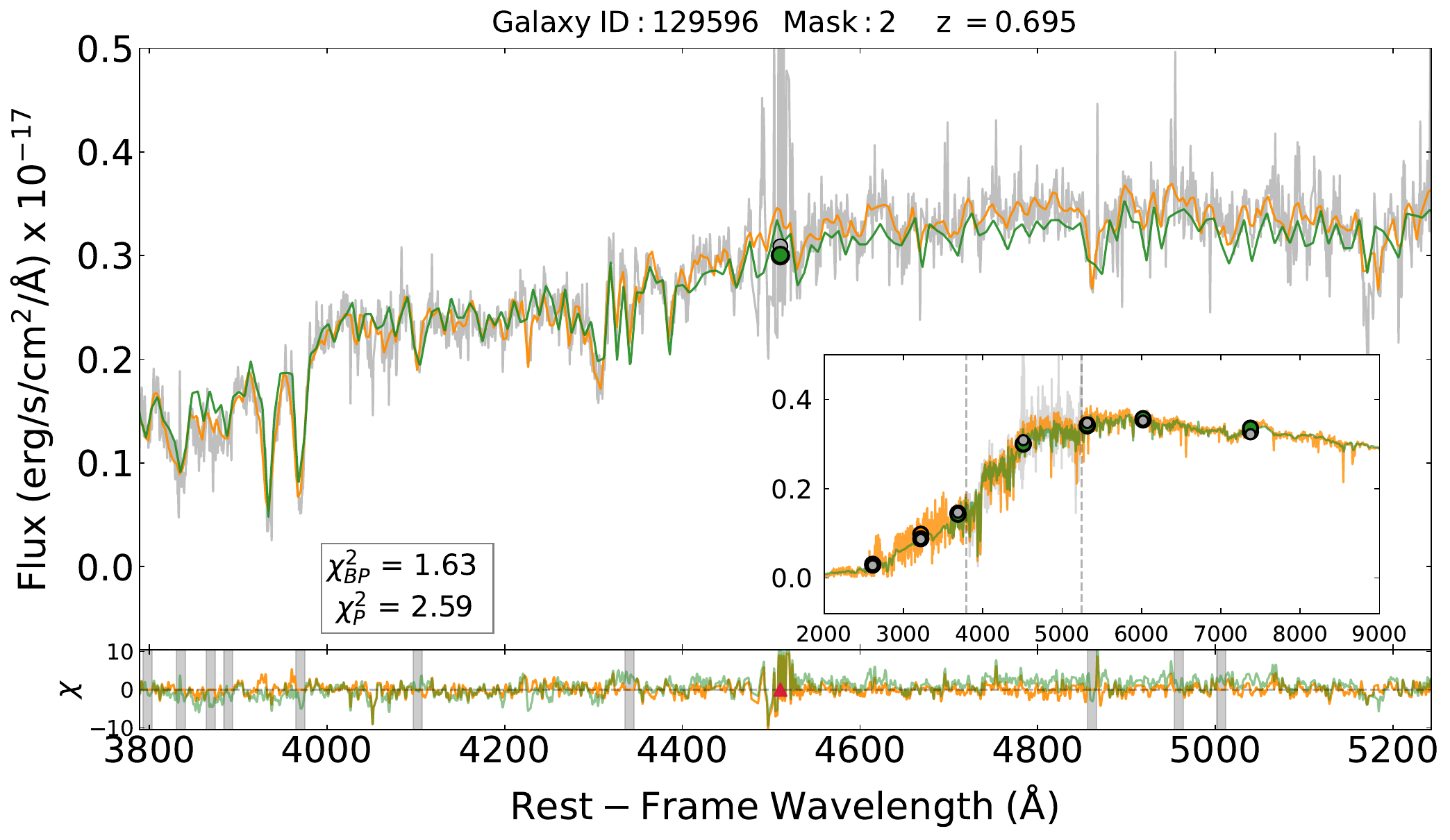}
    \includegraphics[width=0.48\textwidth,height=135pt]{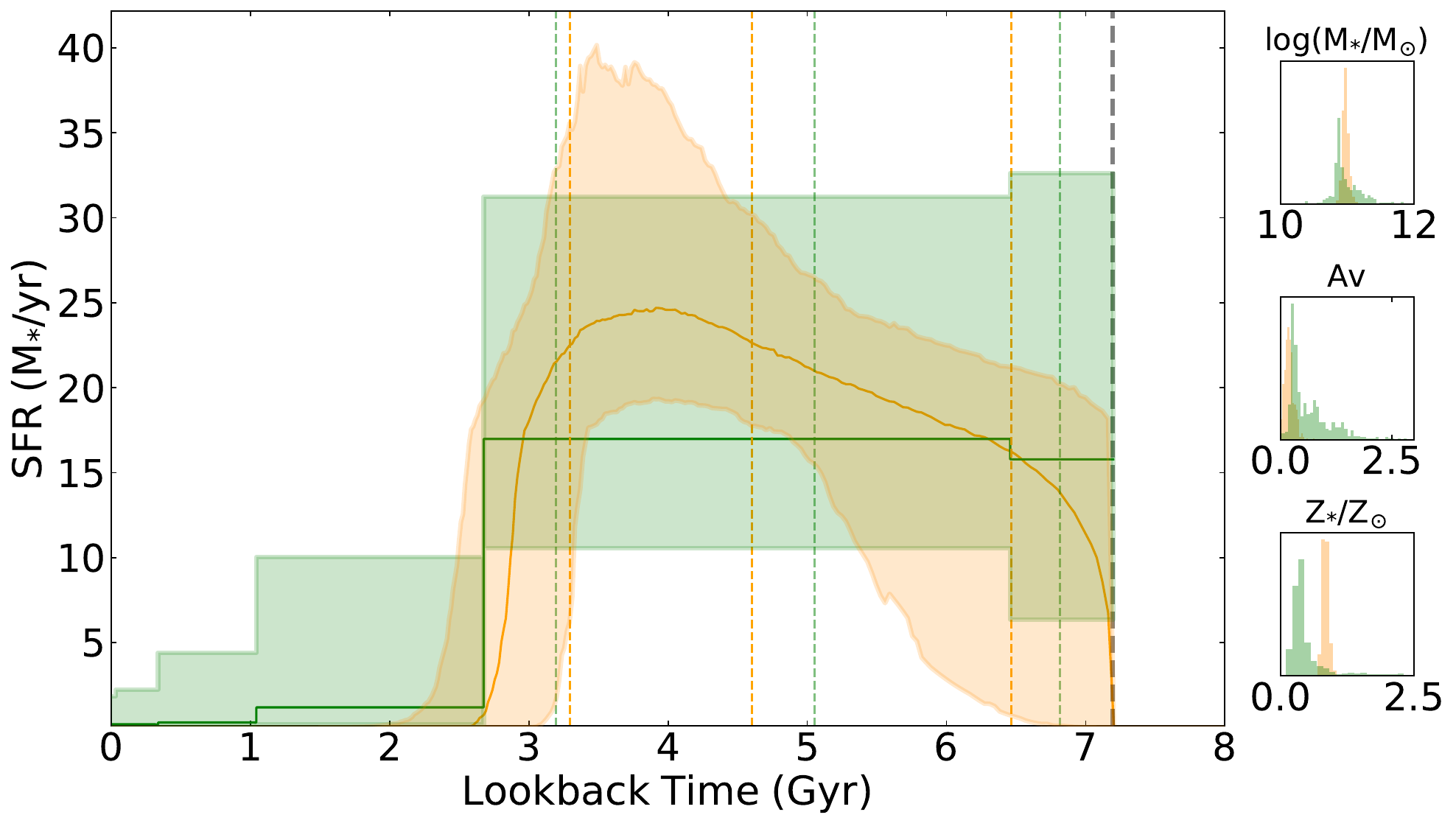}
    
\caption{Spectro-photometric modeling (left) of four representative quiescent galaxies using LEGA-C spectra and UltraVISTA photometry (grey) along with best fit models from \texttt{Bagpipes} (orange) and \texttt{Prospector} (green). Lower sub-panels show $\chi$ values (data-model/error) with grey bands showing regions masked in parametric modeling. Reduced chi-squared values are quoted in inset boxes. The right panels show median SFHs (solid lines) and 16th - 84th percentile distributions of the posteriors (shaded regions). The vertical dashed lines from left to right correspond to 90\%, 50\% and 10\% formation time stamps for each tool (color coded accordingly). The right most column shows the posterior distributions of stellar mass (without mass loss), dust Av, and stellar metallicity. Both modeling techniques yield good fits to the data and overall the SFHs agree reasonably well in late times ($t_{90}$), with some divergence at early times ($t_{10}$).}
    \label{fig:quiescent-demo}
\end{figure*}


\begin{figure*}[!t]
\centering

    \includegraphics[width=0.48\textwidth]{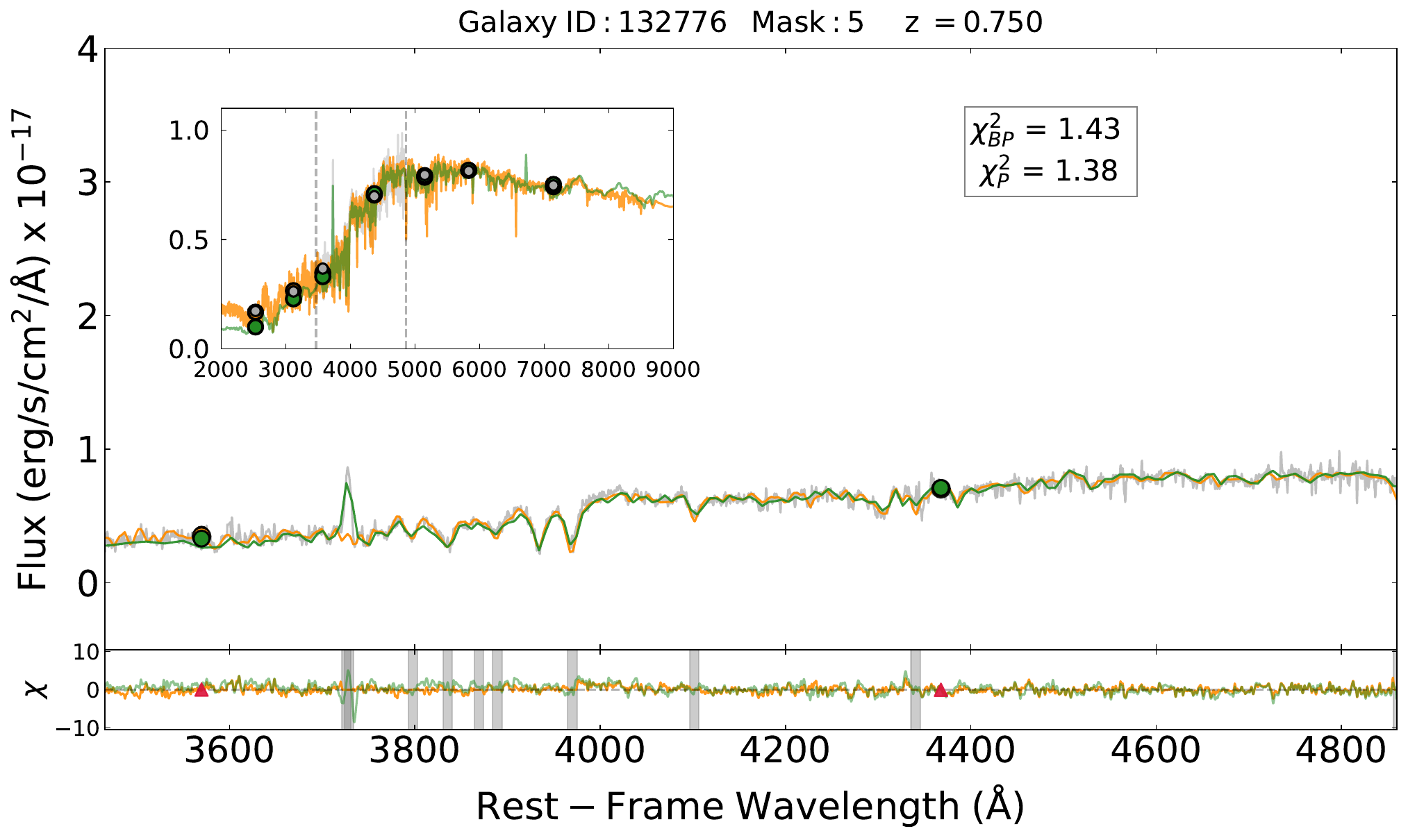}
    \includegraphics[width=0.48\textwidth,height=140pt]{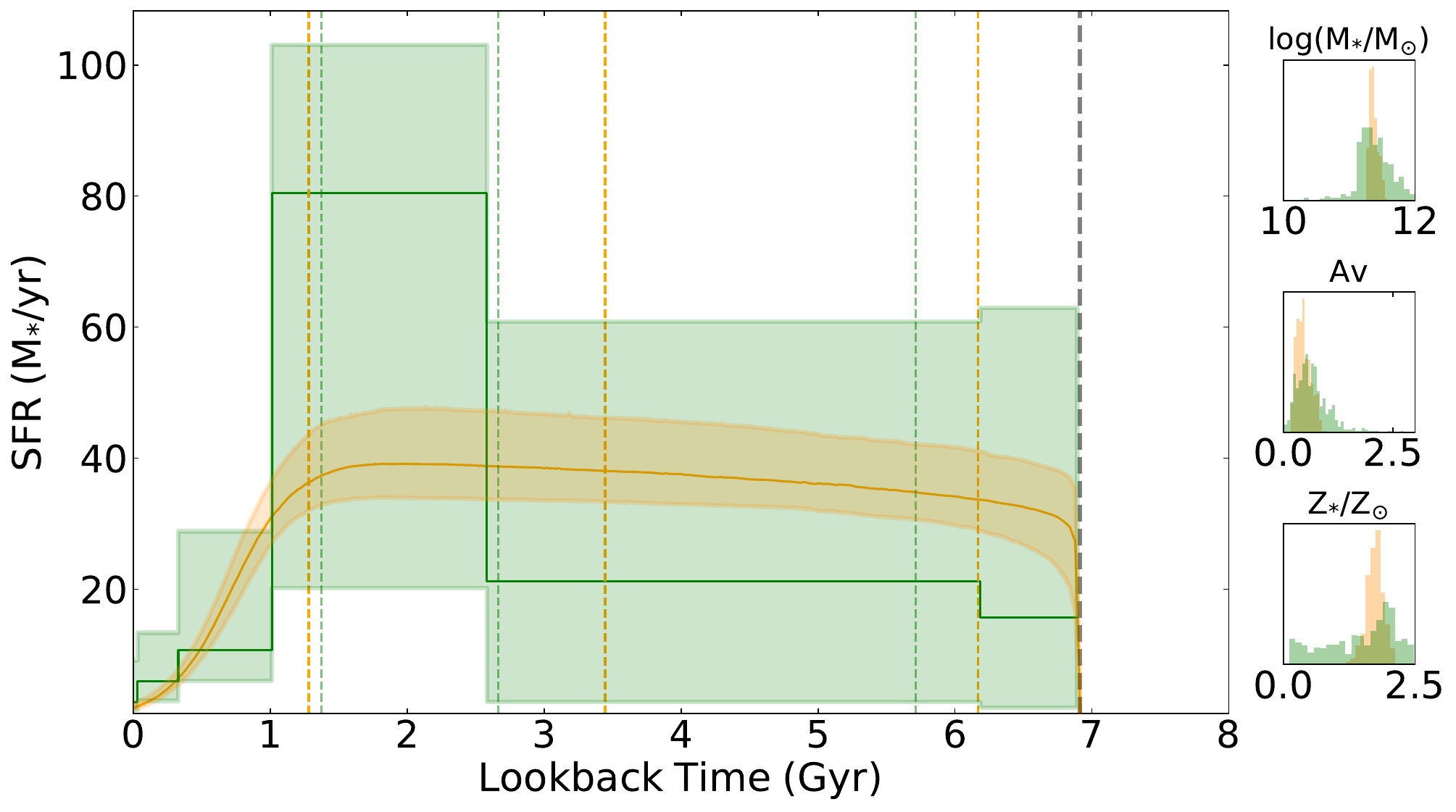}

    \includegraphics[width=0.48\textwidth]{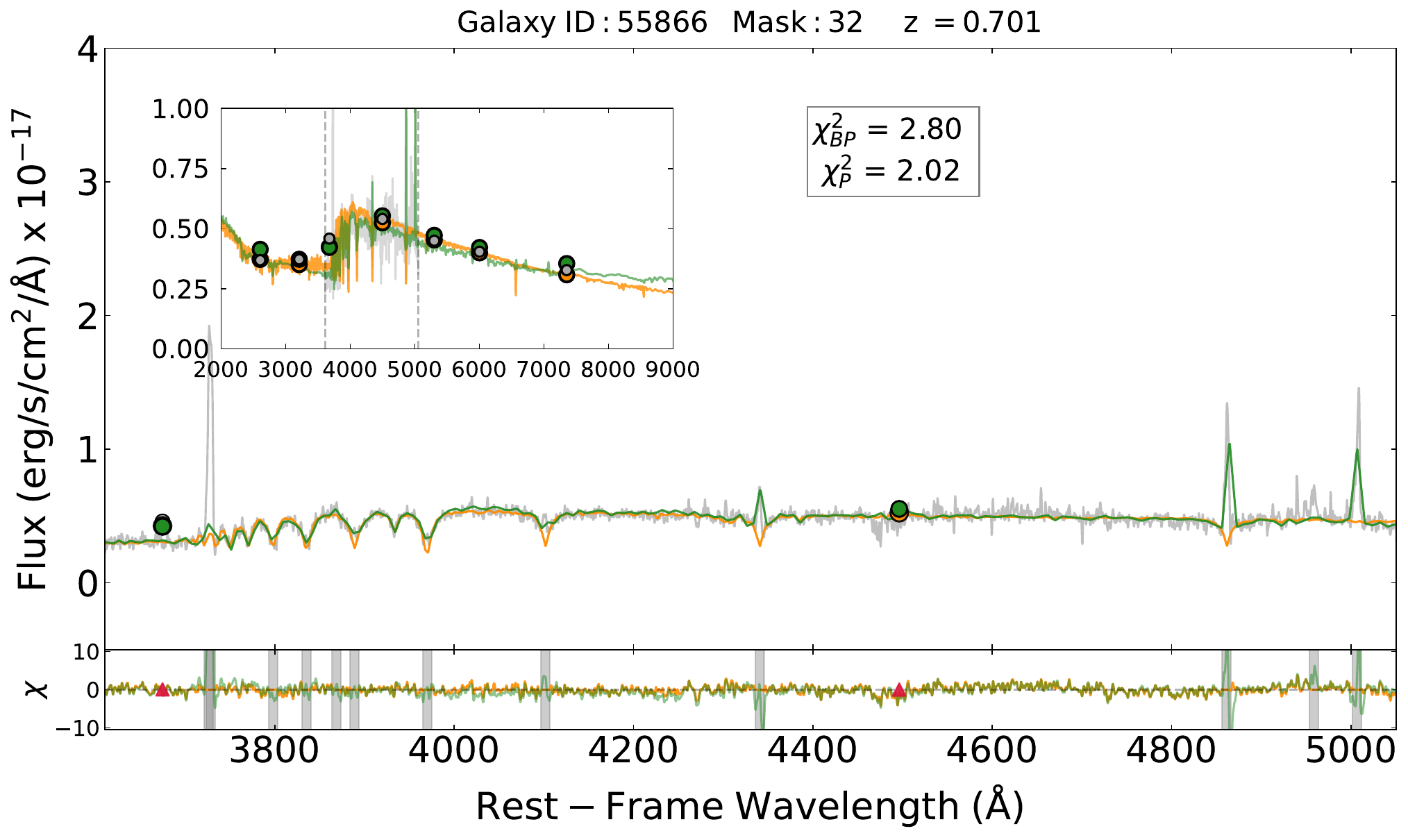}
    \includegraphics[width=0.48\textwidth,height=140pt]{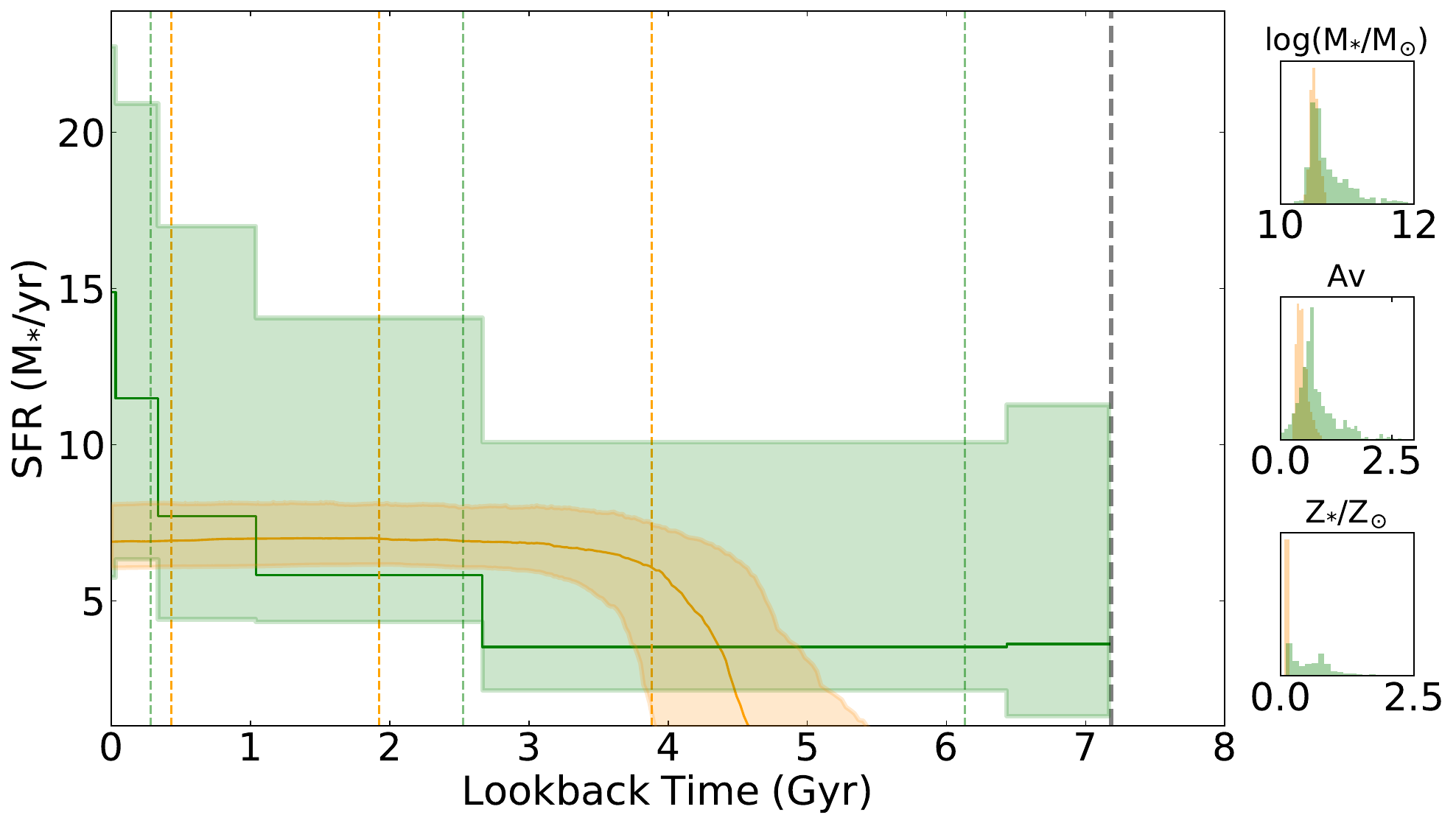}

    \includegraphics[width=0.48\textwidth]{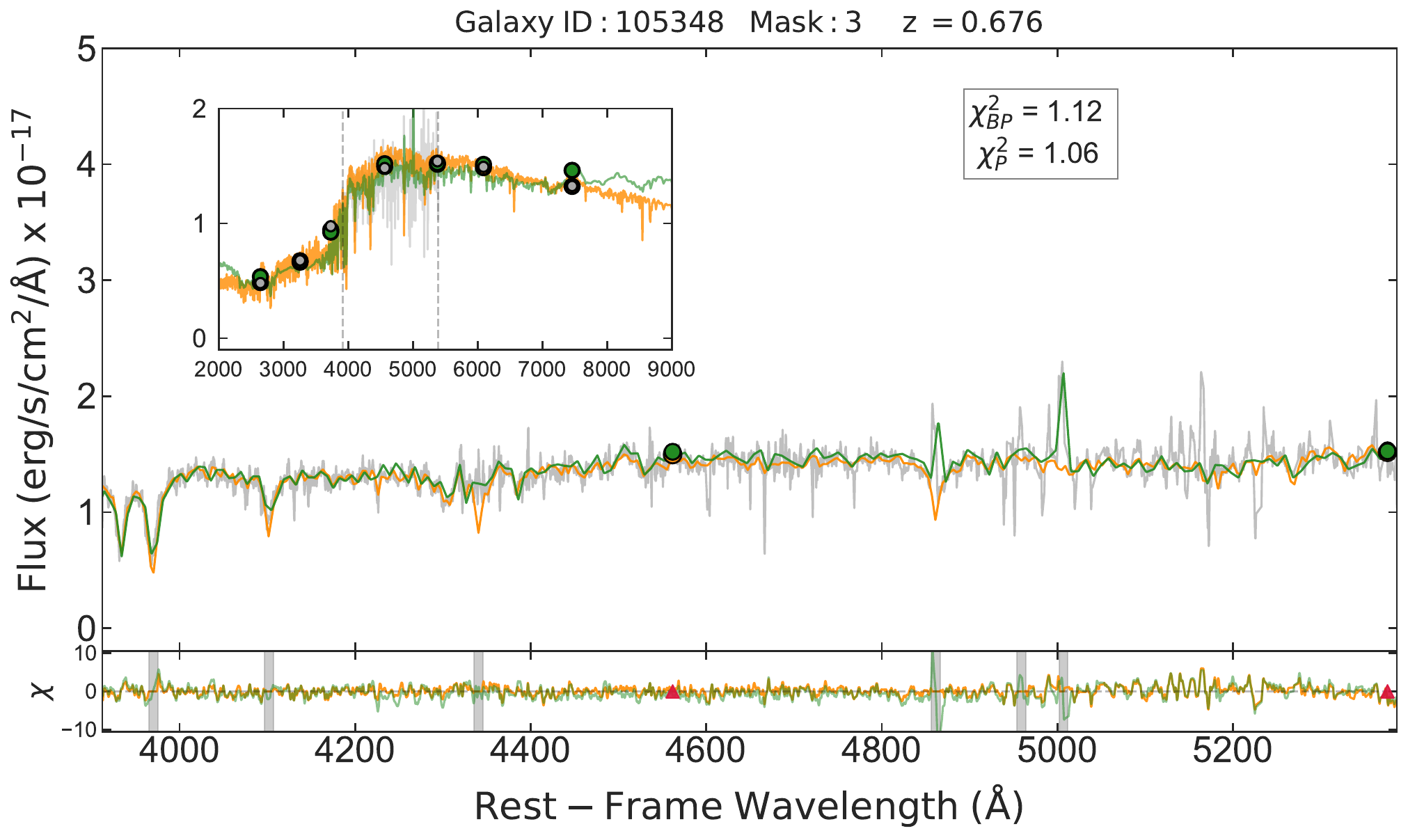}
    \includegraphics[width=0.48\textwidth,height=140pt]{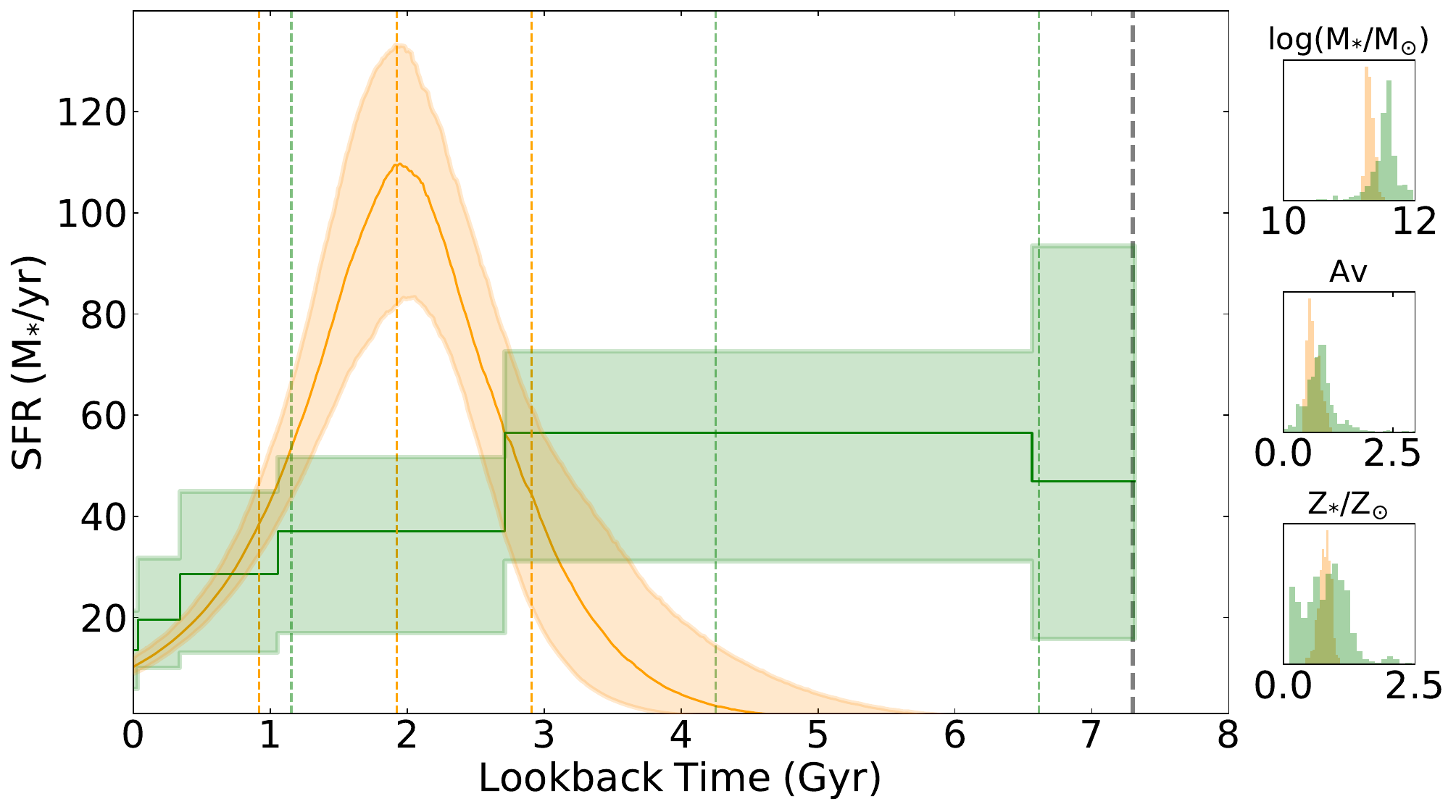}

    \includegraphics[width=0.48\textwidth]{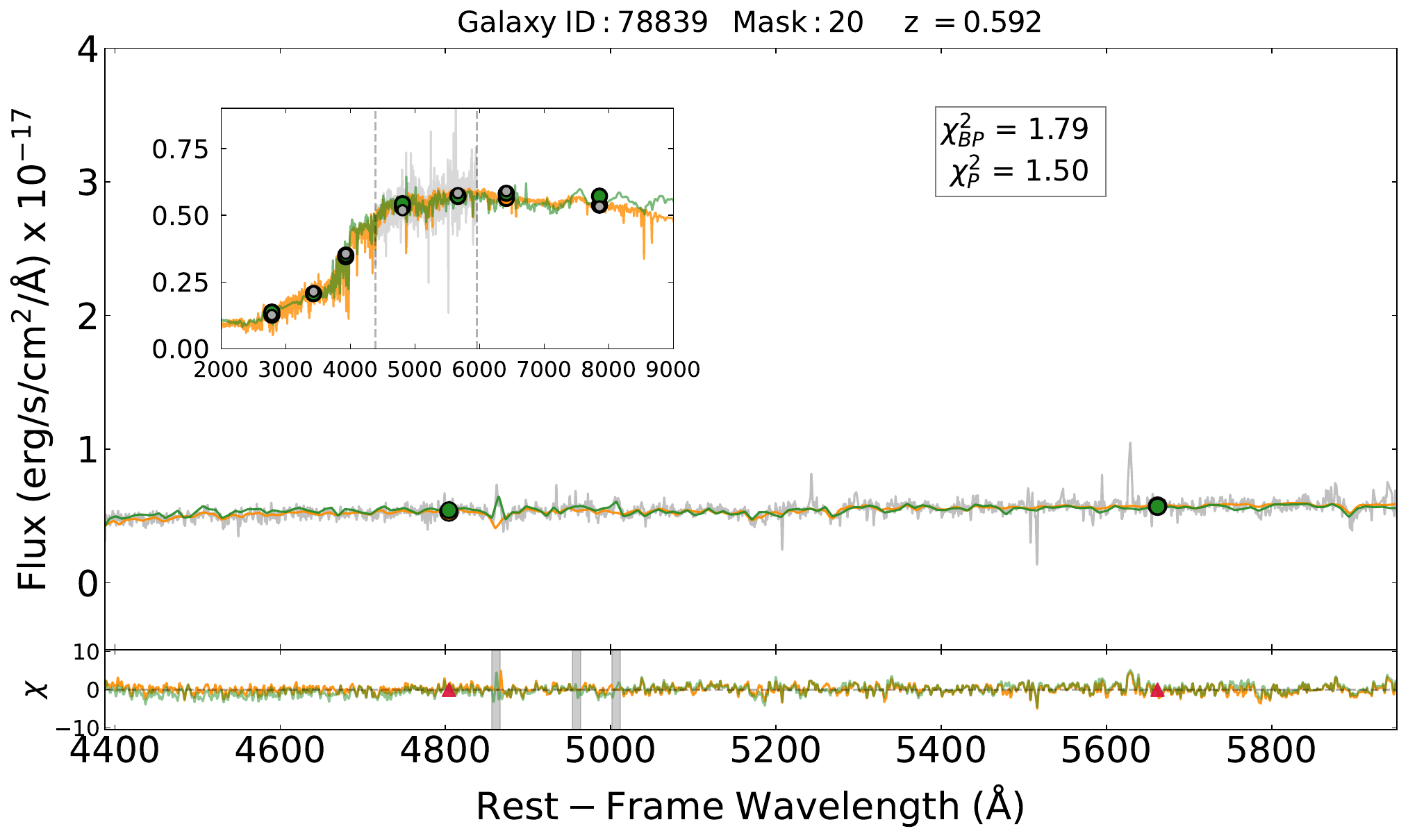}
    \includegraphics[width=0.48\textwidth,height=140pt]{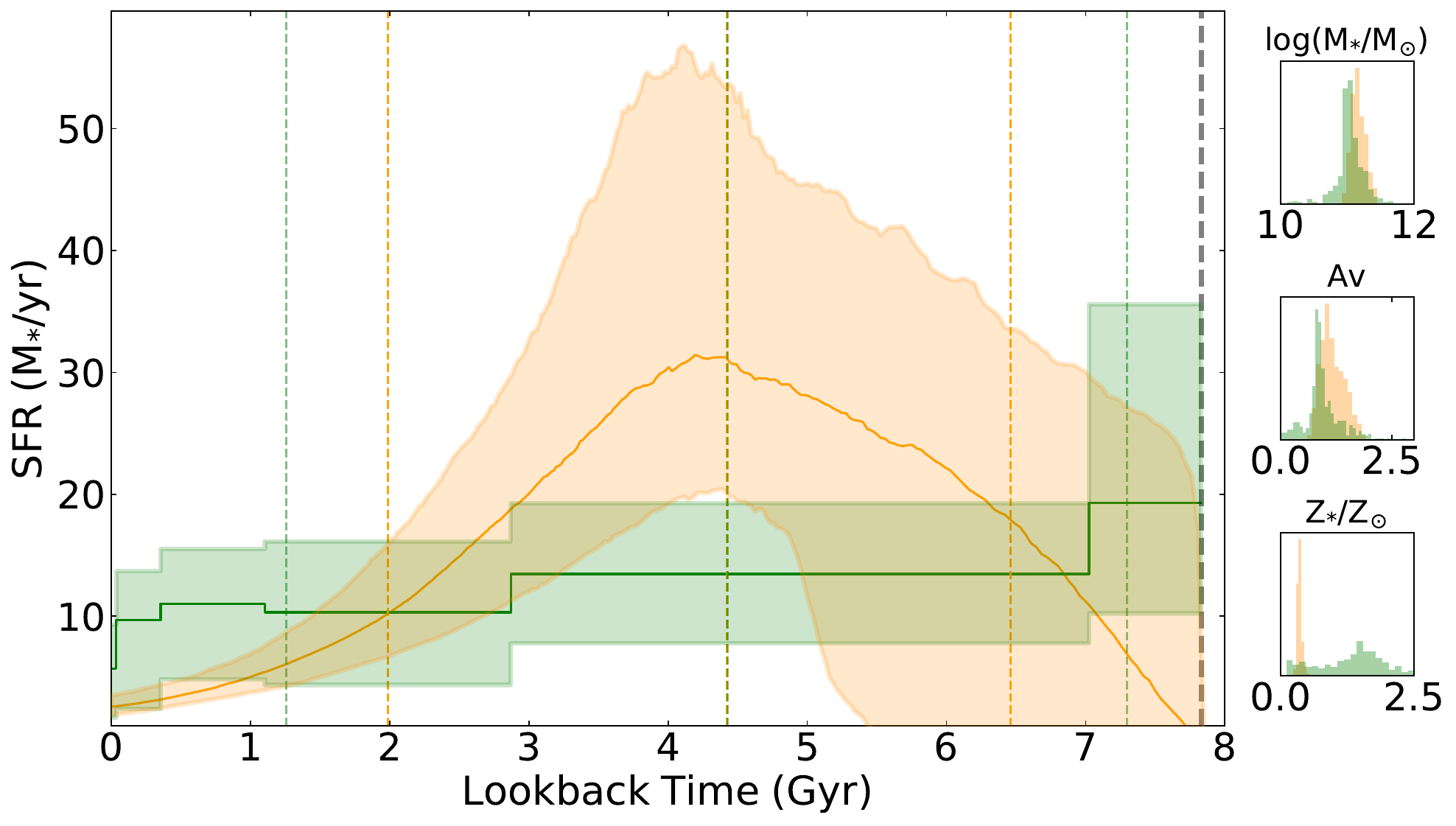}

\caption{Four representative star-forming galaxies with modeling outputs in the same format as described in Figure \ref{fig:quiescent-demo}. In general, parametric and non-parametric SFHs for individual star-forming galaxies agree reasonably well at all times.}
    \label{fig:star-forming-demo}
\end{figure*}

\subsection{Modeling Results and Examples}

First, we emphasize that all SFHs are allowed to start from the Big Bang, however the \textit{analytic} SFHs (\texttt{Bagpipes}) naturally exhibit more flexibility in onset time (with negligible star-formation in early times for some cases) and slope than the early bins in piece-wise constant \textit{non-parametric} SFHs (\texttt{Prospector}). Although in principle the first bin in the latter models could exhibit negligible star formation, the fits prefer at least some non-zero average SFR at the earliest times. Also, since bins represent SFRs averaged over an extended period of time, they are more likely to be non-zero. This discrepancy is partially driven by differences in modeling choices and also likely reflects a lack of constraining power in the data at the earliest times, even with the high S/N of the LEGA-C dataset. Fundamentally such information is in the prior dominated regime; in this paper we quantify the ultimate impact of popular choices on aggregate SFHs of the populations.


Fig.~\ref{fig:quiescent-demo} and Fig.~\ref{fig:star-forming-demo} show representative examples of spectro-photometric modeling and recovered SFHs of four quiescent and four star-forming galaxies respectively. In the left column, each panel shows an observed LEGA-C spectrum and UltraVISTA photometry (grey), along with the best-fitting models from \texttt{Bagpipes} (orange) and \texttt{Prospector} (green). The insets show the full \textit{BvrizYJ} photometric SEDs and models. We note that the models fit both the spectroscopic and photometric data very well. The residuals are quantified in $\chi$ values in the bottom panel ($\chi$ = (observed flux - model flux)/error). Grey bands indicate regions that are masked to avoid emission lines in the \texttt{Bagpipes} modeling. The right large panels show the corresponding SFHs. Note that we chose to logarithmically scale the lookback time (horizontal axis) to highlight the most robustly measured epochs at late times.

For quiescent galaxies (Fig.~\ref{fig:quiescent-demo}), we note that both models provide excellent fits to the observed data and median SFHs agree reasonably well, more so in late times than early times, when quantified in the formation time metrics $t_{10}$, $t_{50}$ and $t_{90}$ ($\Delta t_{10}|_{median}$ = 2.05 Gyr, $\Delta t_{50}|_{median}$ = 1.47 Gyr and $\Delta t_{90}|_{median}$ = 0.22 Gyr with uncertainties (calculated from posteriors) of order of 0.27/0.13/0.20 Gyr in each delta metric respectively. Note that these are representative of the full LEGA-C quiescent population statistics ($\Delta t_{10}|_{median}$ = 2.41 Gyr, $\Delta t_{50}|_{median}$ = 1.38 Gyr and $\Delta t_{90}|_{median}$ = 0.23 Gyr). One subtle difference emerges at the earliest times when \texttt{Prospector} consistently assigns finite star formation in the first time bin leading to older stellar populations, whereas it is not necessarily the case for parametric SFHs in \texttt{Bagpipes}. For star-forming galaxies (Fig.~\ref{fig:star-forming-demo}), we see better agreement in median SFRs at later times compared to quiescent examples above, with differences in timescales within the uncertainties. ($\Delta t_{10}|_{median}$ = 0.49 Gyr, $\Delta t_{50}|_{median}$ = 0.28 Gyr and $\Delta t_{90}|_{median}$ = 0.14 Gyr with uncertainties of order of 1.25/0.84/0.22 Gyr in each delta metric respectively). Note that these are representative of the full LEGA-C star-forming population statistics ($\Delta t_{10}|_{median}$ = 1.30 Gyr, $\Delta t_{50}|_{median}$ = 0.34 Gyr and $\Delta t_{90}|_{median}$ = 0.17 Gyr). Further mass and sigma trends of the full LEGA-C sample would be analyzed in the following sections.

\section{The Star Formation Histories of Massive Galaxies} \label{section3}
\begin{figure*}[t!]
\centering
    \includegraphics[width=\textwidth]{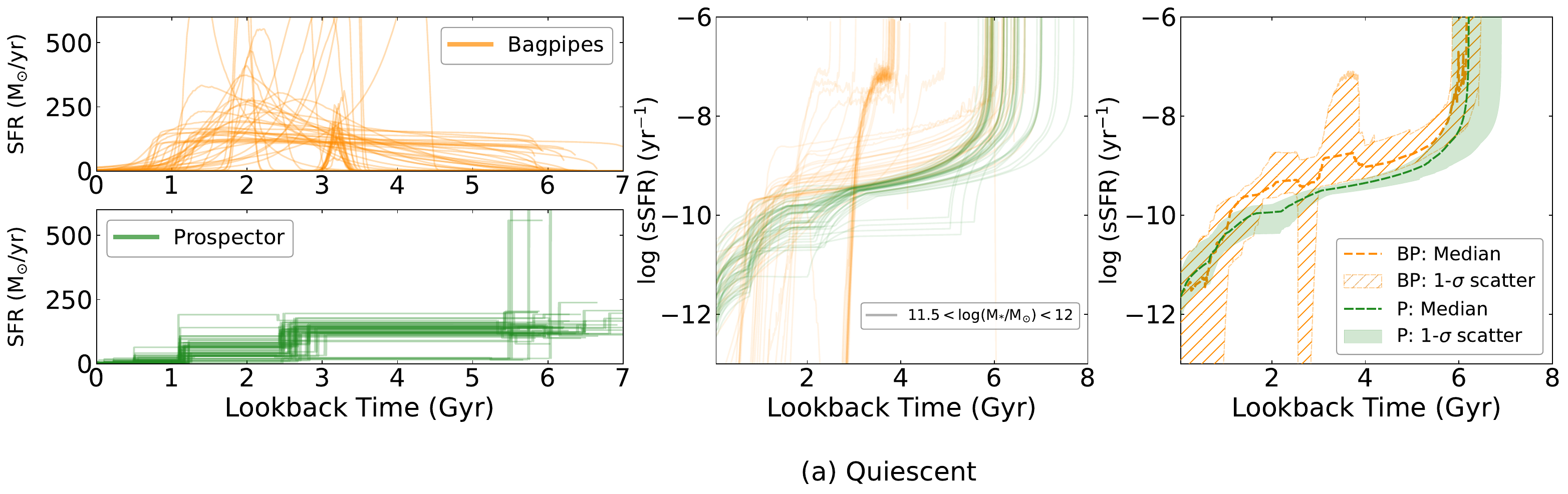}
    \includegraphics[width=\textwidth]{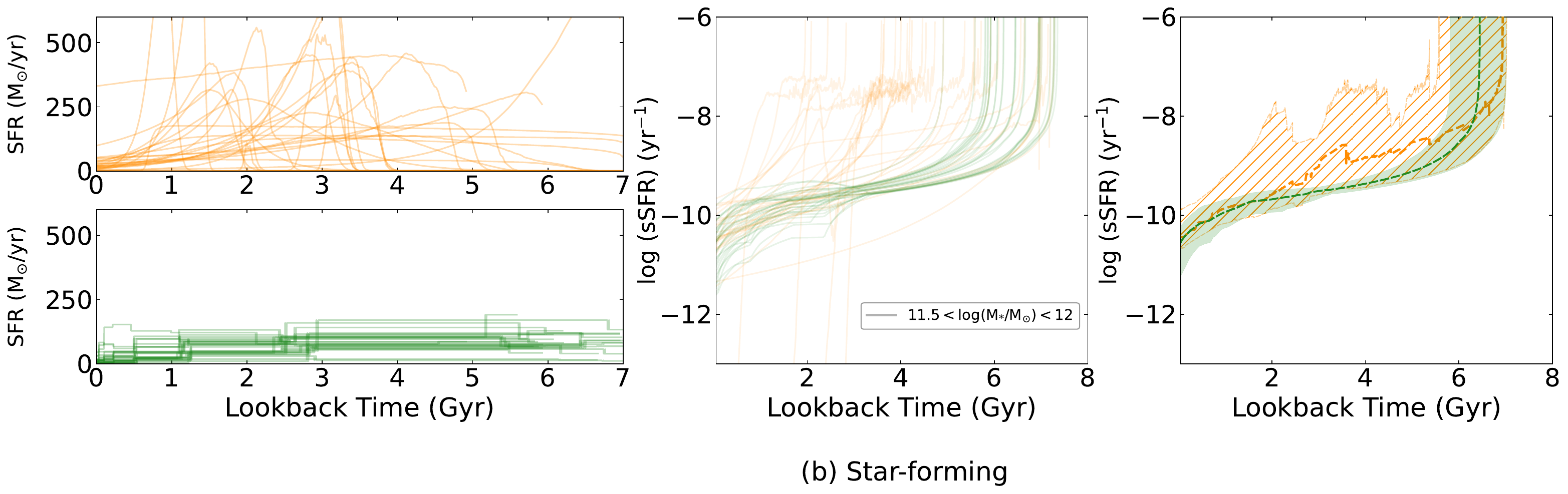}
    \caption{Individual (left) and population (center and right) SFHs recovered from \texttt{Bagpipes} and \texttt{Prospector} for two example samples of massive (11.5 $<$ log($\mathrm{M_*/M_{\odot}}$) $<$ 12) quiescent galaxies (top row (a)) and star-forming galaxies (bottom row (b)). The middle panel shows individual SFHs in the sSFR and the right panel shows 16th, 50th and 84th percentile population distributions. Note that \texttt{Bagpipes} generates relatively higher population scatter due to variation in the onset and truncation of the double power law parametrization, in contrast with the early onset of star formation inferred from non-parametric models.}
    \label{fig:ssfr-example}
\end{figure*}

In this section, we combine the median star-formation histories derived for each galaxy in the LEGA-C sample to characterize the overall growth histories of massive galaxies. We first interpolate all SFHs on a uniform age grid from (0.01, 8) Gyr and impose a minimum star-formation rate (floor) of 0.001 $M_\odot\;yr^{-1}$ to arbitrarily small SFR values. For each galaxy, we take the median SFH from the posterior distributions and combine these medians to calculate population medians in bins of - (1) stellar mass and (2) stellar velocity dispersions from LEGA-C DR3 catalog. For ease of presentation, we first focus on three \textit{coarse} stellar mass bins representative of LEGA-C primary sample - a) 10.5 < log($\mathrm{M_*/M_{\odot}}$) < 11 b) 11 < log($\mathrm{M_*/M_{\odot}}$) < 11.5 c) 11.5 < log($\mathrm{M_*/M_{\odot}}$) < 12 and four coarse stellar velocity dispersion bins - a) 100 < $\mathrm{\sigma_*}$(km/s) < 150 b) 150 < $\mathrm{\sigma_*}$(km/s) < 200 c) 200 < $\mathrm{\sigma_*}$(km/s) < 250 and d) 250 < $\mathrm{\sigma_*}$(km/s) < 300 and later increase the resolution to \textit{finer} bins to characterize population trends. Note that the mass and sigma values used to bin galaxies are taken from the LEGA-C DR3 catalog (LOGM\_MEDIAN,SIGMA\_STARS)and are mass loss corrected, described in detail in \S{2}. These catalog values have mean offsets of +0.05 dex and -0.02 dex with \texttt{Bagpipes} and \texttt{Prospector} modeling results respectively. In each mass/velocity dispersion bin, each individual SFH is weighted by $T_{cor}$ (which includes both a volume and sample correction factor, for details see \citet{vanderwel:21} Appendix A) in the calculations of population median and scatter to account for the LEGA-C survey targeting strategy. 

To demonstrate our approach to combining the posteriors from individual fits, we show the individual median and population trends in an example subset of LEGA-C data in Figure \ref{fig:ssfr-example}. The top row shows individual median SFHs from the two methods for the most massive quiescent galaxies in the stellar mass bin 11.5 < log($\mathrm{M_*/M_{\odot}}$) < 12. \texttt{Bagpipes} SFHs are shown in orange and \texttt{Prospector} SFHs are shown in green. The middle and right panels show these SFHs in specific star-formation rate (log (sSFR)) as a function of time. sSFR at an epoch is obtained by dividing SFR at that epoch by the total stellar mass formed up until that epoch excluding mass loss. The center panel includes median SFHs for individual galaxies and the right panel collapses these to show only the population distributions. The bottom row shows these trends for most massive star-forming population. Note that \texttt{Bagpipes} has more variety in onset, duration and quenching of star-formation for individual galaxies, which are often imprinted on the population trends even though the individual parametric SFHs are smooth. In addition to the 126 galaxies with questionable fits that are discussed in Section \S{2}, there is a significant number of quiescent galaxies in \texttt{Bagpipes} with $t_{50}$ formation times between 3 Gyr to 4 Gyr, as evident in the SFHs in Figure \ref{fig:ssfr-example} (similarly for other mass bins, see Appendix Figure \ref{fig:ssfr1}). 

\begin{figure*}[!t]
  \centering
    \includegraphics[width=\textwidth]{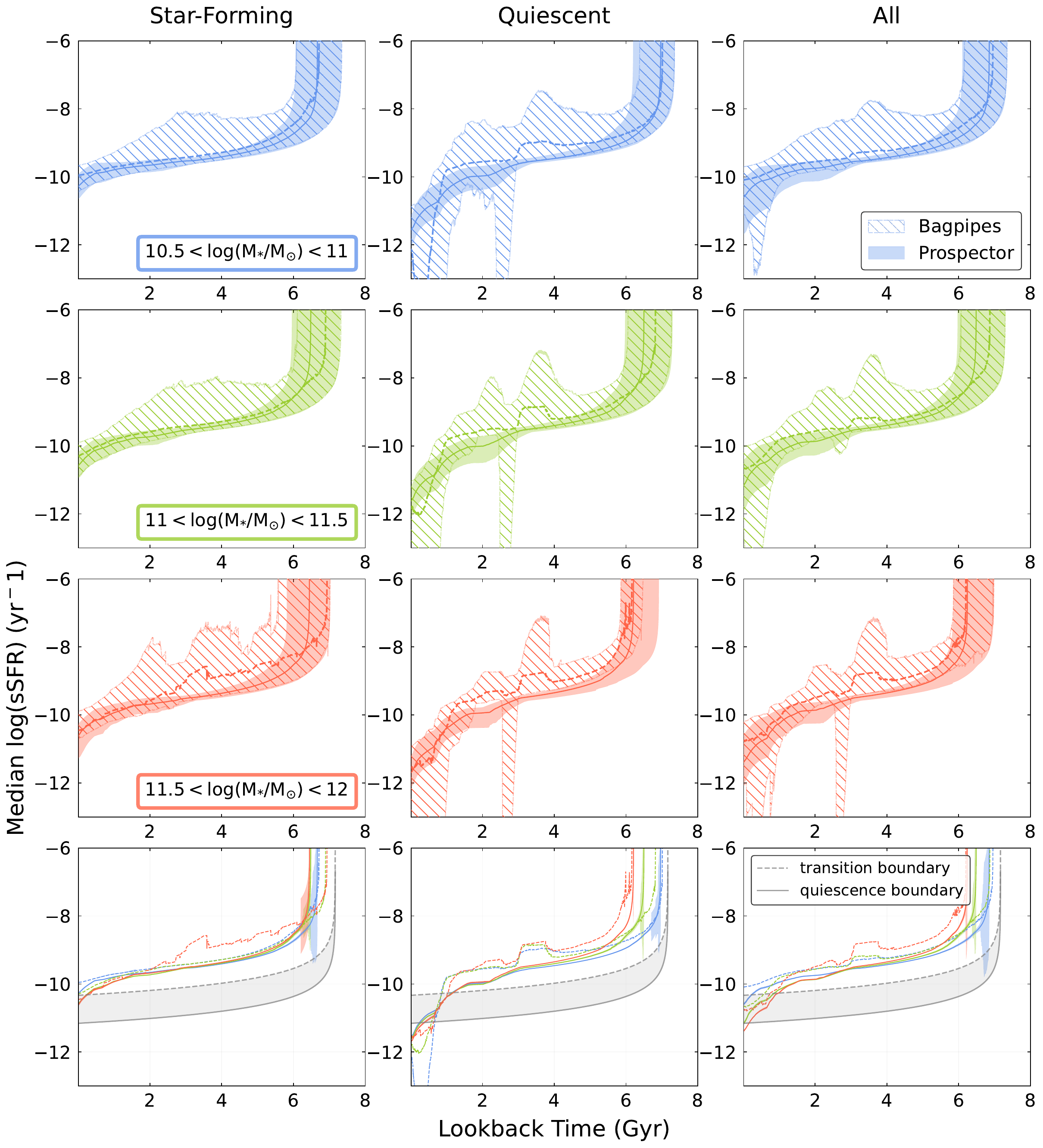}
    \caption{Median specific star-formation rates (log($\mathrm{sSFR}$)) with lookback time for populations of star-forming (left), quiescent (center) and all (right) galaxies sorted in bins of their stellar mass. Colors indicate stellar mass bins - blue < green < red in increasing mass. \texttt{Bagpipes}/ \texttt{Prospector} results are shown in dotted-lines/solid-lines with hashed/shaded regions representing 16th-84th percentile population scatter. Bottom most row shows median trends of all mass bins with thin shaded regions representing standard error on the medians via bootstrapping. Transition region from star forming to quiescent is shown in grey band. Star-forming population shows overall better agreement between two methods than quiescent population. Note that \texttt{Bagpipes} has more bursty pathways to reach quiescence.}
    \label{fig:raw-sfh}
  \hfill
\end{figure*}

\begin{figure*}[!h]
  \centering
    \includegraphics[width=0.98\textwidth]{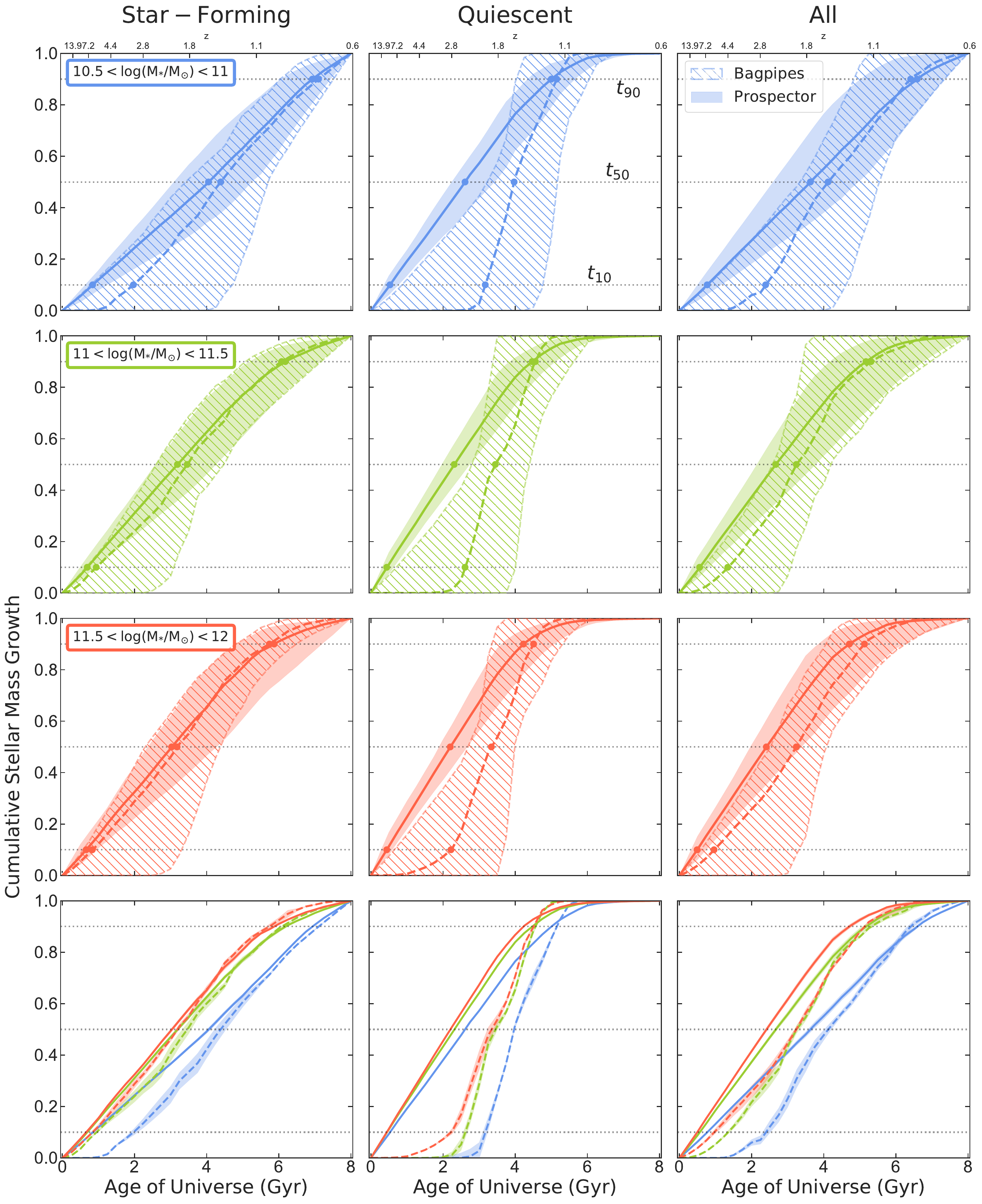}
    \caption{The cumulative stellar mass fractions versus time for star-forming (left), quiescent (center) and all (right) galaxies sorted by stellar mass. The color and plotting scheme is the same as in Fig 4. This represents a complete census as individual galaxies are weighted by correction factors to reflect the LEGA-C targeting. Lower mass galaxies grow more slowly than their massive counterparts. Non-parametric SFHs (Prospector, solid) start earlier and are more extended, especially for quiescent galaxies.}
    \label{fig:cumulative-sfh}
  \hfill
\end{figure*}

Figure \ref{fig:raw-sfh} expands the right panel in Figure \ref{fig:ssfr-example} to show the $\mathrm{sSFR}$ evolution in three stellar mass bins (rows) for: star-forming galaxies (left), quiescent galaxies (center), and the full population (right). Each panel shows the population medians (dashed lines/solid lines) and 16th - 84th percentile population scatter (hashed/filled regions) for \texttt{Bagpipes} and \texttt{Prospector} respectively. The bottom row combines the median trends of all mass bins with thin shaded regions showing 1-$\sigma$ error on the medians calculated from bootstrapping. We adopt the transition and quiescence boundaries as (1/3 $t_H$) and (1/20 $t_H$) from \citet{tacchella:22}, where $t_H$ is the Hubble time at $z_{median} = 0.8$. Although individual galaxies show a variety of SFHs, the population median trends are largely independent of stellar mass, with slight trends at early times that are dominated by the demographics of the LEGA-C sample and modeling priors. In part, differences in the onset of star formation can be partially due to slight differences in the mass-dependent redshift distributions of the LEGA-C sample. However, as discussed in \S2, the modeling priors also introduce subtle differences. \texttt{Bagpipes} consistently measures slightly higher median sSFRs, especially within the quiescent population and shows higher population scatter compared to Prospector. Median sSFR trends within the star-forming population show better agreement between the two codes. To quantify this, for each mass bin, we calculate the median of the differences between the curves. In Figure \ref{fig:raw-sfh}, from top to bottom - median $\Delta sSFR(t)|_{SF}$ = 0.08, 0.12, 0.31 $yr^{-1}$ (left column), median $\Delta sSFR(t)|_{Q}$ = 0.22, 0.25, 0.29  $yr^{-1}$ (middle column) and median $\Delta sSFR(t)|_{ALL}$ = 0.11, 0.18, 0.30  $yr^{-1}$ (right column). For star-forming galaxies, both codes infer that massive galaxies fall to lower sSFRs at the point of observation than the less massive ones. Some differences are seen in the median trends of the most massive galaxies (red) from the two codes that could be attributed to small sample size. For quiescent galaxies, we see a large diversity of \texttt{Bagpipes} sSFR tracks. We do not find that the population of massive quiescent galaxies shut off before lower mass counterparts. Note that this is different from local universe findings \citep{gallazzi:06,nelan:06}. 

Figure \ref{fig:cumulative-sfh} focuses on the median trends in the cumulative stellar mass growth and adopts the same plotting conventions as Figure \ref{fig:raw-sfh}, except the horizontal axis now shows the age of the Universe, starting at the Big Bang. The hashed and filled regions show the 16th - 84th percentile distributions of individual median SFHs from \texttt{Bagpipes} and \texttt{Prospector} respectively. Horizontal grey dotted lines correspond to 10\%, 50\% and 90\% formation thresholds. These timescales, which we hereby define as $t_{10}$, $t_{50}$, $t_{90}$, are widely used metrics to quantify the timescales of stellar mass growth in a galaxy including all progenitors up to the point of observation \citep{behroozi:19,pacifici:16:b,weisz:11,weisz:08}. These values are quantified for the full primary sample in Table \ref{tab:Prospector} and Table \ref{tab:Prospector-results}. As we find in the previous two figures, parametric models in general show higher population scatter than non-parametric ones due to galaxy to galaxy variations. The bottom row compiles all median trends to compare the rate of stellar mass growth across different masses. This cumulative view highlights both mass trends within the populations and discrepancies at early times due to modeling degeneracies and prior assumptions. At any given time, low mass galaxies have grown less than massive galaxies. Significant discrepancies in $t_{10}$, and to some degree in $t_{50}$, suggest that SFHs are prior dominated at large lookback times. Perhaps similar deep spectroscopic observations of galaxies at even earlier times (e.g. with JWST) will help to break modeling degeneracies and resolve this issue, but at this epoch, the measurements of the earliest SFHs of massive galaxies will remain prior dominated, even with high-S/N spectroscopic data.

\begin{figure*}
\centering
    \includegraphics[width=\textwidth]{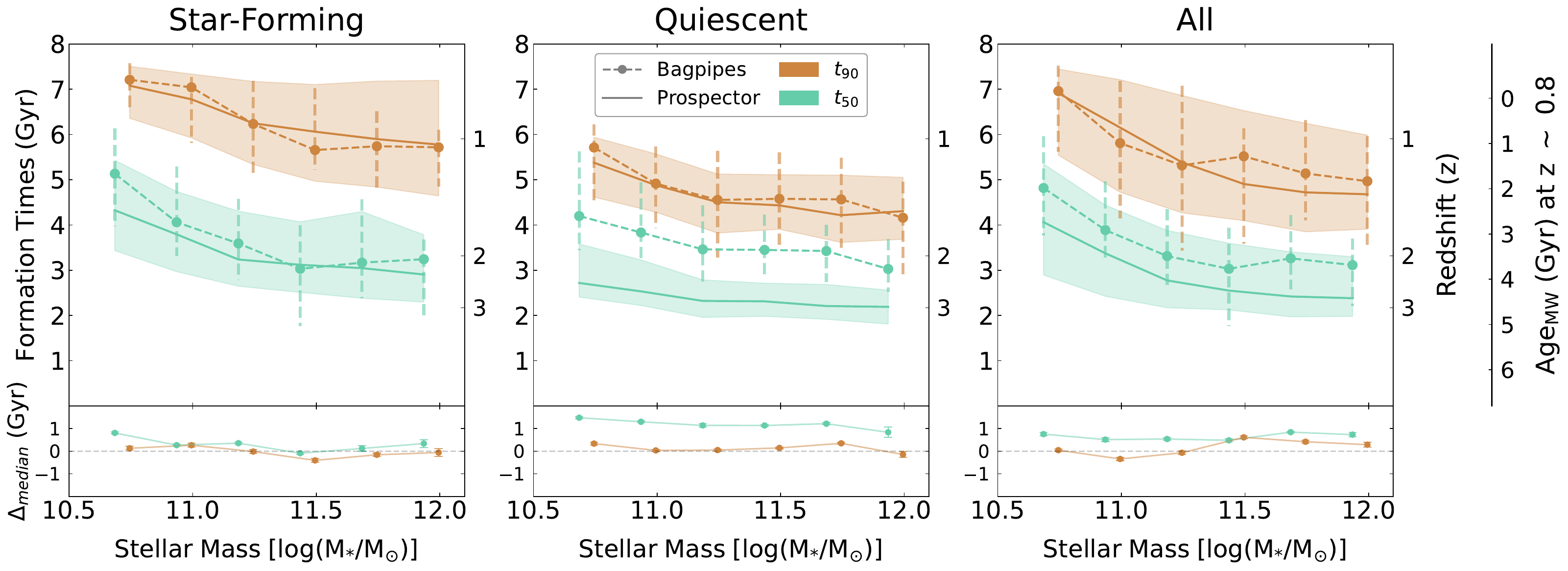}
    \caption{$t_{90}$ (brown) and $t_{50}$ (teal) formation times/redshifts of star-forming, quiescent and total galaxy populations versus stellar mass. Error bars/shaded regions indicate the 16th - 84th percentile population scatter. The bottom row represents the difference in median values of the Bagpipes and Prospector formation times. The far right panel shows the vertical axis in equivalent mass-weighted ages. As the redshift of individual galaxies is different, this is a first order age estimate assuming all galaxies are observed at median LEGA-C redshift of z $\sim$ 0.8. Late formation times ($t_{90}$) are consistent between models, with greater disagreement at earlier times ($t_{50}$), especially for quiescent galaxies. The strong mass trends within the full population (right panel) are primarily driven by star forming galaxies that dominate the low-mass end.}
    \label{fig:tform-mass}
  \hfill
\end{figure*}

\begin{figure*}[ht!]
  \centering
    \includegraphics[width=\textwidth]{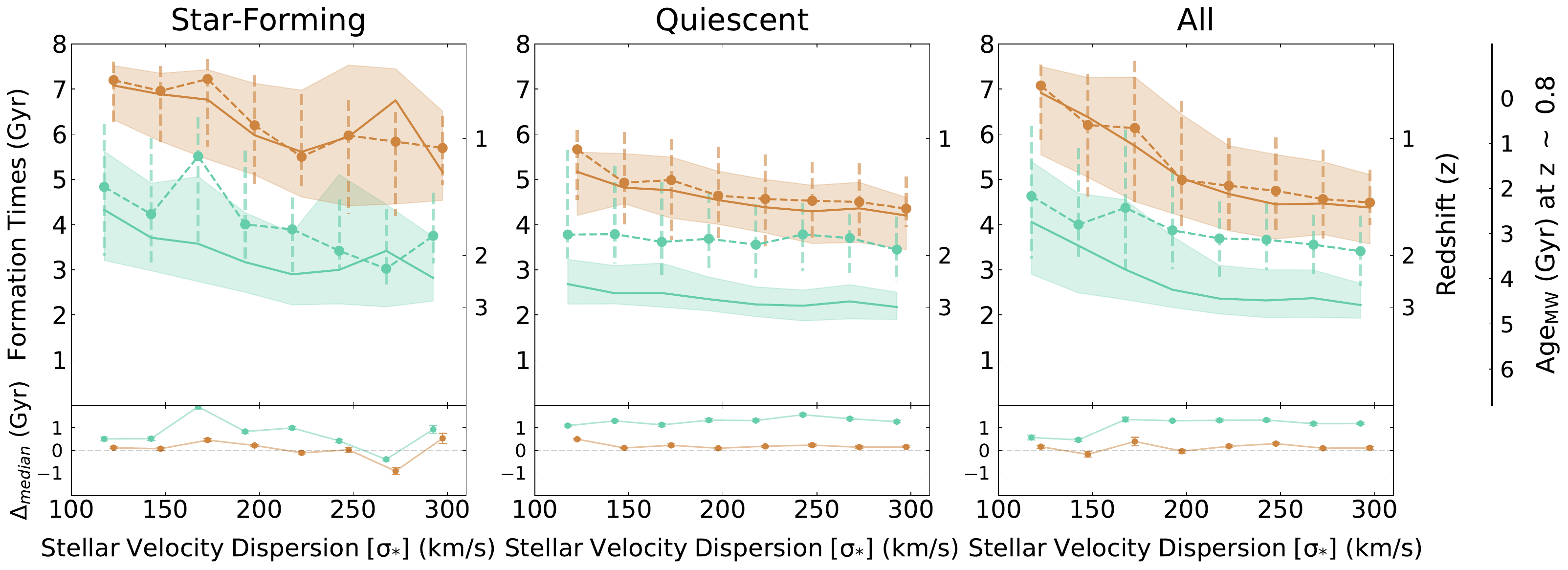}
    \caption{Similar to Fig.~\ref{fig:tform-mass}, now binned in stellar velocity dispersion. We find a striking correlation between the formation timescales and stellar velocity dispersion in full population (right). This trend is in part driven by demographics; the quiescent fraction depends more strongly on stellar velocity dispersion than stellar mass \citep[e.g.,][]{taylor:22}.}
    \label{fig:tform-sigma}
  \hfill
\end{figure*}

Figures \ref{fig:tform-mass} and \ref{fig:tform-sigma} show $t_{50}$ (teal) and $t_{90}$ (brown) the formation time scales for the populations of star-forming (left), quiescent (center) and all galaxies (right) versus stellar mass (6) and stellar velocity dispersion (8). We exclude $t_{10}$ because of the dramatic disagreement between the two sets of models at early times, suggesting that its measurements are completely prior dominated. \texttt{Bagpipes} median trends and population scatter are shown with dotted lines and error bars and \texttt{Prospector} as solid lines with shaded bands. The bottom row shows the difference between the median values derived from the two methods with (very small) error bars indicating the standard error in the medians. The measured late formation times ($t_{90}$) agree well across all stellar masses, with a slight offset in early formation times ($t_{50}$) especially of quiescent systems ($t_{50}$ - $\Delta_{median,SF}$ < 0.9 Gyr,  $\Delta_{median,Q}$ < 1.8 Gyr,  $\Delta_{median,ALL}$ < 0.9 Gyr and $t_{90}$ -   $\Delta_{median,SF}$ < 0.4 Gyr,  $\Delta_{median,Q}$ < 0.3 Gyr,  $\Delta_{median,ALL}$ <  0.6 Gyr). This consistent offset in $t_{50}$ for quiescent systems could be partially driven by preference for more flexible onset of star-formation in \texttt{Bagpipes} compared to \texttt{Prospector}. We find clear trends in both $t_{50}$ and $t_{90}$ with stellar mass amongst star-forming galaxies. In contrast, the median formation times are mostly independent of mass amongst the quiescent galaxies. The overall correlation with mass is reflected by the full population, in which the younger star-forming galaxies dominate at low masses and are rare at the massive end. We note that similar correlations with stellar mass remain within finer redshift bins ($\Delta z =0.1$). Correlations of formation times with stellar velocity dispersion qualitatively similar to those with stellar mass (Figure \ref{fig:tform-sigma}). Again, the median $t_{50}$ and $t_{90}$ formation times are essentially uniform for the quiescent population. However, because the fraction of star-forming galaxies plummets at high stellar velocity dispersion (especially above $\sigma_{\star}\gtrsim200$ km s$^{-1}$) \citep{taylor:22}, the formation timescales of the full population exhibits an even stronger correlation with stellar velocity dispersion (Fig.~\ref{fig:tform-sigma},right panel).

Finally, we investigate the duration of late time star formation ($t_{90}$ - $t_{50}$) from \texttt{Prospector} (green) and \texttt{Bagpipes} (orange) in Figure \ref{fig:tscale-90-50}. Some studies have used ($t_{80}$ - $t_{20}$) and ($t_{90}$ - $t_{10}$) timescales \citep{pacifici:16:b,tacchella:22} to quantify full duration of star-formation, but we choose to avoid large uncertainties at earlier times and quantify late-time star-formation as $\tau_{SF,late}\equiv t_{90}$ - $t_{50}$. Figure \ref{fig:tscale-90-50} shows these timescales versus stellar mass (top row) and stellar velocity dispersion (bottom row) with error bars/shaded regions representing the population scatter, following the plotting conventions of Figures \ref{fig:tform-mass} and \ref{fig:tform-sigma}. Here we see that while agreement is reasonably good between e.g., median $t_{50}$ in the two models, systematic offsets in the duration of star-formation remain. Non-parametric models from \texttt{Prospector} exhibit consistently more extended star formation histories than those derived with double-power law models. This discrepancy at the earliest times has been well-documented \citep[e.g.,][]{carnall:19:a,leja:19:a}. Notably, the median $\tau_{SF,late}$ from \texttt{Bagpipes} is dramatically ($\sim$2 Gyr) more rapid for quiescent galaxies, with an enormous population scatter, reflecting the larger variety of SFHs (see e.g., Fig.~\ref{fig:ssfr-example}). Mock recovery testing of $\tau_{SF,late}$ using spectro-photometric data could reveal if this timescale is recoverable. \citet{leja:19:a} and \citet{carnall:19:b} have shown that with mock photometric data, \texttt{Prospector} was able to recover late-time star-formation better than \texttt{Bagpipes}. Note that low $\tau_{SF,late}$ values in \texttt{Bagpipes} are partially driven by flexibility in onset times as well as in the late time SFH parameter (e.g. falling slope ‘alpha’) to adapt steep/shallow values to give a range of $t_{90}$ - $t_{50}$ values. In fact, the population scatter on \texttt{Bagpipes} star formation duration detracts from the informative nature of the median trends. However, significant offset between $\tau_{SF,late}$ of star-forming and quiescent galaxies does imprint a trend in the full population median value with stellar velocity dispersion (bottom row, right-most panel), where demographics are more cleanly separated than in stellar mass \citep[e.g.,][Fig.~\ref{fig:tform-sigma}]{franx:08,belli:14,taylor:22}. The median duration of star-formation derived by \texttt{Prospector} does not vary significantly within the star-forming population, but quiescent galaxies exhibit a slight decrease with stellar mass and stellar velocity dispersion ($\lesssim 0.5$ Gyr across the full range). This is consistent with weak correlations between alpha enhancement and stellar velocity dispersion for a subset of quiescent galaxies from the same LEGA-C dataset \cite{beverage:23}.

\section{Discussion} \label{section4}

\begin{figure*}[t!]
  \centering
    \includegraphics[width=\textwidth,height=11.5cm]{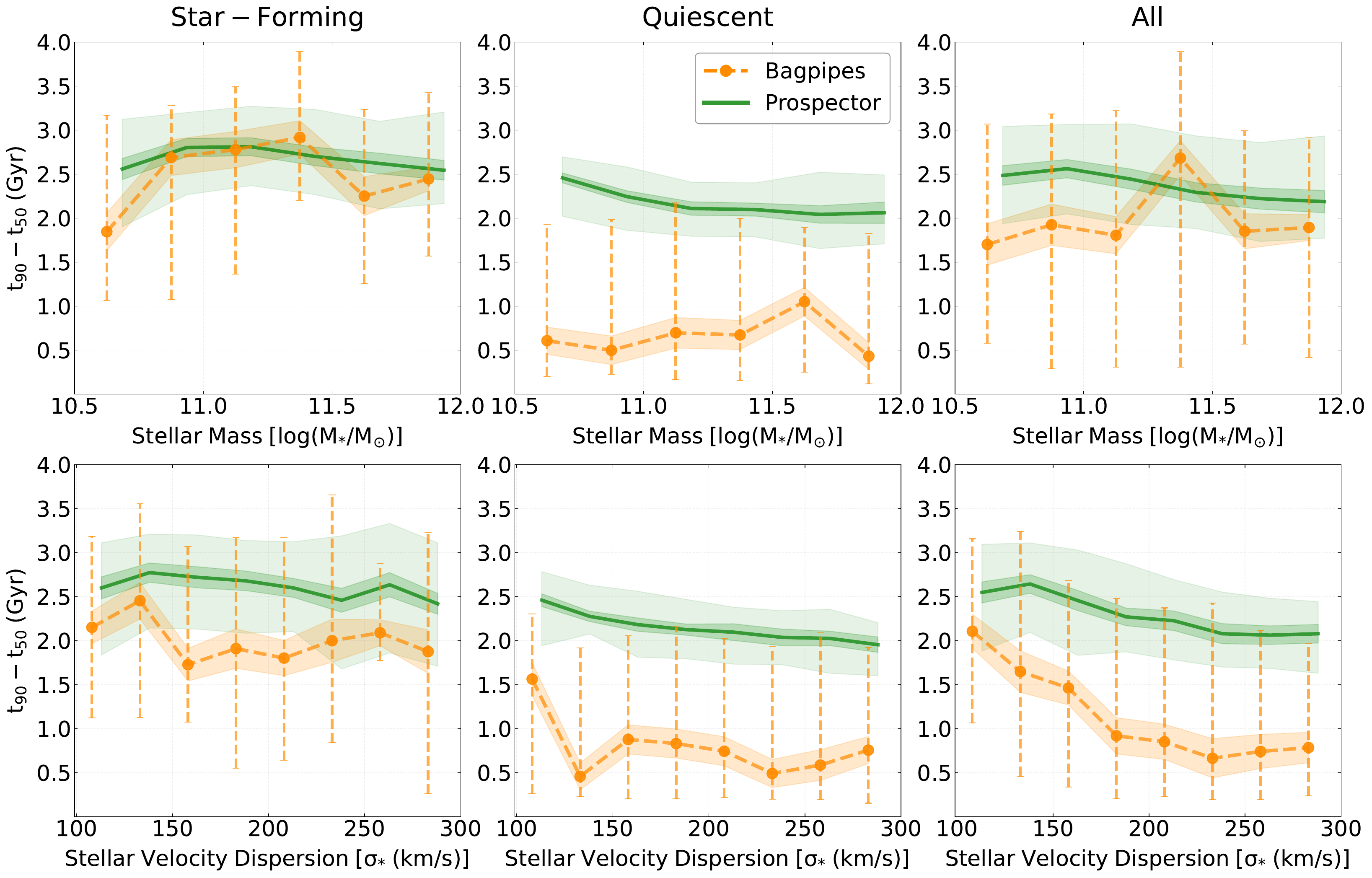}
    \caption{The distribution of duration of star-formation at late times ($\tau_{SF,late}\equiv t_{90}$ - $t_{50}$) in stellar mass (top row) and stellar velocity dispersion (bottom row) bins for population of star-forming (left), quiescent (center) and all (right) as recovered from \texttt{Bagpipes} and \texttt{Prospector}. Green solid lines and orange solid points represent 50th percentile values (median) and broad green shaded regions and orange error bars represent 16th-84th percentile population scatter in that bin. The standard error on the medians is represented in slightly darker thin shaded regions calculated from bootstrapping. Non-parametric models are consistently more extended than the double-power law SFHs. We find no strong correlation between $\tau_{SF,late}$ and stellar mass or stellar velocity dispersion. This lack of correlation is in contrast with trends found amongst analogous massive elliptical galaxies today \citep[e.g.,][]{graves:09,greene:22}. The significant offset between $\tau_{SF,late}$ of star-forming and quiescent galaxies does imprint a trend in the full population median value with stellar velocity dispersion (bottom row, right-most panel), where demographics are more cleanly separated than in stellar mass.}
    \label{fig:tscale-90-50}
  \hfill
\end{figure*}
Star-formation activity has been demonstrated to depend both on stellar mass \citep{calvi:18,pacifici:16:b,kauffmann:04} and on stellar velocity dispersion \citep{franx:08}. It has been long debated whether stellar velocity dispersion is a more fundamental property than stellar mass \citep{sharma:21,zahid:18,wake:12}, as it is directly connected to the total gravitational potential well (including effects of dark matter halo and supermassive black hole at galactic center) in which a galaxy resides \citep{elahi:18, dutton:10}. 
  
This census of star formation histories builds on extensive studies based on complete photometric \citep[e.g.,][]{olsen:21,aufort:20,iyer:19,iyer:17,pacifici:16:a,pacifici:16:b,dye:08} and smaller and/or less complete, and more observationally expensive, spectroscopic dataset \citep[e.g.,][]{gallazzi:08,sanchez-blazquez:11,carnall:19:b,carnall:22,tacchella:22}. For the first time at significant lookback time, we present a comprehensive view of the star-formation histories of the full population of massive galaxies (not just quiescent or star forming galaxies separately). This builds on a preliminary study by \citet{chauke:18} based on early LEGA-C data (607 galaxies) with simpler modeling assumptions: adopting fixed solar metallicity and piece-wise constant star-formation rates using CSP templates from FSPS package \citep{conroy:09}. This current paper expands in both scope and detail to yield a comprehensive analysis of the full LEGA-C sample. 

\begin{figure*}[t!]
  \centering
    \includegraphics[height=0.4\textwidth]{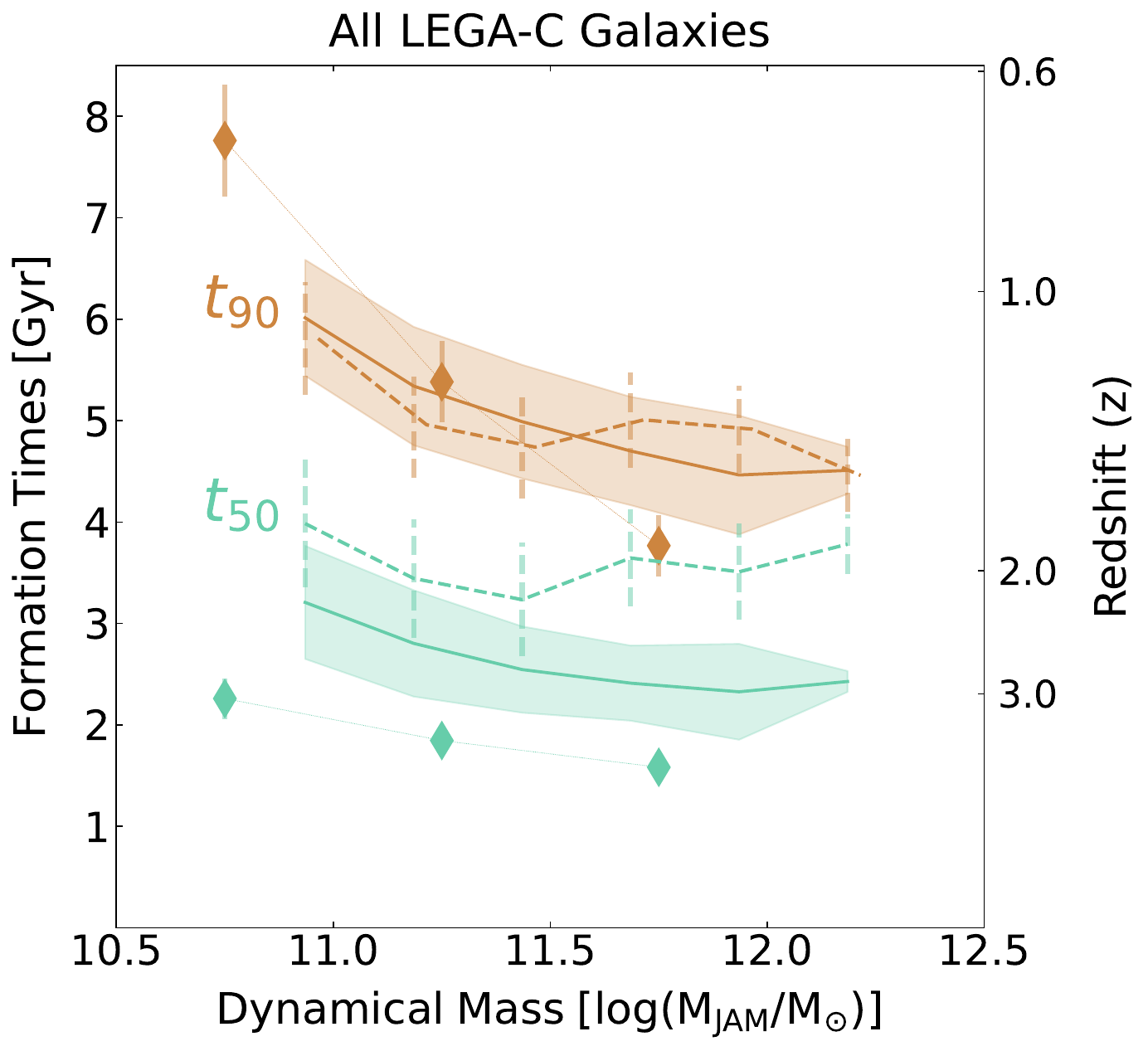}
    \hspace{1cm}
    \includegraphics[height=0.4\textwidth]{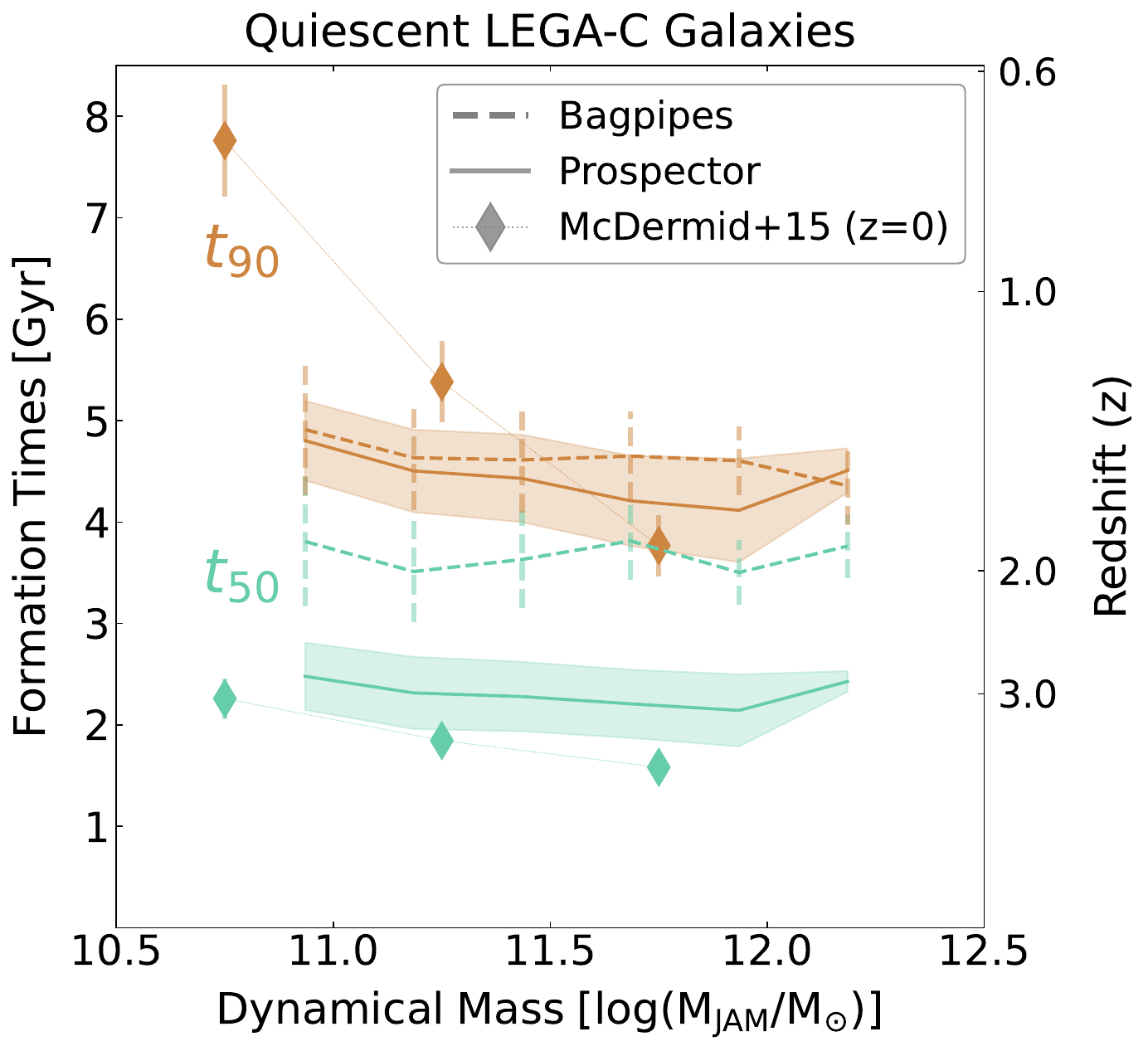}
    \caption{Comparison of median formation times $t_{50}$ (teal) and $t_{90}$ (brown) measured forward in the direction of the age of the universe in bins of dynamical masses. These are derived for the progenitor populations from the LEGA-C sample and compared to those inferred from local elliptical galaxies from ATLAS3D (after correcting for stellar mass growth between $z\sim0.8$ and $z\sim0$) in \citet{mcDermid:15}. The left panel shows the comparison with the full LEGA-C population (star-forming and quiescent progenitors) and the right panel compares only quiescent galaxies to the local ellipticals. Error bars and shaded regions represent 1-$\sigma$ dispersion in the population medians. SPS modeling of local ellipticals finds ubiquitously earlier ($t_{50}$ $\sim$ 2 Gyr) star formation (diamonds), which is even more extreme than even non-parametric SFHs for LEGA-C progenitors. The strong trends in later star formation ($t_{90}$) suggest the importance of star forming progenitors joining the population at $M_{dyn}\sim10^{11}M_{\odot}$ since $z\sim0.8$.
    }
    \label{fig:mcdermid}
  \hfill
\end{figure*}

Our results are in qualitative agreement with previous analyses, and quantitative differences can generally be attributed to modeling systematics. Previous studies have found similar correlations between formation times $t_{50}$ and stellar mass and stellar velocity dispersion e.g. for 10.5 < log($\mathrm{M_*/M_{\odot}}$) < 12 linear relation goes from (forward in time) 1.8 Gyr to 0.9 Gyr for quiescent sample in \citet{tacchella:22}, 1.9 Gyr to 1.4 Gyr for quiescent and 3 Gyr to 2.5 Gyr for star-forming in \citep{ferreras:19}, $\sim$ 1 Gyr for quiescent sample and $\sim$ 4 Gyr to 2 Gyr for star forming sample across full mass range in \citep{chauke:18}, which is similar to what we find in our analysis in Fig \ref{fig:tform-mass} and Fig \ref{fig:tform-sigma}. Interestingly, both of our modeling methods recover significant population scatter in the late time formation timescales of star forming galaxies, however the median $\tau_{SF,late}$ is almost independent of stellar mass. This result is in contrast with previous studies. One interesting comparison is with \citet{estrada-carpenter:20}, in which $\tau_{SF,late}$ is defined as \textit{quenching timescale ($t_{Q}$)}. Using \texttt{Prospector} to fit non-parametric SFHs to a much smaller sample of $\sim$30 quiescent systems at 0.7 < $z$ < 1, finding median formation redshift $z50$ roughly corresponding to 3.3 Gyr - 1.5 Gyr formation times for 10.5 < log($\mathrm{M_*/M_{\odot}}$) < 12 which is consistent within our measured population scatter (Fig.\ref{fig:tform-mass}: \texttt{Bagpipes} - 4.2 Gyr - 3.0  Gyr, \texttt{Prospector} - 2.8 Gyr - 2.3 Gyr). However, they find a broader range of quenching timescales varying from 0.4 Gyr to 2.2 Gyr which is different from our $\tau_{SF,late}$ derived using \texttt{Prospector} (Fig.\ref{fig:tscale-90-50}: 1.7 Gyr to 2.7 Gyr) but similar to what we find in \texttt{Bagpipes} (Fig.\ref{fig:tscale-90-50}: 0.2 Gyr to 2.3 Gyr). We can only speculate that this subtle difference could be driven by priors, sample size or differences in data characteristics (e.g., the use of low-resolution grism spectra). \citet{beverage:23} study 135 LEGA-C massive quiescent galaxies based on elemental abundance patterns and found a slightly stronger correlation between formation times and stellar velocity dispersion. The highest dispersion galaxies (> 250 km/s ) formed the earliest around 3.1 Gyr and are most metal rich whereas low dispersion galaxies (< 150  Km/s) formed around 4.4 Gyr. This is qualitatively similar to our findings in Fig.\ref{fig:tform-sigma}, middle panel.

Interestingly, although the qualitative trends agree well with cluster galaxies at a similar epoch, there may be some hint of environmental effects on SFHs. \citet{khullar:22} find $t_{50}$ of massive quiescent cluster galaxies to be inversely correlated with stellar mass at 0.3 < $z$ < 1.4 with values within 2 Gyr. This is $\sim$ 1 Gyr/2 Gyr earlier than our Prospector/Bagpipes estimates. Again, this difference could be driven by modeling systematics (e.g. that study adopts a delayed tau SFH) or a physical manifestation of environmental processes that combine to quench galaxies earlier or the generally speed up galaxy evolution in clusters. Even within the LEGA-C dataset age sensitive indices (Dn4000 and H$\delta$) of quiescent galaxies depend on the environment, not just in cluster versus field; galaxies in over-dense regions are older and formed earlier than those in less extreme environments \citep{sobral:22}. We note that direct comparisons between different studies can be challenging; \citet{webb:20} perform a cluster analysis at similar redshifts and find little difference between the ages of galaxies in the field versus cluster environments at fixed mass. 

Next, we investigate whether our results from LEGA-C at $z\sim0.8$ differ significantly from our understanding of this time period as recovered from previous modeling of their presumed descendent galaxies in the local Universe. We start by comparing to SPS modeling of local elliptical galaxies from \citet{mcDermid:15}. Figure \ref{fig:mcdermid} shows a comparison of the median formation times $t_{50}$ and $t_{90}$ as a function of the Virial mass for all galaxies in our sample at < $z$ > = 0.8 from \texttt{Bagpipes} (dashed line) and \texttt{Prospector} (solid-line) relative to quiescent systems at z $\sim$ 0 (diamonds) from \citet{mcDermid:15}, after subtracting the minimal cumulative stellar mass growth between $z\sim0$ and $z\sim0.6$. The error bars and shaded regions represent 1-$\sigma$ dispersion in the population medians. The vertical axis is forward in time with t=0 corresponding to the Big Bang. Note that these elliptical galaxies at $z\sim0$ could either be quiescent or star-forming at our redshift, hence for a first order comparison with LEGA-C, we compare to the full (left) and quiescent only (right) population. Virial masses have been calibrated for the majority of the LEGA-C sample \citep{vanderwel:22} using spatially resolved stellar kinematics \citep{houdt:21}. For this simplistic calculation we adopt a median offset of 0.33 dex between the stellar and dynamical mass for each LEGA-C individual object. \citet{mcDermid:15} find that most massive galaxies (log$\mathrm{(M_*/M_{\odot})}$ > 11) have already accumulated > 95\% of their stellar mass by $z$ $\sim$ 0.6, whereas low mass galaxies have formed $\sim$90\% of their stellar mass. They also find ubiquitously earlier ($t_{50}=2Gyr$) stellar mass growth (diamonds with dotted line), than the view from LEGA-C using either modeling framework. This reveals a natural uncertainty of stellar population modeling for the oldest local stellar systems. 
We see similar dynamical mass trends in the left panel for z$\sim$0.8 galaxies (both star-forming and quiescent) and local ellipticals especially toward the low mass end. This correlation is weaker when only quiescent galaxies from the LEGA-C sample are included. This emphasizes the importance of star-forming progenitors joining the local elliptical population.

\citet{thomas:05,thomas:10} provide clear evidence from a $z$ $\sim$ 0 sample of galaxies that the duration and timing of galaxy formation depends strongly on the stellar mass, arguing that this ultimately reflects halo assembly. However, even at $z\sim0.8$, we do not find these trends, even when star-forming progenitors are included. Some of these differences could be attributed to modeling systematics e.g. use of Gaussian SFH prescription to describe population average, chemical enrichment models/abundance patterns, various environmental dependence and most importantly aperture effects - probing the central few kpc (not the full galaxy) with a 3'' diameter fiber that makes results more mass dependent for older ages and higher metallicities, whereas slit aperture effects are not that extreme. Furthermore, the view from maximally old galaxies in local Universe will always be complicated by the relative challenge of modeling slowly evolving old stellar populations. However, it is hard to escape the conclusion that suggests that the trends found in local ellipticals reflect the transformation and assembly/merger history of massive galaxies and not just their in-situ SFHs. 

\begin{figure}[!t]
    \includegraphics[width=0.45\textwidth]{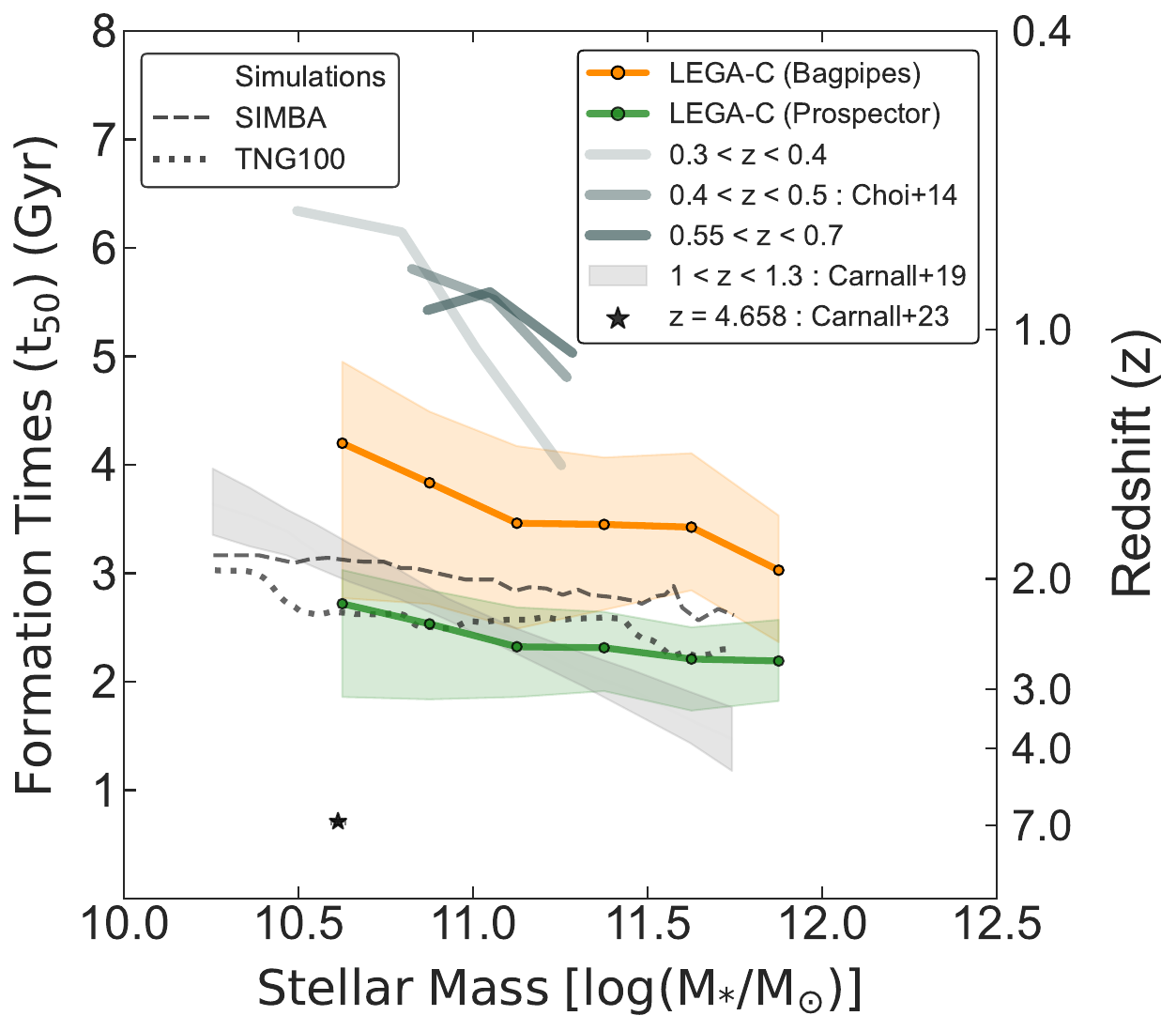}
    \caption{Formation times versus stellar mass for massive quiescent galaxies across redshifts. The grey lines and bands indicate relations at lower and higher redshifts respectively \citep{choi:14,carnall:19:b}. The highest redshift measurement from JWST is shown with black star\citep{carnall:23b}. Our results are shown in orange (parametric SFHs) and green (non-parametric SFHs). Shaded color bands represent 16th - 84th percentile population scatter in the median values. Scaling relations from cosmological simulations \citep[dashed/dotted lines,][]{dave:19,nelson:19} agree well with our \texttt{Prospector} models. The evolution in this relation for quiescent galaxies emphasizes the importance of including star-forming progenitors and ex-situ contributions when comparing the stellar populations of galaxies across cosmic time.}
    \label{fig:tform-comp}
\end{figure}

Finally, we compare our measured formation time trends for quiescent galaxies with similar studies at other epochs. Figure \ref{fig:tform-comp} shows the comparison of median formation times ($t_{50}$) of massive quiescent systems from our study (orange and green bands indicating populations scatter) with other empirical (spectroscopic) studies and simulations across redshifts. \citet{choi:14} results (grey solid lines) are derived from full-spectral fitting based on alpha abundance measurements and SSP-equivalent ages. The grey band indicates mean formation times recovered from the VANDELS survey (1 < $z$ < 1.3) in \citet{carnall:19:b} using \texttt{Bagpipes}. The single JWST spectroscopic observation around $z$ = 4.65 from \citet{carnall:23} is shown with a star-symbol. We note that the incredibly small uncertainty on formation time reflects the incredible data quality and limits imposed by the age of the Universe at that early time. Trends found in cosmological simulations at $z$ = 1 are shown with dashed (SIMBA \citep{dave:19}) and dotted (IllustrisTNG-100 \citep{nelson:19}) lines. For details on the simulation selection, see \citet{carnall:19:b}. Although modeling differences could introduce significant systematic offsets with respect to our fits (e.g. extended SFH versus SSPs, scaled solar metallicity versus alpha enrichment), it is interesting to speculate whether some of the observed offset in $t_{50}$ between epochs is physical, perhaps driven by late additions to the quiescent population. This progenitor bias could exaggerate the evolution of these scaling relations without changing the stellar populations of any individual galaxies. Superficially the older LEGA-C ages from \texttt{Prospector} are roughly consistent with the trend found for quiescent galaxies in VANDELS at higher z, however our  \texttt{Bagpipes} modeling adopts much more similar priors and provide the more consistent comparison. Thus, the observed offset between the grey and orange bands suggests that the relation evolves even between $z\sim1.15$ and $z\sim0.8$. \citet{choi:14} modeled stacked optical spectra of massive quiescent galaxies from 0.1 < $z$ < 0.7 from SDSS and AGES Survey. Although our spectro-photometric modeling of individual galaxies differs from their analysis, the similar offset to later formation times in lower-redshift quiescent populations is consistent with new additions shifting the scaling relations, perhaps more efficiently at lower mass. There is one caveat in comparing with local elliptical galaxies: galaxy mergers are also major avenues of mass growth and therefore will also contribute to the spatially-integrated ages of galaxies (e.g. $t_{50}$ and $t_{90}$). With SPS modeling, we measure the formation times of stars from all progenitors of a galaxy at $t_{obs}$, which can be quite different from assembly times ($t_{a}$). \citet{hill:17a} demonstrate that this can be a significant effect, by $z$$\sim$0.1 assembly times and light-weighted formation times can differ by up to $\sim2.5$ Gyr \citep[see also e.g.,][]{hill:17b, muzzin:17}. We hence note that progenitor-descendant linking is further complicated by uncertain importance of merging.

Note that one of the key findings of this study is the impact of priors of SPS modeling on early formation time recovery at z $\sim$ 0.8, even with high S/N data. A part of this is also driven by the outshining of old populations. A natural conclusion, therefore, is that the earliest SFHs can be better understood by modeling the light from galaxies at progressively earlier times, when this becomes less of a concern. Future spectroscopic studies of higher redshifts can precisely pinpoint the emergence of massive quiescent systems and can help settle uncertainties at the earliest times. Recent results from JWST photometry-only SED analyses suggest that massive galaxies may indeed form efficiently and rapidly within 1 Gyr of the Big Bang (e.g. \citet{labbe:23}). However, given uncertainties in the emission line contributions to photometry (e.g. \citet{mckinney:23}), spectroscopic data are needed more than ever. JWST/NIRSpec has been shown to provide the sensitivity required to constrain the SFHs and and quenching timescales of galaxies at higher redshifts and lower masses: e.g., the  \citet{carnall:23}'s extreme massive quiescent galaxy at $z$ = 4.65 included in Figure \ref{fig:tform-comp}. Other recent studies include \citet{looser:23} and \citet{nanayakkara:23}. Larger, more complete spectroscopic samples from JWST will be crucial to map the early evolution of today's massive quiescent galaxy population. These samples will hopefully shed more light on the modeling degeneracies at the earliest times. 

\section{Conclusions} \label{section5}
In this paper, we analyze the median SFHs of massive (log($\mathrm{M_*/M_{\odot}}$) > 10.5) galaxies at $0.6<z<1.0$ from the LEGA-C survey by quantifying their formation times ($t_{50}$ and $t_{90}$) and investigate the population trends in this census with stellar masses and stellar velocity dispersions. From our spectro-photometric analysis, we conclude that:
\begin{enumerate}
    \item The two modeling methods (parametric and non-parametric) yield consistent late-time star-formation histories for star-forming and quiescent galaxies. Non-parametric SFHs consistently prefer earlier stellar mass formation, especially for quiescent systems. Potential reasons could be - stellar population outshining driving this inference into a systematics-limited regime emphasizing the dominance of modeling priors at the earliest times or a host of different modeling assumption e.g. SPS models, treatment of emission lines, dust, parametrization of SFHs, etc. (Figure \ref{fig:sfh-prior},\ref{fig:ssfr-example}).
    \item Although individual galaxies show a variety of SFH pathways, the median sSFR evolution is similar for all mass bins. \texttt{Bagpipes} SFHs shows population scatter at any time. Both codes infer that massive star-forming galaxies exhibit falling sSFRs with redshift. Neither model finds a mass-dependence in the median time at which quiescent galaxies cross a sSFR threshold (Figure \ref{fig:raw-sfh}).
    \item From median trends, we find that for both non-parametric and parametric SFH modeling, quiescent galaxies formed 90\% of their stars well within $\sim$4 Gyr and $\sim$5.5 Gyr age of the universe for massive and less massive end respectively. Our analysis reflects the established impact of these priors at earlier times, yielding differences in mass-weighted ages $t_{50}$ $\sim$1.5 Gyr) within that population, with SFHs that further diverge at earlier times. (Figures \ref{fig:cumulative-sfh},\ref{fig:tform-mass}).
    \item Lower mass galaxies are slightly younger than massive galaxies (Figure \ref{fig:cumulative-sfh}), however this trend is relatively weak. This is perhaps due to the fact that the full LEGA-C sample only probes a limited dynamic range in stellar mass. This picture is also consistent with simulations \citep{iyer:20,torrey:18,shamshiri:13}. The formation times of star-forming galaxies correlate with stellar mass and stellar velocity dispersion, but similar trends do not exist not amongst the quiescent galaxies. For combined population, the formation times depend strongly on the $M_{*}$ and $\sigma_{*}$ as quiescent fraction increases with each property. (Figure \ref{fig:tform-mass},\ref{fig:tform-sigma}). 
    \item For the star-forming population, both codes recover consistent median formation times ($|t_{Bagpipes}-t_{Prospector}| < 300$ Myr); the most massive systems formed within $t_{50}$/$t_{90}$ $\sim$3 Gyr/5.5 Gyr and the less massive ones within $\sim$5 Gyr/7 Gyr (Figure \ref{fig:tform-mass}, or similarly for high/low $\sigma_{\star}$, Fig.~\ref{fig:tform-sigma}). Even for these galaxies, SFRs were on the decline by the time of observation, corroborating a number of previous studies \citep[e.g.,][]{feulner:05,ilbert:15}. 
    \item The late time duration of star formation ($\tau_{SF,late}$ = $t_{90}-t_{50}$) does not exhibit a significant correlation with $M_{\star}$ or $\sigma_{\star}$ for quiescent or star-forming galaxy populations. At face value, this is in contrast with expectations from the local universe (e.g. \citet{thomas:05}). We posit that this suggests the importance of transformation, quenching and/or ex-situ evolution even at late times ($z$ < 1) (Figure \ref{fig:tscale-90-50}).
\end{enumerate}

Remaining open questions include connecting quenching timescales (defined from recovered SFHs), environment, and processes driving quenching in quiescent systems to stellar mass and stellar velocity dispersion. A number of studies have begun to answer those questions using photometric data \citep{mao:22} and age sensitive spectral indices \citep{sobral:22,wu:21}, but not much effort has been put towards understanding this evolution from the SFH of individual galaxies. Another interesting comparison would be to compare our SFHs to those derived  from simulation using PCA analysis \citep[e.g.,][]{sparre:14}. 

This SFH census of a representative sample of $\sim3000$ massive galaxies using two state-of-the-art Bayesian modeling tools will set the benchmark at $z\sim1$. It is a "cosmic mid-point" in time, providing a stepping stone to future spectroscopic studies of high redshift galaxies (e.g., from JWST/NIRSpec, VLT/MOONS \citep{maiolino:20} or Subaru/PFS \citep{greene:22}). By including two independent sets of models, we hope that this intermediate redshift study would be a link to other similar studies and help connect our understanding of evolution of stellar mass growth and assembly in galaxies from cosmic noon and beyond to present day universe.

\section{Acknowledgements}
RB and YK gratefully acknowledge funding for project KA2019-105551 provided by the Robert C. Smith Fund and the Betsy R. Clark Fund of The Pittsburgh Foundation. RB acknowledges support from the Research Corporation for Scientific Advancement (RCSA) Cottrell Scholar Award ID No: 27587. RB and YK also acknowledge support from NSF grant AST-2144314. AG acknowledges support from INAF-Minigrant-2022 "LEGA-C" 1.05.12.04.01. This research was supported in part by the University of Pittsburgh Center for Research Computing, RRID:SCR\_022735, through the resources provided. Specifically, this work used the H2P cluster, which is supported by NSF award number OAC-2117681. We specifically acknowledge the assistance of Prof. Kim Wong to setup the framework at the cluster. YK thanks Brett Andrews, Alan Pearl, David Setton and Biprateep Dey for their constructive feedback and suggestions that greatly improved this work. 

This work made extensive use of the publicly available tools \href{https://bagpipes.readthedocs.io/en/latest/}{\texttt{Bagpipes}} \citep{carnall:18}, \href{https://prospect.readthedocs.io/en/latest/}{\texttt{Prospector}} \citep{Johnson:21,johnson:17},
\href{http://www.python.org}{Python} programming language
\citep{vanrossum1995},
{\sc \href{https://pypi.org/project/astropy/}{astropy}} \citep{astropyco+2013},
{\sc \href{https://pypi.org/project/matplotlib/}{matplotlib}} \citep{hunter2007},
{\sc \href{https://pypi.org/project/numpy/}{numpy}} \citep{harris+2020},
{\sc \href{https://pypi.org/project/scipy/}{scipy}} \citep{jones+2001}.

\bibliography{sample631}

\begin{thebibliography}{}
\expandafter\ifx\csname natexlab\endcsname\relax\def\natexlab#1{#1}\fi
\providecommand{\url}[1]{\href{#1}{#1}}
\providecommand{\dodoi}[1]{doi:~\href{http://doi.org/#1}{\nolinkurl{#1}}}
\providecommand{\doeprint}[1]{\href{http://ascl.net/#1}{\nolinkurl{http://ascl.net/#1}}}
\providecommand{\doarXiv}[1]{\href{https://arxiv.org/abs/#1}{\nolinkurl{https://arxiv.org/abs/#1}}}

\bibitem[{{Abramson} {et~al.}(2015){Abramson}, {Gladders}, {Dressler},
  {Oemler}, {Poggianti}, \& {Vulcani}}]{abramson:15}
{Abramson}, L.~E., {Gladders}, M.~D., {Dressler}, A., {et~al.} 2015, \apjl,
  801, L12, \dodoi{10.1088/2041-8205/801/1/L12}

\bibitem[{{Astropy Collaboration} {et~al.}(2013){Astropy Collaboration},
  {Robitaille}, {Tollerud}, {Greenfield}, {Droettboom}, {Bray}, {Aldcroft},
  {Davis}, {Ginsburg}, {Price-Whelan}, {Kerzendorf}, {Conley}, {Crighton},
  {Barbary}, {Muna}, {Ferguson}, {Grollier}, {Parikh}, {Nair}, {Unther},
  {Deil}, {Woillez}, {Conseil}, {Kramer}, {Turner}, {Singer}, {Fox}, {Weaver},
  {Zabalza}, {Edwards}, {Azalee Bostroem}, {Burke}, {Casey}, {Crawford},
  {Dencheva}, {Ely}, {Jenness}, {Labrie}, {Lim}, {Pierfederici}, {Pontzen},
  {Ptak}, {Refsdal}, {Servillat}, \& {Streicher}}]{astropyco+2013}
{Astropy Collaboration}, {Robitaille}, T.~P., {Tollerud}, E.~J., {et~al.} 2013,
  \aap, 558, A33, \dodoi{10.1051/0004-6361/201322068}

\bibitem[{{Aufort} {et~al.}(2020){Aufort}, {Ciesla}, {Pudlo}, \&
  {Buat}}]{aufort:20}
{Aufort}, G., {Ciesla}, L., {Pudlo}, P., \& {Buat}, V. 2020, \aap, 635, A136,
  \dodoi{10.1051/0004-6361/201936788}

\bibitem[{{Behroozi} {et~al.}(2019){Behroozi}, {Wechsler}, {Hearin}, \&
  {Conroy}}]{behroozi:19}
{Behroozi}, P., {Wechsler}, R.~H., {Hearin}, A.~P., \& {Conroy}, C. 2019,
  \mnras, 488, 3143, \dodoi{10.1093/mnras/stz1182}

\bibitem[{{Behroozi} {et~al.}(2013){Behroozi}, {Wechsler}, \&
  {Conroy}}]{behroozi:13}
{Behroozi}, P.~S., {Wechsler}, R.~H., \& {Conroy}, C. 2013, \apj, 770, 57,
  \dodoi{10.1088/0004-637X/770/1/57}

\bibitem[{{Bell} {et~al.}(2004){Bell}, {Wolf}, {Meisenheimer}, {Rix}, {Borch},
  {Dye}, {Kleinheinrich}, {Wisotzki}, \& {McIntosh}}]{bell:04}
{Bell}, E.~F., {Wolf}, C., {Meisenheimer}, K., {et~al.} 2004, \apj, 608, 752,
  \dodoi{10.1086/420778}

\bibitem[{{Belli} {et~al.}(2018){Belli}, {Contursi}, \& {Davies}}]{belli:18}
{Belli}, S., {Contursi}, A., \& {Davies}, R.~I. 2018, \mnras, 478, 2097,
  \dodoi{10.1093/mnras/sty1236}

\bibitem[{{Belli} {et~al.}(2014){Belli}, {Newman}, \& {Ellis}}]{belli:14}
{Belli}, S., {Newman}, A.~B., \& {Ellis}, R.~S. 2014, \apj, 783, 117,
  \dodoi{10.1088/0004-637X/783/2/117}

\bibitem[{{Beverage} {et~al.}(2023){Beverage}, {Kriek}, {Conroy}, {Sandford},
  {Bezanson}, {Franx}, {van der Wel}, \& {Weisz}}]{beverage:23}
{Beverage}, A.~G., {Kriek}, M., {Conroy}, C., {et~al.} 2023, \apj, 948, 140,
  \dodoi{10.3847/1538-4357/acc176}

\bibitem[{{Bezanson} {et~al.}(2018){Bezanson}, {van der Wel}, {Straatman},
  {Pacifici}, {Wu}, {Bari{\v{s}}i{\'c}}, {Bell}, {Conroy}, {D'Eugenio},
  {Franx}, {Gallazzi}, {van Houdt}, {Maseda}, {Muzzin}, {van de Sande},
  {Sobral}, \& {Spilker}}]{bezanson:18}
{Bezanson}, R., {van der Wel}, A., {Straatman}, C., {et~al.} 2018, \apjl, 868,
  L36, \dodoi{10.3847/2041-8213/aaf16b}

\bibitem[{{Bowman} {et~al.}(2020){Bowman}, {Zeimann}, {Nagaraj}, {Ciardullo},
  {Gronwall}, {McCarron}, {Weiss}, {Molina}, {Belles}, \&
  {Schneider}}]{bowman:20}
{Bowman}, W.~P., {Zeimann}, G.~R., {Nagaraj}, G., {et~al.} 2020, \apj, 899, 7,
  \dodoi{10.3847/1538-4357/ab9f3c}

\bibitem[{{Bruzual} \& {Charlot}(2003)}]{bruzual:03}
{Bruzual}, G., \& {Charlot}, S. 2003, \mnras, 344, 1000,
  \dodoi{10.1046/j.1365-8711.2003.06897.x}

\bibitem[{{Calvi} {et~al.}(2018){Calvi}, {Vulcani}, {Poggianti}, {Moretti},
  {Fritz}, \& {Fasano}}]{calvi:18}
{Calvi}, R., {Vulcani}, B., {Poggianti}, B.~M., {et~al.} 2018, \mnras, 481,
  3456, \dodoi{10.1093/mnras/sty2476}

\bibitem[{{Cappellari}(2017)}]{cappellari:17}
{Cappellari}, M. 2017, \mnras, 466, 798, \dodoi{10.1093/mnras/stw3020}

\bibitem[{{Carnall} {et~al.}(2019{\natexlab{a}}){Carnall}, {Leja}, {Johnson},
  {McLure}, {Dunlop}, \& {Conroy}}]{carnall:19:a}
{Carnall}, A.~C., {Leja}, J., {Johnson}, B.~D., {et~al.} 2019{\natexlab{a}},
  \apj, 873, 44, \dodoi{10.3847/1538-4357/ab04a2}

\bibitem[{{Carnall} {et~al.}(2018){Carnall}, {McLure}, {Dunlop}, \&
  {Dav{\'e}}}]{carnall:18}
{Carnall}, A.~C., {McLure}, R.~J., {Dunlop}, J.~S., \& {Dav{\'e}}, R. 2018,
  \mnras, 480, 4379, \dodoi{10.1093/mnras/sty2169}

\bibitem[{{Carnall} {et~al.}(2019{\natexlab{b}}){Carnall}, {McLure}, {Dunlop},
  {Cullen}, {McLeod}, {Wild}, {Johnson}, {Appleby}, {Dav{\'e}}, {Amorin},
  {Bolzonella}, {Castellano}, {Cimatti}, {Cucciati}, {Gargiulo}, {Garilli},
  {Marchi}, {Pentericci}, {Pozzetti}, {Schreiber}, {Talia}, \&
  {Zamorani}}]{carnall:19:b}
{Carnall}, A.~C., {McLure}, R.~J., {Dunlop}, J.~S., {et~al.}
  2019{\natexlab{b}}, \mnras, 490, 417, \dodoi{10.1093/mnras/stz2544}

\bibitem[{{Carnall} {et~al.}(2022){Carnall}, {McLure}, {Dunlop}, {Hamadouche},
  {Cullen}, {McLeod}, {Begley}, {Amorin}, {Bolzonella}, {Castellano},
  {Cimatti}, {Fontanot}, {Gargiulo}, {Garilli}, {Mannucci}, {Pentericci},
  {Talia}, {Zamorani}, {Calabro}, {Cresci}, \& {Hathi}}]{carnall:22}
---. 2022, \apj, 929, 131, \dodoi{10.3847/1538-4357/ac5b62}

\bibitem[{{Carnall} {et~al.}(2023{\natexlab{a}}){Carnall}, {McLure}, {Dunlop},
  {McLeod}, {Wild}, {Cullen}, {Magee}, {Begley}, {Cimatti}, {Donnan},
  {Hamadouche}, {Jewell}, \& {Walker}}]{carnall:23b}
---. 2023{\natexlab{a}}, \nat, 619, 716, \dodoi{10.1038/s41586-023-06158-6}

\bibitem[{{Carnall} {et~al.}(2023{\natexlab{b}}){Carnall}, {McLeod}, {McLure},
  {Dunlop}, {Begley}, {Cullen}, {Donnan}, {Hamadouche}, {Jewell}, {Jones},
  {Pollock}, \& {Wild}}]{carnall:23}
{Carnall}, A.~C., {McLeod}, D.~J., {McLure}, R.~J., {et~al.}
  2023{\natexlab{b}}, \mnras, 520, 3974, \dodoi{10.1093/mnras/stad369}

\bibitem[{{Chabrier}(2003)}]{chabrier:09}
{Chabrier}, G. 2003, \pasp, 115, 763, \dodoi{10.1086/376392}

\bibitem[{{Charlot} \& {Fall}(2000)}]{Charlot:00}
{Charlot}, S., \& {Fall}, S.~M. 2000, \apj, 539, 718, \dodoi{10.1086/309250}

\bibitem[{{Chauke} {et~al.}(2018){Chauke}, {van der Wel}, {Pacifici},
  {Bezanson}, {Wu}, {Gallazzi}, {Noeske}, {Straatman}, {Mu{\~n}os-Mateos},
  {Franx}, {Bari{\v{s}}i{\'c}}, {Bell}, {Brammer}, {Calhau}, {van Houdt},
  {Labb{\'e}}, {Maseda}, {Muzzin}, {Rix}, \& {Sobral}}]{chauke:18}
{Chauke}, P., {van der Wel}, A., {Pacifici}, C., {et~al.} 2018, \apj, 861, 13,
  \dodoi{10.3847/1538-4357/aac324}

\bibitem[{{Chaves-Montero} \& {Hearin}(2020)}]{chaves-montero:19}
{Chaves-Montero}, J., \& {Hearin}, A. 2020, \mnras, 495, 2088,
  \dodoi{10.1093/mnras/staa1230}

\bibitem[{{Chevallard} \& {Charlot}(2016)}]{chevallard:16}
{Chevallard}, J., \& {Charlot}, S. 2016, \mnras, 462, 1415,
  \dodoi{10.1093/mnras/stw1756}

\bibitem[{{Chevallard} {et~al.}(2019){Chevallard}, {Curtis-Lake}, {Charlot},
  {Ferruit}, {Giardino}, {Franx}, {Maseda}, {Amorin}, {Arribas}, {Bunker},
  {Carniani}, {Husemann}, {Jakobsen}, {Maiolino}, {Pforr}, {Rawle}, {Rix},
  {Smit}, \& {Willott}}]{chevallard:19}
{Chevallard}, J., {Curtis-Lake}, E., {Charlot}, S., {et~al.} 2019, \mnras, 483,
  2621, \dodoi{10.1093/mnras/sty2426}

\bibitem[{{Choi} {et~al.}(2014){Choi}, {Conroy}, {Moustakas}, {Graves},
  {Holden}, {Brodwin}, {Brown}, \& {van Dokkum}}]{choi:14}
{Choi}, J., {Conroy}, C., {Moustakas}, J., {et~al.} 2014, \apj, 792, 95,
  \dodoi{10.1088/0004-637X/792/2/95}

\bibitem[{{Cid Fernandes}(2007)}]{cis-fernandes:07}
{Cid Fernandes}, R. 2007, in Stellar Populations as Building Blocks of
  Galaxies, ed. A.~{Vazdekis} \& R.~{Peletier}, Vol. 241, 461--469,
  \dodoi{10.1017/S1743921307008794}

\bibitem[{{Cid Fernandes} {et~al.}(2005){Cid Fernandes}, {Mateus}, {Sodr{\'e}},
  {Stasi{\'n}ska}, \& {Gomes}}]{cid-fernandes:05}
{Cid Fernandes}, R., {Mateus}, A., {Sodr{\'e}}, L., {Stasi{\'n}ska}, G., \&
  {Gomes}, J.~M. 2005, \mnras, 358, 363,
  \dodoi{10.1111/j.1365-2966.2005.08752.x}

\bibitem[{{Ciesla} {et~al.}(2017){Ciesla}, {Elbaz}, \& {Fensch}}]{ciesla:17}
{Ciesla}, L., {Elbaz}, D., \& {Fensch}, J. 2017, \aap, 608, A41,
  \dodoi{10.1051/0004-6361/201731036}

\bibitem[{{Citro} {et~al.}(2016){Citro}, {Pozzetti}, {Moresco}, \&
  {Cimatti}}]{citro:16}
{Citro}, A., {Pozzetti}, L., {Moresco}, M., \& {Cimatti}, A. 2016, \aap, 592,
  A19, \dodoi{10.1051/0004-6361/201527772}

\bibitem[{{Cohn}(2018)}]{cohn:18}
{Cohn}, J.~D. 2018, \mnras, 478, 2291, \dodoi{10.1093/mnras/sty1148}

\bibitem[{{Conroy} {et~al.}(2009){Conroy}, {Gunn}, \& {White}}]{conroy:09}
{Conroy}, C., {Gunn}, J.~E., \& {White}, M. 2009, \apj, 699, 486,
  \dodoi{10.1088/0004-637X/699/1/486}

\bibitem[{{Cullen} {et~al.}(2019){Cullen}, {McLure}, {Dunlop}, {Khochfar},
  {Dav{\'e}}, {Amor{\'\i}n}, {Bolzonella}, {Carnall}, {Castellano}, {Cimatti},
  {Cirasuolo}, {Cresci}, {Fynbo}, {Fontanot}, {Gargiulo}, {Garilli}, {Guaita},
  {Hathi}, {Hibon}, {Mannucci}, {Marchi}, {McLeod}, {Pentericci}, {Pozzetti},
  {Shapley}, {Talia}, \& {Zamorani}}]{cullen:19}
{Cullen}, F., {McLure}, R.~J., {Dunlop}, J.~S., {et~al.} 2019, \mnras, 487,
  2038, \dodoi{10.1093/mnras/stz1402}

\bibitem[{{Dav{\'e}} {et~al.}(2019){Dav{\'e}}, {Angl{\'e}s-Alc{\'a}zar},
  {Narayanan}, {Li}, {Rafieferantsoa}, \& {Appleby}}]{dave:19}
{Dav{\'e}}, R., {Angl{\'e}s-Alc{\'a}zar}, D., {Narayanan}, D., {et~al.} 2019,
  \mnras, 486, 2827, \dodoi{10.1093/mnras/stz937}

\bibitem[{{Diemer} {et~al.}(2017){Diemer}, {Sparre}, {Abramson}, \&
  {Torrey}}]{diemer:17}
{Diemer}, B., {Sparre}, M., {Abramson}, L.~E., \& {Torrey}, P. 2017, \apj, 839,
  26, \dodoi{10.3847/1538-4357/aa68e5}

\bibitem[{{Draine} \& {Li}(2007)}]{Draine:07}
{Draine}, B.~T., \& {Li}, A. 2007, \apj, 657, 810, \dodoi{10.1086/511055}

\bibitem[{{Dutton} {et~al.}(2010){Dutton}, {Conroy}, {van den Bosch}, {Prada},
  \& {More}}]{dutton:10}
{Dutton}, A.~A., {Conroy}, C., {van den Bosch}, F.~C., {Prada}, F., \& {More},
  S. 2010, \mnras, 407, 2, \dodoi{10.1111/j.1365-2966.2010.16911.x}

\bibitem[{{Dye}(2008)}]{dye:08}
{Dye}, S. 2008, \mnras, 389, 1293, \dodoi{10.1111/j.1365-2966.2008.13639.x}

\bibitem[{{Elahi} {et~al.}(2018){Elahi}, {Power}, {Lagos}, {Poulton}, \&
  {Robotham}}]{elahi:18}
{Elahi}, P.~J., {Power}, C., {Lagos}, C. d.~P., {Poulton}, R., \& {Robotham},
  A. S.~G. 2018, \mnras, 477, 616, \dodoi{10.1093/mnras/sty590}

\bibitem[{{Estrada-Carpenter} {et~al.}(2020){Estrada-Carpenter}, {Papovich},
  {Momcheva}, {Brammer}, {Simons}, {Bridge}, {Cleri}, {Ferguson},
  {Finkelstein}, {Giavalisco}, {Jung}, {Matharu}, {Trump}, \&
  {Weiner}}]{estrada-carpenter:20}
{Estrada-Carpenter}, V., {Papovich}, C., {Momcheva}, I., {et~al.} 2020, \apj,
  898, 171, \dodoi{10.3847/1538-4357/aba004}

\bibitem[{{Falc{\'o}n-Barroso} {et~al.}(2011){Falc{\'o}n-Barroso},
  {S{\'a}nchez-Bl{\'a}zquez}, {Vazdekis}, {Ricciardelli}, {Cardiel}, {Cenarro},
  {Gorgas}, \& {Peletier}}]{falcon-barroso:11}
{Falc{\'o}n-Barroso}, J., {S{\'a}nchez-Bl{\'a}zquez}, P., {Vazdekis}, A.,
  {et~al.} 2011, \aap, 532, A95, \dodoi{10.1051/0004-6361/201116842}

\bibitem[{{Ferland} {et~al.}(2013){Ferland}, {Porter}, {van Hoof}, {Williams},
  {Abel}, {Lykins}, {Shaw}, {Henney}, \& {Stancil}}]{ferland:13}
{Ferland}, G.~J., {Porter}, R.~L., {van Hoof}, P.~A.~M., {et~al.} 2013, \rmxaa,
  49, 137.
\newblock \doarXiv{1302.4485}

\bibitem[{{Feroz} \& {Hobson}(2008)}]{feroz:08}
{Feroz}, F., \& {Hobson}, M.~P. 2008, \mnras, 384, 449,
  \dodoi{10.1111/j.1365-2966.2007.12353.x}

\bibitem[{{Feroz} {et~al.}(2019){Feroz}, {Hobson}, {Cameron}, \&
  {Pettitt}}]{feroz:19}
{Feroz}, F., {Hobson}, M.~P., {Cameron}, E., \& {Pettitt}, A.~N. 2019, The Open
  Journal of Astrophysics, 2, 10, \dodoi{10.21105/astro.1306.2144}

\bibitem[{{Feroz} \& {Skilling}(2013)}]{feroz:13}
{Feroz}, F., \& {Skilling}, J. 2013, in American Institute of Physics
  Conference Series, Vol. 1553, Bayesian Inference and Maximum Entropy Methods
  in Science and Engineering: 32nd International Workshop on Bayesian Inference
  and Maximum Entropy Methods in Science and Engineering, ed. U.~{von
  Toussaint}, 106--113, \dodoi{10.1063/1.4819989}

\bibitem[{{Ferreras} {et~al.}(2019){Ferreras}, {Pasquali}, {Pirzkal}, {Pharo},
  {Malhotra}, {Rhoads}, {Hathi}, {Windhorst}, {Cimatti}, {Christensen},
  {Finkelstein}, {Grogin}, {Joshi}, {Kim}, {Koekemoer}, {O'Connell},
  {{\"O}stlin}, {Rothberg}, \& {Ryan}}]{ferreras:19}
{Ferreras}, I., {Pasquali}, A., {Pirzkal}, N., {et~al.} 2019, \mnras, 486,
  1358, \dodoi{10.1093/mnras/stz849}

\bibitem[{{Feulner} {et~al.}(2005){Feulner}, {Goranova}, {Drory}, {Hopp}, \&
  {Bender}}]{feulner:05}
{Feulner}, G., {Goranova}, Y., {Drory}, N., {Hopp}, U., \& {Bender}, R. 2005,
  \mnras, 358, L1, \dodoi{10.1111/j.1745-3933.2005.00012.x}

\bibitem[{{Forrest} {et~al.}(2020){Forrest}, {Marsan}, {Annunziatella},
  {Wilson}, {Muzzin}, {Marchesini}, {Cooper}, {Chan}, {McConachie}, {Gomez},
  {Kado-Fong}, {La Barbera}, {Lange-Vagle}, {Nantais}, {Nonino}, {Saracco},
  {Stefanon}, \& {van der Burg}}]{forrest:20}
{Forrest}, B., {Marsan}, Z.~C., {Annunziatella}, M., {et~al.} 2020, \apj, 903,
  47, \dodoi{10.3847/1538-4357/abb819}

\bibitem[{{Franx} {et~al.}(2008){Franx}, {van Dokkum}, {F{\"o}rster Schreiber},
  {Wuyts}, {Labb{\'e}}, \& {Toft}}]{franx:08}
{Franx}, M., {van Dokkum}, P.~G., {F{\"o}rster Schreiber}, N.~M., {et~al.}
  2008, \apj, 688, 770, \dodoi{10.1086/592431}

\bibitem[{{Gallazzi} {et~al.}(2008){Gallazzi}, {Brinchmann}, {Charlot}, \&
  {White}}]{gallazzi:08}
{Gallazzi}, A., {Brinchmann}, J., {Charlot}, S., \& {White}, S. D.~M. 2008,
  \mnras, 383, 1439, \dodoi{10.1111/j.1365-2966.2007.12632.x}

\bibitem[{{Gallazzi} {et~al.}(2006){Gallazzi}, {Charlot}, {Brinchmann}, \&
  {White}}]{gallazzi:06}
{Gallazzi}, A., {Charlot}, S., {Brinchmann}, J., \& {White}, S. D.~M. 2006,
  \mnras, 370, 1106, \dodoi{10.1111/j.1365-2966.2006.10548.x}

\bibitem[{{Gladders} {et~al.}(2013){Gladders}, {Oemler}, {Dressler},
  {Poggianti}, {Vulcani}, \& {Abramson}}]{gladders:13}
{Gladders}, M.~D., {Oemler}, A., {Dressler}, A., {et~al.} 2013, \apj, 770, 64,
  \dodoi{10.1088/0004-637X/770/1/64}

\bibitem[{{Graves} {et~al.}(2009){Graves}, {Faber}, \& {Schiavon}}]{graves:09}
{Graves}, G.~J., {Faber}, S.~M., \& {Schiavon}, R.~P. 2009, \apj, 698, 1590,
  \dodoi{10.1088/0004-637X/698/2/1590}

\bibitem[{{Greene} {et~al.}(2022){Greene}, {Bezanson}, {Ouchi}, {Silverman}, \&
  {the PFS Galaxy Evolution Working Group}}]{greene:22}
{Greene}, J., {Bezanson}, R., {Ouchi}, M., {Silverman}, J., \& {the PFS Galaxy
  Evolution Working Group}. 2022, arXiv e-prints, arXiv:2206.14908,
  \dodoi{10.48550/arXiv.2206.14908}

\bibitem[{{Hamadouche} {et~al.}(2023){Hamadouche}, {Carnall}, {McLure},
  {Dunlop}, {Begley}, {Cullen}, {McLeod}, {Donnan}, \&
  {Stanton}}]{hamadouche:23}
{Hamadouche}, M.~L., {Carnall}, A.~C., {McLure}, R.~J., {et~al.} 2023, \mnras,
  521, 5400, \dodoi{10.1093/mnras/stad773}

\bibitem[{{Harris} {et~al.}(2020){Harris}, {Millman}, {van der Walt},
  {Gommers}, {Virtanen}, {Cournapeau}, {Wieser}, {Taylor}, {Berg}, {Smith},
  {Kern}, {Picus}, {Hoyer}, {van Kerkwijk}, {Brett}, {Haldane}, {del R{\'\i}o},
  {Wiebe}, {Peterson}, {G{\'e}rard-Marchant}, {Sheppard}, {Reddy}, {Weckesser},
  {Abbasi}, {Gohlke}, \& {Oliphant}}]{harris+2020}
{Harris}, C.~R., {Millman}, K.~J., {van der Walt}, S.~J., {et~al.} 2020, \nat,
  585, 357, \dodoi{10.1038/s41586-020-2649-2}

\bibitem[{{Hill} {et~al.}(2017{\natexlab{a}}){Hill}, {Muzzin}, {Franx}, \&
  {Marchesini}}]{hill:17b}
{Hill}, A.~R., {Muzzin}, A., {Franx}, M., \& {Marchesini}, D.
  2017{\natexlab{a}}, \apjl, 849, L26, \dodoi{10.3847/2041-8213/aa951a}

\bibitem[{{Hill} {et~al.}(2017{\natexlab{b}}){Hill}, {Muzzin}, {Franx},
  {Clauwens}, {Schreiber}, {Marchesini}, {Stefanon}, {Labbe}, {Brammer},
  {Caputi}, {Fynbo}, {Milvang-Jensen}, {Skelton}, {van Dokkum}, \&
  {Whitaker}}]{hill:17a}
{Hill}, A.~R., {Muzzin}, A., {Franx}, M., {et~al.} 2017{\natexlab{b}}, \apj,
  837, 147, \dodoi{10.3847/1538-4357/aa61fe}

\bibitem[{{Hunter}(2007)}]{hunter2007}
{Hunter}, J.~D. 2007, Computing in Science and Engineering, 9, 90,
  \dodoi{10.1109/MCSE.2007.55}

\bibitem[{{Ibarra-Medel} {et~al.}(2016){Ibarra-Medel}, {S{\'a}nchez},
  {Avila-Reese}, {Hern{\'a}ndez-Toledo}, {Gonz{\'a}lez}, {Drory}, {Bundy},
  {Bizyaev}, {Cano-D{\'\i}az}, {Malanushenko}, {Pan}, {Roman-Lopes}, \&
  {Thomas}}]{ibarra-medel:16}
{Ibarra-Medel}, H.~J., {S{\'a}nchez}, S.~F., {Avila-Reese}, V., {et~al.} 2016,
  \mnras, 463, 2799, \dodoi{10.1093/mnras/stw2126}

\bibitem[{{Ilbert} {et~al.}(2015){Ilbert}, {Arnouts}, {Le Floc'h}, {Aussel},
  {Bethermin}, {Capak}, {Hsieh}, {Kajisawa}, {Karim}, {Le F{\`e}vre}, {Lee},
  {Lilly}, {McCracken}, {Michel-Dansac}, {Moutard}, {Renzini}, {Salvato},
  {Sanders}, {Scoville}, {Sheth}, {Silverman}, {Smol{\v{c}}i{\'c}},
  {Taniguchi}, \& {Tresse}}]{ilbert:15}
{Ilbert}, O., {Arnouts}, S., {Le Floc'h}, E., {et~al.} 2015, \aap, 579, A2,
  \dodoi{10.1051/0004-6361/201425176}

\bibitem[{{Iyer}(2019)}]{iyer:19}
{Iyer}, K. 2019, in American Astronomical Society Meeting Abstracts, Vol. 233,
  American Astronomical Society Meeting Abstracts \#233, 429.05

\bibitem[{{Iyer} \& {Gawiser}(2017)}]{iyer:17}
{Iyer}, K., \& {Gawiser}, E. 2017, \apj, 838, 127,
  \dodoi{10.3847/1538-4357/aa63f0}

\bibitem[{{Iyer} {et~al.}(2020){Iyer}, {Tacchella}, {Genel}, {Hayward},
  {Hernquist}, {Brooks}, {Caplar}, {Dav{\'e}}, {Diemer}, {Forbes}, {Gawiser},
  {Somerville}, \& {Starkenburg}}]{iyer:20}
{Iyer}, K.~G., {Tacchella}, S., {Genel}, S., {et~al.} 2020, \mnras, 498, 430,
  \dodoi{10.1093/mnras/staa2150}

\bibitem[{{Johnson} \& {Leja}(2017)}]{johnson:17}
{Johnson}, B., \& {Leja}, J. 2017, {Bd-J/Prospector: Initial Release}, v0.1,
  Zenodo,  Zenodo, \dodoi{10.5281/zenodo.1116491}

\bibitem[{{Johnson} {et~al.}(2021){Johnson}, {Leja}, {Conroy}, \&
  {Speagle}}]{Johnson:21}
{Johnson}, B.~D., {Leja}, J., {Conroy}, C., \& {Speagle}, J.~S. 2021, \apjs,
  254, 22, \dodoi{10.3847/1538-4365/abef67}

\bibitem[{Jones {et~al.}(2001)Jones, Oliphant, Peterson, {et~al.}}]{jones+2001}
Jones, E., Oliphant, T., Peterson, P., {et~al.} 2001, {SciPy}: Open source
  scientific tools for {Python}.
\newblock \url{http://www.scipy.org/}

\bibitem[{{Kauffmann} {et~al.}(2004){Kauffmann}, {White}, {Heckman},
  {M{\'e}nard}, {Brinchmann}, {Charlot}, {Tremonti}, \&
  {Brinkmann}}]{kauffmann:04}
{Kauffmann}, G., {White}, S. D.~M., {Heckman}, T.~M., {et~al.} 2004, \mnras,
  353, 713, \dodoi{10.1111/j.1365-2966.2004.08117.x}

\bibitem[{{Khullar} {et~al.}(2022){Khullar}, {Bayliss}, {Gladders}, {Kim},
  {Calzadilla}, {Strazzullo}, {Bleem}, {Mahler}, {McDonald}, {Floyd},
  {Reichardt}, {Ruppin}, {Saro}, {Sharon}, {Somboonpanyakul}, {Stalder}, \&
  {Stark}}]{khullar:22}
{Khullar}, G., {Bayliss}, M.~B., {Gladders}, M.~D., {et~al.} 2022, \apj, 934,
  177, \dodoi{10.3847/1538-4357/ac7c0c}

\bibitem[{{Kriek} \& {Conroy}(2013)}]{Kriek_2013ApJ...775L..16K}
{Kriek}, M., \& {Conroy}, C. 2013, \apjl, 775, L16,
  \dodoi{10.1088/2041-8205/775/1/L16}

\bibitem[{{Kriek} {et~al.}(2015){Kriek}, {Shapley}, {Reddy}, {Siana}, {Coil},
  {Mobasher}, {Freeman}, {de Groot}, {Price}, {Sanders}, {Shivaei}, {Brammer},
  {Momcheva}, {Skelton}, {van Dokkum}, {Whitaker}, {Aird}, {Azadi}, {Kassis},
  {Bullock}, {Conroy}, {Dav{\'e}}, {Kere{\v{s}}}, \& {Krumholz}}]{kriek:15}
{Kriek}, M., {Shapley}, A.~E., {Reddy}, N.~A., {et~al.} 2015, \apjs, 218, 15,
  \dodoi{10.1088/0067-0049/218/2/15}

\bibitem[{{Kroupa}(2001)}]{kroupa:01}
{Kroupa}, P. 2001, \mnras, 322, 231, \dodoi{10.1046/j.1365-8711.2001.04022.x}

\bibitem[{{Labb{\'e}} {et~al.}(2023){Labb{\'e}}, {van Dokkum}, {Nelson},
  {Bezanson}, {Suess}, {Leja}, {Brammer}, {Whitaker}, {Mathews}, {Stefanon}, \&
  {Wang}}]{labbe:23}
{Labb{\'e}}, I., {van Dokkum}, P., {Nelson}, E., {et~al.} 2023, \nat, 616, 266,
  \dodoi{10.1038/s41586-023-05786-2}

\bibitem[{{Le Fevre} {et~al.}(2000){Le Fevre}, {Saisse}, {Mancini},
  {Vettolani}, {Maccagni}, {Picat}, {Mellier}, {Mazure}, {Cuby}, {Delabre},
  {Garilli}, {Hill}, {Prieto}, {Voet}, {Arnold}, {Brau-Nogue}, {Cascone},
  {Conconi}, {Finger}, {Huster}, {Laloge}, {Lucuix}, {Mattaini}, {Schipani},
  {Waultier}, {Zerbi}, {Avila}, {Beletic}, {D'Odorico}, {Moorwood}, {Monnet},
  \& {Reyes Moreno}}]{lefevre:03}
{Le Fevre}, O., {Saisse}, M., {Mancini}, D., {et~al.} 2000, in Society of
  Photo-Optical Instrumentation Engineers (SPIE) Conference Series, Vol. 4008,
  Optical and IR Telescope Instrumentation and Detectors, ed. M.~{Iye} \& A.~F.
  {Moorwood}, 546--557, \dodoi{10.1117/12.395513}

\bibitem[{{Leitner}(2012)}]{leitner:12}
{Leitner}, S.~N. 2012, \apj, 745, 149, \dodoi{10.1088/0004-637X/745/2/149}

\bibitem[{{Leja} {et~al.}(2019){Leja}, {Carnall}, {Johnson}, {Conroy}, \&
  {Speagle}}]{leja:19:a}
{Leja}, J., {Carnall}, A.~C., {Johnson}, B.~D., {Conroy}, C., \& {Speagle},
  J.~S. 2019, \apj, 876, 3, \dodoi{10.3847/1538-4357/ab133c}

\bibitem[{{Leja} {et~al.}(2017){Leja}, {Johnson}, {Conroy}, {van Dokkum}, \&
  {Byler}}]{leja:17}
{Leja}, J., {Johnson}, B.~D., {Conroy}, C., {van Dokkum}, P.~G., \& {Byler}, N.
  2017, \apj, 837, 170, \dodoi{10.3847/1538-4357/aa5ffe}

\bibitem[{{Looser} {et~al.}(2023){Looser}, {D'Eugenio}, {Maiolino}, {Witstok},
  {Sandles}, {Curtis-Lake}, {Chevallard}, {Tacchella}, {Johnson}, {Baker},
  {Suess}, {Carniani}, {Ferruit}, {Arribas}, {Bonaventura}, {Bunker},
  {Cameron}, {Charlot}, {Curti}, {de Graaff}, {Maseda}, {Rawle}, {Rix},
  {Rodriguez Del Pino}, {Smit}, {{\"U}bler}, {Willott}, {Alberts}, {Egami},
  {Eisenstein}, {Endsley}, {Hausen}, {Rieke}, {Robertson}, {Shivaei},
  {Williams}, {Boyett}, {Chen}, {Ji}, {Jones}, {Kumari}, {Nelson}, {Perna},
  {Saxena}, \& {Scholtz}}]{looser:23}
{Looser}, T.~J., {D'Eugenio}, F., {Maiolino}, R., {et~al.} 2023, arXiv
  e-prints, arXiv:2302.14155, \dodoi{10.48550/arXiv.2302.14155}

\bibitem[{{Lower} {et~al.}(2020){Lower}, {Narayanan}, {Leja}, {Johnson},
  {Conroy}, \& {Dav{\'e}}}]{lower:20}
{Lower}, S., {Narayanan}, D., {Leja}, J., {et~al.} 2020, \apj, 904, 33,
  \dodoi{10.3847/1538-4357/abbfa7}

\bibitem[{{Madau} \& {Dickinson}(2014)}]{madau:14}
{Madau}, P., \& {Dickinson}, M. 2014, \araa, 52, 415,
  \dodoi{10.1146/annurev-astro-081811-125615}

\bibitem[{{Maiolino} {et~al.}(2020){Maiolino}, {Cirasuolo}, {Afonso}, {Bauer},
  {Bowler}, {Cucciati}, {Daddi}, {De Lucia}, {Evans}, {Flores}, {Gargiulo},
  {Garilli}, {Jablonka}, {Jarvis}, {Kneib}, {Lilly}, {Looser}, {Magliocchetti},
  {Man}, {Mannucci}, {Maurogordato}, {McLure}, {Norberg}, {Oesch}, {Oliva},
  {Paltani}, {Pappalardo}, {Peng}, {Pentericci}, {Pozzetti}, {Renzini},
  {Rodrigues}, {Royer}, {Serjeant}, {Vanzi}, {Wild}, \&
  {Zamorani}}]{maiolino:20}
{Maiolino}, R., {Cirasuolo}, M., {Afonso}, J., {et~al.} 2020, The Messenger,
  180, 24, \dodoi{10.18727/0722-6691/5197}

\bibitem[{{Maltby} {et~al.}(2018){Maltby}, {Almaini}, {Wild}, {Hatch},
  {Hartley}, {Simpson}, {Rowlands}, \& {Socolovsky}}]{maltby:18}
{Maltby}, D.~T., {Almaini}, O., {Wild}, V., {et~al.} 2018, \mnras, 480, 381,
  \dodoi{10.1093/mnras/sty1794}

\bibitem[{{Mao} {et~al.}(2022){Mao}, {Kodama}, {P{\'e}rez-Mart{\'\i}nez},
  {Suzuki}, {Yamamoto}, \& {Adachi}}]{mao:22}
{Mao}, Z., {Kodama}, T., {P{\'e}rez-Mart{\'\i}nez}, J.~M., {et~al.} 2022, \aap,
  666, A141, \dodoi{10.1051/0004-6361/202243733}

\bibitem[{{Martins}(2021)}]{martins:20}
{Martins}, L.~P. 2021, in Galaxy Evolution and Feedback across Different
  Environments, ed. T.~{Storchi Bergmann}, W.~{Forman}, R.~{Overzier}, \&
  R.~{Riffel}, Vol. 359, 386--390, \dodoi{10.1017/S1743921320001647}

\bibitem[{{McDermid} {et~al.}(2015){McDermid}, {Alatalo}, {Blitz}, {Bournaud},
  {Bureau}, {Cappellari}, {Crocker}, {Davies}, {Davis}, {de Zeeuw}, {Duc},
  {Emsellem}, {Khochfar}, {Krajnovi{\'c}}, {Kuntschner}, {Morganti}, {Naab},
  {Oosterloo}, {Sarzi}, {Scott}, {Serra}, {Weijmans}, \& {Young}}]{mcDermid:15}
{McDermid}, R.~M., {Alatalo}, K., {Blitz}, L., {et~al.} 2015, \mnras, 448,
  3484, \dodoi{10.1093/mnras/stv105}

\bibitem[{{McKinney} {et~al.}(2023){McKinney}, {Finnerty}, {Casey}, {Franco},
  {Long}, {Fujimoto}, {Zavala}, {Cooper}, {Akins}, {Pope}, {Armus}, {Soifer},
  {Larson}, {Matthews}, {Melbourne}, \& {Cushing}}]{mckinney:23}
{McKinney}, J., {Finnerty}, L., {Casey}, C.~M., {et~al.} 2023, \apjl, 946, L39,
  \dodoi{10.3847/2041-8213/acc322}

\bibitem[{{McLure} {et~al.}(2018){McLure}, {Pentericci}, {Cimatti}, {Dunlop},
  {Elbaz}, {Fontana}, {Nandra}, {Amorin}, {Bolzonella}, {Bongiorno}, {Carnall},
  {Castellano}, {Cirasuolo}, {Cucciati}, {Cullen}, {De Barros}, {Finkelstein},
  {Fontanot}, {Franzetti}, {Fumana}, {Gargiulo}, {Garilli}, {Guaita},
  {Hartley}, {Iovino}, {Jarvis}, {Juneau}, {Karman}, {Maccagni}, {Marchi},
  {M{\'a}rmol-Queralt{\'o}}, {Pompei}, {Pozzetti}, {Scodeggio}, {Sommariva},
  {Talia}, {Almaini}, {Balestra}, {Bardelli}, {Bell}, {Bourne}, {Bowler},
  {Brusa}, {Buitrago}, {Caputi}, {Cassata}, {Charlot}, {Citro}, {Cresci},
  {Cristiani}, {Curtis-Lake}, {Dickinson}, {Fazio}, {Ferguson}, {Fiore},
  {Franco}, {Fynbo}, {Galametz}, {Georgakakis}, {Giavalisco}, {Grazian},
  {Hathi}, {Jung}, {Kim}, {Koekemoer}, {Khusanova}, {Le F{\`e}vre}, {Lotz},
  {Mannucci}, {Maltby}, {Matsuoka}, {McLeod}, {Mendez-Hernandez},
  {Mendez-Abreu}, {Mignoli}, {Moresco}, {Mortlock}, {Nonino}, {Pannella},
  {Papovich}, {Popesso}, {Rosario}, {Salvato}, {Santini}, {Schaerer},
  {Schreiber}, {Stark}, {Tasca}, {Thomas}, {Treu}, {Vanzella}, {Wild},
  {Williams}, {Zamorani}, \& {Zucca}}]{mclure:18}
{McLure}, R.~J., {Pentericci}, L., {Cimatti}, A., {et~al.} 2018, \mnras, 479,
  25, \dodoi{10.1093/mnras/sty1213}

\bibitem[{{Mortlock} {et~al.}(2017){Mortlock}, {McLure}, {Bowler}, {McLeod},
  {M{\'a}rmol-Queralt{\'o}}, {Parsa}, {Dunlop}, \& {Bruce}}]{mortlock:17}
{Mortlock}, A., {McLure}, R.~J., {Bowler}, R. A.~A., {et~al.} 2017, \mnras,
  465, 672, \dodoi{10.1093/mnras/stw2728}

\bibitem[{{Muzzin}(2017)}]{muzzin:17}
{Muzzin}, A. 2017, in Early stages of Galaxy Cluster Formation, 24,
  \dodoi{10.5281/zenodo.833363}

\bibitem[{{Muzzin} {et~al.}(2013{\natexlab{a}}){Muzzin}, {Marchesini},
  {Stefanon}, {Franx}, {Milvang-Jensen}, {Dunlop}, {Fynbo}, {Brammer},
  {Labb{\'e}}, \& {van Dokkum}}]{muzzin:13}
{Muzzin}, A., {Marchesini}, D., {Stefanon}, M., {et~al.} 2013{\natexlab{a}},
  \apjs, 206, 8, \dodoi{10.1088/0067-0049/206/1/8}

\bibitem[{{Muzzin} {et~al.}(2013{\natexlab{b}}){Muzzin}, {Marchesini},
  {Stefanon}, {Franx}, {Milvang-Jensen}, {Dunlop}, {Fynbo}, {Brammer},
  {Labb{\'e}}, \& {van Dokkum}}]{muzzin:13:a}
---. 2013{\natexlab{b}}, \apjs, 206, 8, \dodoi{10.1088/0067-0049/206/1/8}

\bibitem[{{Muzzin} {et~al.}(2013{\natexlab{c}}){Muzzin}, {Marchesini},
  {Stefanon}, {Franx}, {McCracken}, {Milvang-Jensen}, {Dunlop}, {Fynbo},
  {Brammer}, {Labb{\'e}}, \& {van Dokkum}}]{muzzin+13:b}
---. 2013{\natexlab{c}}, \apj, 777, 18, \dodoi{10.1088/0004-637X/777/1/18}

\bibitem[{{Nanayakkara} {et~al.}(2023){Nanayakkara}, {Glazebrook}, {Jacobs},
  {Bonchi}, {Castellano}, {Fontana}, {Mason}, {Merlin}, {Morishita}, {Paris},
  {Trenti}, {Treu}, {Calabr{\`o}}, {Boyett}, {Bradac}, {Leethochawalit},
  {Marchesini}, {Santini}, {Strait}, {Vanzella}, {Vulcani}, {Wang}, \&
  {Yang}}]{nanayakkara:23}
{Nanayakkara}, T., {Glazebrook}, K., {Jacobs}, C., {et~al.} 2023, \apjl, 947,
  L26, \dodoi{10.3847/2041-8213/acbfb9}

\bibitem[{{Nelan} {et~al.}(2005){Nelan}, {Smith}, {Hudson}, {Wegner}, {Lucey},
  {Moore}, {Quinney}, \& {Suntzeff}}]{nelan:06}
{Nelan}, J.~E., {Smith}, R.~J., {Hudson}, M.~J., {et~al.} 2005, \apj, 632, 137,
  \dodoi{10.1086/431962}

\bibitem[{{Nelson} {et~al.}(2019){Nelson}, {Springel}, {Pillepich},
  {Rodriguez-Gomez}, {Torrey}, {Genel}, {Vogelsberger}, {Pakmor}, {Marinacci},
  {Weinberger}, {Kelley}, {Lovell}, {Diemer}, \& {Hernquist}}]{nelson:19}
{Nelson}, D., {Springel}, V., {Pillepich}, A., {et~al.} 2019, Computational
  Astrophysics and Cosmology, 6, 2, \dodoi{10.1186/s40668-019-0028-x}

\bibitem[{{Newman} {et~al.}(2013){Newman}, {Cooper}, {Davis}, {Faber}, {Coil},
  {Guhathakurta}, {Koo}, {Phillips}, {Conroy}, {Dutton}, {Finkbeiner}, {Gerke},
  {Rosario}, {Weiner}, {Willmer}, {Yan}, {Harker}, {Kassin}, {Konidaris},
  {Lai}, {Madgwick}, {Noeske}, {Wirth}, {Connolly}, {Kaiser}, {Kirby},
  {Lemaux}, {Lin}, {Lotz}, {Luppino}, {Marinoni}, {Matthews}, {Metevier}, \&
  {Schiavon}}]{newman:12}
{Newman}, J.~A., {Cooper}, M.~C., {Davis}, M., {et~al.} 2013, \apjs, 208, 5,
  \dodoi{10.1088/0067-0049/208/1/5}

\bibitem[{{Ocvirk} {et~al.}(2006){Ocvirk}, {Pichon}, {Lan{\c{c}}on}, \&
  {Thi{\'e}baut}}]{ocvirk:06}
{Ocvirk}, P., {Pichon}, C., {Lan{\c{c}}on}, A., \& {Thi{\'e}baut}, E. 2006,
  \mnras, 365, 46, \dodoi{10.1111/j.1365-2966.2005.09182.x}

\bibitem[{{Olsen} {et~al.}(2021){Olsen}, {Gawiser}, {Iyer}, {McQuinn},
  {Johnson}, {Telford}, {Wright}, {Broussard}, \& {Kurczynski}}]{olsen:21}
{Olsen}, C., {Gawiser}, E., {Iyer}, K., {et~al.} 2021, \apj, 913, 45,
  \dodoi{10.3847/1538-4357/abf3c2}

\bibitem[{{Pacifici} {et~al.}(2012){Pacifici}, {Charlot}, {Blaizot}, \&
  {Brinchmann}}]{pacifici:12}
{Pacifici}, C., {Charlot}, S., {Blaizot}, J., \& {Brinchmann}, J. 2012, \mnras,
  421, 2002, \dodoi{10.1111/j.1365-2966.2012.20431.x}

\bibitem[{{Pacifici} {et~al.}(2016{\natexlab{a}}){Pacifici}, {Oh}, {Oh}, {Lee},
  \& {Yi}}]{pacifici:16:b}
{Pacifici}, C., {Oh}, S., {Oh}, K., {Lee}, J., \& {Yi}, S.~K.
  2016{\natexlab{a}}, \apj, 824, 45, \dodoi{10.3847/0004-637X/824/1/45}

\bibitem[{{Pacifici} {et~al.}(2016{\natexlab{b}}){Pacifici}, {Kassin},
  {Weiner}, {Holden}, {Gardner}, {Faber}, {Ferguson}, {Koo}, {Primack}, {Bell},
  {Dekel}, {Gawiser}, {Giavalisco}, {Rafelski}, {Simons}, {Barro}, {Croton},
  {Dav{\'e}}, {Fontana}, {Grogin}, {Koekemoer}, {Lee}, {Salmon}, {Somerville},
  \& {Behroozi}}]{pacifici:16:a}
{Pacifici}, C., {Kassin}, S.~A., {Weiner}, B.~J., {et~al.} 2016{\natexlab{b}},
  \apj, 832, 79, \dodoi{10.3847/0004-637X/832/1/79}

\bibitem[{{Pacifici} {et~al.}(2023){Pacifici}, {Iyer}, {Mobasher}, {da Cunha},
  {Acquaviva}, {Burgarella}, {Calistro Rivera}, {Carnall}, {Chang}, {Chartab},
  {Cooke}, {Fairhurst}, {Kartaltepe}, {Leja}, {Ma{\l}ek}, {Salmon}, {Torelli},
  {Vidal-Garc{\'\i}a}, {Boquien}, {Brammer}, {Brown}, {Capak}, {Chevallard},
  {Circosta}, {Croton}, {Davidzon}, {Dickinson}, {Duncan}, {Faber}, {Ferguson},
  {Fontana}, {Guo}, {Haeussler}, {Hemmati}, {Jafariyazani}, {Kassin}, {Larson},
  {Lee}, {Mantha}, {Marchi}, {Nayyeri}, {Newman}, {Pandya}, {Pforr}, {Reddy},
  {Sanders}, {Shah}, {Shahidi}, {Stevans}, {Triani}, {Tyler}, {Vanderhoof}, {de
  la Vega}, {Wang}, \& {Weston}}]{pacifici:22}
{Pacifici}, C., {Iyer}, K.~G., {Mobasher}, B., {et~al.} 2023, \apj, 944, 141,
  \dodoi{10.3847/1538-4357/acacff}

\bibitem[{{Panter} {et~al.}(2007){Panter}, {Jimenez}, {Heavens}, \&
  {Charlot}}]{panter:07}
{Panter}, B., {Jimenez}, R., {Heavens}, A.~F., \& {Charlot}, S. 2007, \mnras,
  378, 1550, \dodoi{10.1111/j.1365-2966.2007.11909.x}

\bibitem[{{Pforr} {et~al.}(2012){Pforr}, {Maraston}, \& {Tonini}}]{pforr:12}
{Pforr}, J., {Maraston}, C., \& {Tonini}, C. 2012, \mnras, 422, 3285,
  \dodoi{10.1111/j.1365-2966.2012.20848.x}

\bibitem[{{Rudie} {et~al.}(2012){Rudie}, {Steidel}, {Trainor}, {Rakic},
  {Bogosavljevi{\'c}}, {Pettini}, {Reddy}, {Shapley}, {Erb}, \&
  {Law}}]{rudie:12}
{Rudie}, G.~C., {Steidel}, C.~C., {Trainor}, R.~F., {et~al.} 2012, \apj, 750,
  67, \dodoi{10.1088/0004-637X/750/1/67}

\bibitem[{{S{\'a}nchez-Bl{\'a}zquez} {et~al.}(2011){S{\'a}nchez-Bl{\'a}zquez},
  {Ocvirk}, {Gibson}, {P{\'e}rez}, \& {Peletier}}]{sanchez-blazquez:11}
{S{\'a}nchez-Bl{\'a}zquez}, P., {Ocvirk}, P., {Gibson}, B.~K., {P{\'e}rez}, I.,
  \& {Peletier}, R.~F. 2011, \mnras, 415, 709,
  \dodoi{10.1111/j.1365-2966.2011.18749.x}

\bibitem[{{Schawinski} {et~al.}(2014){Schawinski}, {Urry}, {Simmons},
  {Fortson}, {Kaviraj}, {Keel}, {Lintott}, {Masters}, {Nichol}, {Sarzi},
  {Skibba}, {Treister}, {Willett}, {Wong}, \& {Yi}}]{schawinski:14}
{Schawinski}, K., {Urry}, C.~M., {Simmons}, B.~D., {et~al.} 2014, \mnras, 440,
  889, \dodoi{10.1093/mnras/stu327}

\bibitem[{{Schiavon} {et~al.}(2006){Schiavon}, {Faber}, {Konidaris}, {Graves},
  {Willmer}, {Weiner}, {Coil}, {Cooper}, {Davis}, {Harker}, {Koo}, {Newman}, \&
  {Yan}}]{schiavon:06}
{Schiavon}, R.~P., {Faber}, S.~M., {Konidaris}, N., {et~al.} 2006, \apjl, 651,
  L93, \dodoi{10.1086/509074}

\bibitem[{{Schreiber} {et~al.}(2016){Schreiber}, {Elbaz}, {Pannella}, {Ciesla},
  {Wang}, {Koekemoer}, {Rafelski}, \& {Daddi}}]{schreiber:16}
{Schreiber}, C., {Elbaz}, D., {Pannella}, M., {et~al.} 2016, \aap, 589, A35,
  \dodoi{10.1051/0004-6361/201527200}

\bibitem[{{Shamshiri} {et~al.}(2015){Shamshiri}, {Thomas}, {Henriques},
  {Tojeiro}, {Lemson}, {Oliver}, \& {Wilkins}}]{shamshiri:13}
{Shamshiri}, S., {Thomas}, P.~A., {Henriques}, B.~M., {et~al.} 2015, \mnras,
  451, 2681, \dodoi{10.1093/mnras/stv883}

\bibitem[{{Sharma} {et~al.}(2021){Sharma}, {Hayden}, {Bland-Hawthorn},
  {Stello}, {Buder}, {Zinn}, {Kallinger}, {Asplund}, {De Silva}, {D'Orazi},
  {Freeman}, {Kos}, {Lewis}, {Lin}, {Lind}, {Martell}, {Simpson}, {Wittenmyer},
  {Zucker}, {Zwitter}, {Chen}, {Cotar}, {Esdaile}, {Hon}, {Horner}, {Huber},
  {Kafle}, {Khanna}, {Ting}, {Nataf}, {Nordlander}, {Saadon}, {Tepper-Garcia},
  {Tinney}, {Traven}, {Watson}, {Wright}, \& {Wyse}}]{sharma:21}
{Sharma}, S., {Hayden}, M.~R., {Bland-Hawthorn}, J., {et~al.} 2021, \mnras,
  506, 1761, \dodoi{10.1093/mnras/stab1086}

\bibitem[{{Simha} {et~al.}(2014){Simha}, {Weinberg}, {Conroy}, {Dave},
  {Fardal}, {Katz}, \& {Oppenheimer}}]{simha:14}
{Simha}, V., {Weinberg}, D.~H., {Conroy}, C., {et~al.} 2014, arXiv e-prints,
  arXiv:1404.0402.
\newblock \doarXiv{1404.0402}

\bibitem[{{Siudek} {et~al.}(2017){Siudek}, {Ma{\l}ek}, {Scodeggio}, {Garilli},
  {Pollo}, {Haines}, {Fritz}, {Bolzonella}, {de la Torre}, {Granett}, {Guzzo},
  {Abbas}, {Adami}, {Bottini}, {Cappi}, {Cucciati}, {De Lucia}, {Davidzon},
  {Franzetti}, {Iovino}, {Krywult}, {Le Brun}, {Le F{\`e}vre}, {Maccagni},
  {Marchetti}, {Marulli}, {Polletta}, {Tasca}, {Tojeiro}, {Vergani},
  {Zanichelli}, {Arnouts}, {Bel}, {Branchini}, {Ilbert}, {Gargiulo},
  {Moscardini}, {Takeuchi}, \& {Zamorani}}]{siudek:17}
{Siudek}, M., {Ma{\l}ek}, K., {Scodeggio}, M., {et~al.} 2017, \aap, 597, A107,
  \dodoi{10.1051/0004-6361/201628951}

\bibitem[{Skilling(2006)}]{10.1214/06-BA127}
Skilling, J. 2006, Bayesian Analysis, 1, 833 , \dodoi{10.1214/06-BA127}

\bibitem[{{Smethurst} {et~al.}(2015){Smethurst}, {Lintott}, {Simmons},
  {Schawinski}, {Marshall}, {Bamford}, {Fortson}, {Kaviraj}, {Masters},
  {Melvin}, {Nichol}, {Skibba}, \& {Willett}}]{smethurst:15}
{Smethurst}, R.~J., {Lintott}, C.~J., {Simmons}, B.~D., {et~al.} 2015, \mnras,
  450, 435, \dodoi{10.1093/mnras/stv161}

\bibitem[{{Sobral} {et~al.}(2022){Sobral}, {van der Wel}, {Bezanson}, {Bell},
  {Muzzin}, {D'Eugenio}, {Darvish}, {Gallazzi}, {Wu}, {Maseda}, {Matthee},
  {Paulino-Afonso}, {Straatman}, \& {van Dokkum}}]{sobral:22}
{Sobral}, D., {van der Wel}, A., {Bezanson}, R., {et~al.} 2022, \apj, 926, 117,
  \dodoi{10.3847/1538-4357/ac4419}

\bibitem[{{Sparre} {et~al.}(2015){Sparre}, {Hayward}, {Springel},
  {Vogelsberger}, {Genel}, {Torrey}, {Nelson}, {Sijacki}, \&
  {Hernquist}}]{sparre:14}
{Sparre}, M., {Hayward}, C.~C., {Springel}, V., {et~al.} 2015, \mnras, 447,
  3548, \dodoi{10.1093/mnras/stu2713}

\bibitem[{{Speagle}(2020)}]{speagle:20}
{Speagle}, J.~S. 2020, \mnras, 493, 3132, \dodoi{10.1093/mnras/staa278}

\bibitem[{{Steidel} {et~al.}(2014){Steidel}, {Rudie}, {Strom}, {Pettini},
  {Reddy}, {Shapley}, {Trainor}, {Erb}, {Turner}, {Konidaris}, {Kulas}, {Mace},
  {Matthews}, \& {McLean}}]{steidel:14}
{Steidel}, C.~C., {Rudie}, G.~C., {Strom}, A.~L., {et~al.} 2014, \apj, 795,
  165, \dodoi{10.1088/0004-637X/795/2/165}

\bibitem[{{Straatman} {et~al.}(2018){Straatman}, {van der Wel}, {Bezanson},
  {Pacifici}, {Gallazzi}, {Wu}, {Noeske}, {Bari{\v{s}}i{\'c}}, {Bell},
  {Brammer}, {Calhau}, {Chauke}, {Franx}, {van Houdt}, {Labb{\'e}}, {Maseda},
  {Mu{\~n}oz-Mateos}, {Muzzin}, {van de Sande}, {Sobral}, \&
  {Spilker}}]{straatman:18}
{Straatman}, C. M.~S., {van der Wel}, A., {Bezanson}, R., {et~al.} 2018, \apjs,
  239, 27, \dodoi{10.3847/1538-4365/aae37a}

\bibitem[{{Strom} {et~al.}(2017){Strom}, {Steidel}, {Rudie}, {Trainor},
  {Pettini}, \& {Reddy}}]{strom:17}
{Strom}, A.~L., {Steidel}, C.~C., {Rudie}, G.~C., {et~al.} 2017, \apj, 836,
  164, \dodoi{10.3847/1538-4357/836/2/164}

\bibitem[{{Suess} {et~al.}(2022{\natexlab{a}}){Suess}, {Leja}, {Johnson},
  {Bezanson}, {Greene}, {Kriek}, {Lower}, {Narayanan}, {Setton}, \&
  {Spilker}}]{suess:22a}
{Suess}, K.~A., {Leja}, J., {Johnson}, B.~D., {et~al.} 2022{\natexlab{a}},
  \apj, 935, 146, \dodoi{10.3847/1538-4357/ac82b0}

\bibitem[{{Suess} {et~al.}(2022{\natexlab{b}}){Suess}, {Kriek}, {Bezanson},
  {Greene}, {Setton}, {Spilker}, {Feldmann}, {Goulding}, {Johnson}, {Leja},
  {Narayanan}, {Hall-Hooper}, {Hunt}, {Lower}, \& {Verrico}}]{suess:22b}
{Suess}, K.~A., {Kriek}, M., {Bezanson}, R., {et~al.} 2022{\natexlab{b}}, \apj,
  926, 89, \dodoi{10.3847/1538-4357/ac404a}

\bibitem[{{Tacchella} {et~al.}(2022){Tacchella}, {Conroy}, {Faber}, {Johnson},
  {Leja}, {Barro}, {Cunningham}, {Deason}, {Guhathakurta}, {Guo}, {Hernquist},
  {Koo}, {McKinnon}, {Rockosi}, {Speagle}, {van Dokkum}, \&
  {Yesuf}}]{tacchella:22}
{Tacchella}, S., {Conroy}, C., {Faber}, S.~M., {et~al.} 2022, \apj, 926, 134,
  \dodoi{10.3847/1538-4357/ac449b}

\bibitem[{{Taylor} {et~al.}(2015){Taylor}, {Hopkins}, {Baldry},
  {Bland-Hawthorn}, {Brown}, {Colless}, {Driver}, {Norberg}, {Robotham},
  {Alpaslan}, {Brough}, {Cluver}, {Gunawardhana}, {Kelvin}, {Liske},
  {Conselice}, {Croom}, {Foster}, {Jarrett}, {Lara-Lopez}, \&
  {Loveday}}]{taylor:15}
{Taylor}, E.~N., {Hopkins}, A.~M., {Baldry}, I.~K., {et~al.} 2015, \mnras, 446,
  2144, \dodoi{10.1093/mnras/stu1900}

\bibitem[{{Taylor} {et~al.}(2022){Taylor}, {Bezanson}, {van der Wel}, {Pearl},
  {Bell}, {D'Eugenio}, {Franx}, {Maseda}, {Muzzin}, {Sobral}, {Straatman},
  {Whitaker}, \& {Wu}}]{taylor:22}
{Taylor}, L., {Bezanson}, R., {van der Wel}, A., {et~al.} 2022, \apj, 939, 90,
  \dodoi{10.3847/1538-4357/ac9796}

\bibitem[{{Thomas} {et~al.}(2005){Thomas}, {Maraston}, {Bender}, \& {Mendes de
  Oliveira}}]{thomas:05}
{Thomas}, D., {Maraston}, C., {Bender}, R., \& {Mendes de Oliveira}, C. 2005,
  \apj, 621, 673, \dodoi{10.1086/426932}

\bibitem[{{Thomas} {et~al.}(2010){Thomas}, {Maraston}, {Schawinski}, {Sarzi},
  \& {Silk}}]{thomas:10}
{Thomas}, D., {Maraston}, C., {Schawinski}, K., {Sarzi}, M., \& {Silk}, J.
  2010, \mnras, 404, 1775, \dodoi{10.1111/j.1365-2966.2010.16427.x}

\bibitem[{{Thomas} {et~al.}(2017){Thomas}, {Le F{\`e}vre}, {Scodeggio},
  {Cassata}, {Garilli}, {Le Brun}, {Lemaux}, {Maccagni}, {Pforr}, {Tasca},
  {Zamorani}, {Bardelli}, {Hathi}, {Tresse}, {Zucca}, \&
  {Koekemoer}}]{thomas:17}
{Thomas}, R., {Le F{\`e}vre}, O., {Scodeggio}, M., {et~al.} 2017, \aap, 602,
  A35, \dodoi{10.1051/0004-6361/201628141}

\bibitem[{{Tojeiro} {et~al.}(2009){Tojeiro}, {Wilkins}, {Heavens}, {Panter}, \&
  {Jimenez}}]{tojeiro:09}
{Tojeiro}, R., {Wilkins}, S., {Heavens}, A.~F., {Panter}, B., \& {Jimenez}, R.
  2009, \apjs, 185, 1, \dodoi{10.1088/0067-0049/185/1/1}

\bibitem[{{Torrey} {et~al.}(2018){Torrey}, {Vogelsberger}, {Hernquist},
  {McKinnon}, {Marinacci}, {Simcoe}, {Springel}, {Pillepich}, {Naiman},
  {Pakmor}, {Weinberger}, {Nelson}, \& {Genel}}]{torrey:18}
{Torrey}, P., {Vogelsberger}, M., {Hernquist}, L., {et~al.} 2018, \mnras, 477,
  L16, \dodoi{10.1093/mnrasl/sly031}

\bibitem[{{van der Wel} {et~al.}(2016){van der Wel}, {Noeske}, {Bezanson},
  {Pacifici}, {Gallazzi}, {Franx}, {Mu{\~n}oz-Mateos}, {Bell}, {Brammer},
  {Charlot}, {Chauk{\'e}}, {Labb{\'e}}, {Maseda}, {Muzzin}, {Rix}, {Sobral},
  {van de Sande}, {van Dokkum}, {Wild}, \& {Wolf}}]{vanderwel:16}
{van der Wel}, A., {Noeske}, K., {Bezanson}, R., {et~al.} 2016, \apjs, 223, 29,
  \dodoi{10.3847/0067-0049/223/2/29}

\bibitem[{{van der Wel} {et~al.}(2021){van der Wel}, {Bezanson}, {D'Eugenio},
  {Straatman}, {Franx}, {van Houdt}, {Maseda}, {Gallazzi}, {Wu}, {Pacifici},
  {Barisic}, {Brammer}, {Munoz-Mateos}, {Vervalcke}, {Zibetti}, {Sobral}, {de
  Graaff}, {Calhau}, {Kaushal}, {Muzzin}, {Bell}, \& {van
  Dokkum}}]{vanderwel:21}
{van der Wel}, A., {Bezanson}, R., {D'Eugenio}, F., {et~al.} 2021, \apjs, 256,
  44, \dodoi{10.3847/1538-4365/ac1356}

\bibitem[{{van der Wel} {et~al.}(2022){van der Wel}, {van Houdt}, {Bezanson},
  {Franx}, {D'Eugenio}, {Straatman}, {Bell}, {Muzzin}, {Sobral}, {Maseda}, {de
  Graaff}, \& {Holden}}]{vanderwel:22}
{van der Wel}, A., {van Houdt}, J., {Bezanson}, R., {et~al.} 2022, \apj, 936,
  9, \dodoi{10.3847/1538-4357/ac83c5}

\bibitem[{{van Houdt} {et~al.}(2021){van Houdt}, {van der Wel}, {Bezanson},
  {Franx}, {d'Eugenio}, {Barisic}, {Bell}, {Gallazzi}, {de Graaff}, {Maseda},
  {Pacifici}, {van de Sande}, {Sobral}, {Straatman}, \& {Wu}}]{houdt:21}
{van Houdt}, J., {van der Wel}, A., {Bezanson}, R., {et~al.} 2021, \apj, 923,
  11, \dodoi{10.3847/1538-4357/ac1f29}

\bibitem[{{van Rossum}(1995)}]{vanrossum1995}
{van Rossum}, G. 1995, CWI Technical Report, CS-R9526

\bibitem[{{Vulcani} {et~al.}(2014){Vulcani}, {Bamford}, {H{\"a}u{\ss}ler},
  {Vika}, {Rojas}, {Agius}, {Baldry}, {Bauer}, {Brown}, {Driver}, {Graham},
  {Kelvin}, {Liske}, {Loveday}, {Popescu}, {Robotham}, \& {Tuffs}}]{vulcani:14}
{Vulcani}, B., {Bamford}, S.~P., {H{\"a}u{\ss}ler}, B., {et~al.} 2014, \mnras,
  441, 1340, \dodoi{10.1093/mnras/stu632}

\bibitem[{{Wake} {et~al.}(2012){Wake}, {van Dokkum}, \& {Franx}}]{wake:12}
{Wake}, D.~A., {van Dokkum}, P.~G., \& {Franx}, M. 2012, \apjl, 751, L44,
  \dodoi{10.1088/2041-8205/751/2/L44}

\bibitem[{{Webb} {et~al.}(2020){Webb}, {Balogh}, {Leja}, {van der Burg},
  {Rudnick}, {Muzzin}, {Boak}, {Cerulo}, {Gilbank}, {Lidman}, {Old},
  {Pintos-Castro}, {McGee}, {Shipley}, {Biviano}, {Chan}, {Cooper}, {De Lucia},
  {Demarco}, {Forrest}, {Jablonka}, {Kukstas}, {McCarthy}, {McNab}, {Nantais},
  {Noble}, {Poggianti}, {Reeves}, {Vulcani}, {Wilson}, {Yee}, \&
  {Zaritsky}}]{webb:20}
{Webb}, K., {Balogh}, M.~L., {Leja}, J., {et~al.} 2020, \mnras, 498, 5317,
  \dodoi{10.1093/mnras/staa2752}

\bibitem[{{Weisz} {et~al.}(2008){Weisz}, {Skillman}, {Cannon}, {Dolphin},
  {Kennicutt}, {Lee}, \& {Walter}}]{weisz:08}
{Weisz}, D.~R., {Skillman}, E.~D., {Cannon}, J.~M., {et~al.} 2008, \apj, 689,
  160, \dodoi{10.1086/592323}

\bibitem[{{Weisz} {et~al.}(2011){Weisz}, {Dalcanton}, {Williams}, {Gilbert},
  {Skillman}, {Seth}, {Dolphin}, {McQuinn}, {Gogarten}, {Holtzman}, {Rosema},
  {Cole}, {Karachentsev}, \& {Zaritsky}}]{weisz:11}
{Weisz}, D.~R., {Dalcanton}, J.~J., {Williams}, B.~F., {et~al.} 2011, \apj,
  739, 5, \dodoi{10.1088/0004-637X/739/1/5}

\bibitem[{{Wild} {et~al.}(2016){Wild}, {Almaini}, {Dunlop}, {Simpson},
  {Rowlands}, {Bowler}, {Maltby}, \& {McLure}}]{wild:16}
{Wild}, V., {Almaini}, O., {Dunlop}, J., {et~al.} 2016, \mnras, 463, 832,
  \dodoi{10.1093/mnras/stw1996}

\bibitem[{{Wu} {et~al.}(2018{\natexlab{a}}){Wu}, {van der Wel}, {Bezanson},
  {Gallazzi}, {Pacifici}, {Straatman}, {Bari{\v{s}}i{\'c}}, {Bell}, {Chauke},
  {van Houdt}, {Franx}, {Muzzin}, {Sobral}, \& {Wild}}]{wu:18}
{Wu}, P.-F., {van der Wel}, A., {Bezanson}, R., {et~al.} 2018{\natexlab{a}},
  \apj, 868, 37, \dodoi{10.3847/1538-4357/aae822}

\bibitem[{{Wu} {et~al.}(2018{\natexlab{b}}){Wu}, {van der Wel}, {Gallazzi},
  {Bezanson}, {Pacifici}, {Straatman}, {Franx}, {Bari{\v{s}}i{\'c}}, {Bell},
  {Brammer}, {Calhau}, {Chauke}, {van Houdt}, {Maseda}, {Muzzin}, {Rix},
  {Sobral}, {Spilker}, {van de Sande}, {van Dokkum}, \& {Wild}}]{wu:21}
{Wu}, P.-F., {van der Wel}, A., {Gallazzi}, A., {et~al.} 2018{\natexlab{b}},
  \apj, 855, 85, \dodoi{10.3847/1538-4357/aab0a6}

\bibitem[{{Wuyts} {et~al.}(2009){Wuyts}, {Franx}, {Cox}, {Hernquist},
  {Hopkins}, {Robertson}, \& {van Dokkum}}]{wuyts:09}
{Wuyts}, S., {Franx}, M., {Cox}, T.~J., {et~al.} 2009, \apj, 696, 348,
  \dodoi{10.1088/0004-637X/696/1/348}

\bibitem[{{Zahid} {et~al.}(2018){Zahid}, {Sohn}, \& {Geller}}]{zahid:18}
{Zahid}, H.~J., {Sohn}, J., \& {Geller}, M.~J. 2018, \apj, 859, 96,
  \dodoi{10.3847/1538-4357/aabe31}

\end{thebibliography}
\bibliographystyle{aasjournal}

\appendix
\renewcommand{\thefigure}{A\arabic{figure}}
\setcounter{figure}{0}

\begin{figure}[!htbp]
  \centering
  \begin{minipage}{.98\textwidth}
    \includegraphics[width=\textwidth]{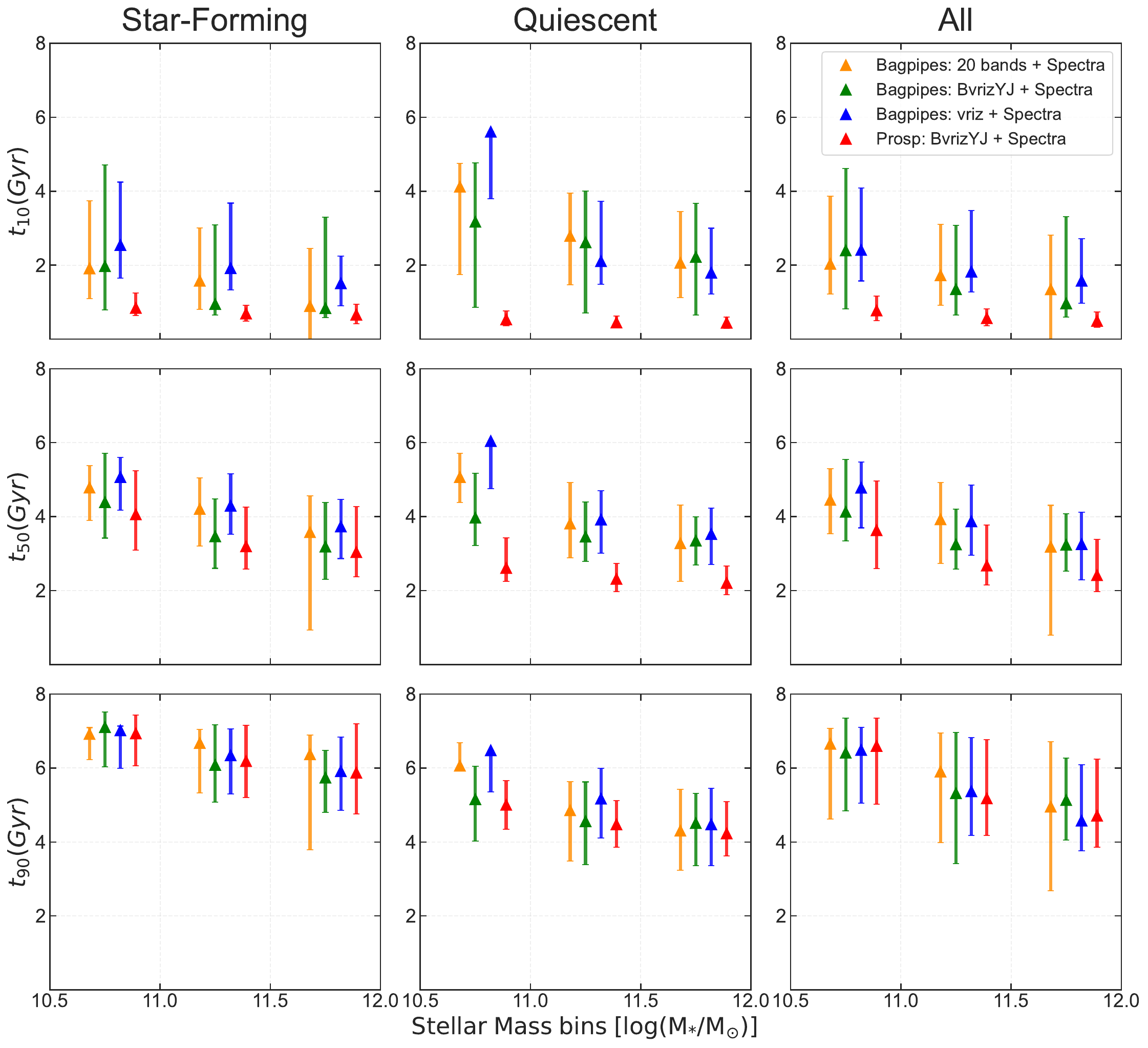}
    \caption{Comparing 10$\%$ (top row), 50$\%$ (middle row) and 90$\%$ (bottom row) formation times for star-forming (left), quiescent (center) and all (right) galaxies from different flavors of spectro-photometric runs using \texttt{Bagpipes}  and \texttt{Prospector}. Our results are robust to the photometric information provided in the fits and are defined by the high-resolution spectral information. Error bars represent 16th - 84th percentile ranges.}
    \label{fig:appendix1}
  \end{minipage}
  \hfill
\end{figure}

\renewcommand{\thefigure}{A\arabic{figure}}
\setcounter{figure}{1}

\begin{figure}[!htbp]
  \centering
  \begin{minipage}{.98\textwidth}
    \includegraphics[width=\textwidth]{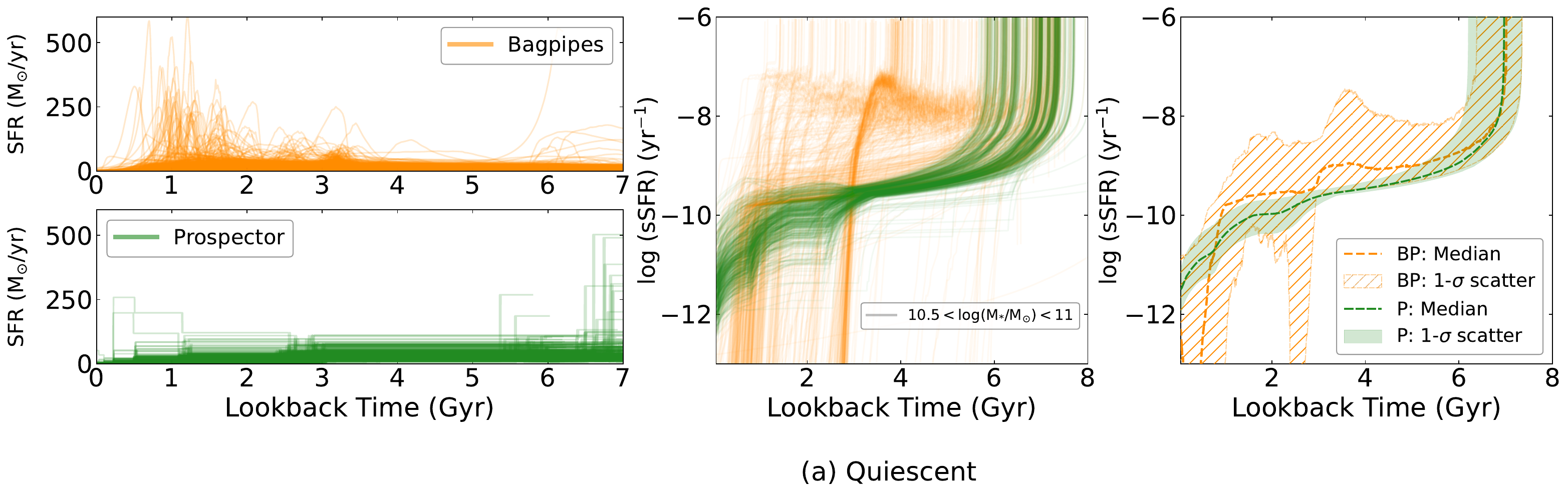}
    \includegraphics[width=\textwidth]{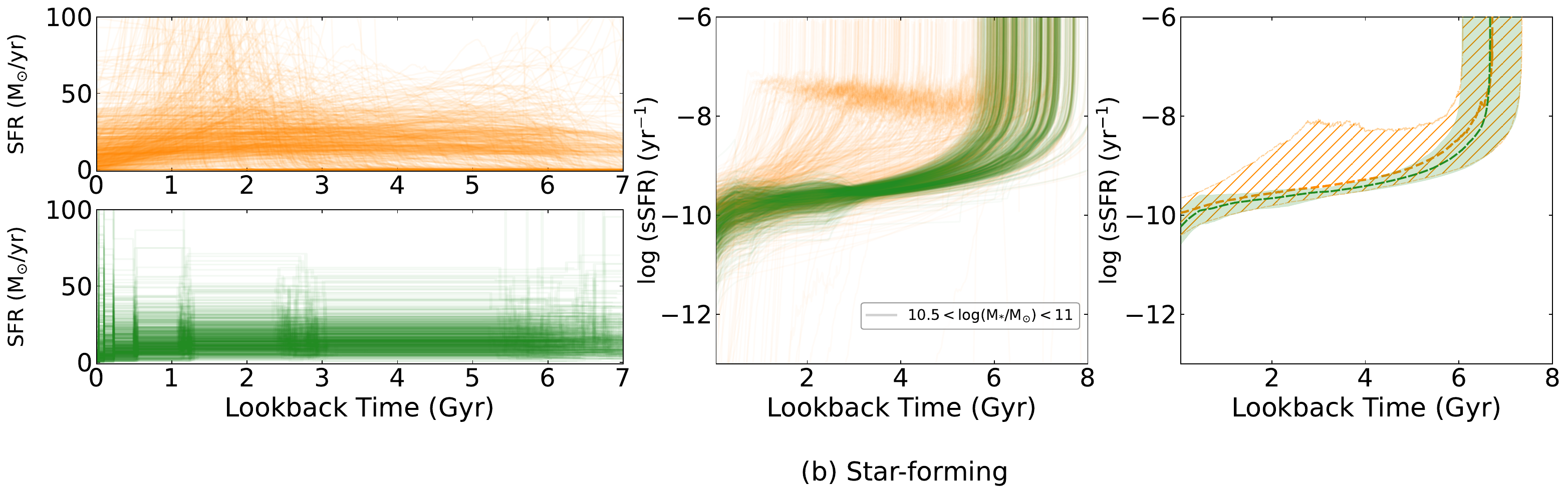}
    \includegraphics[width=\textwidth]{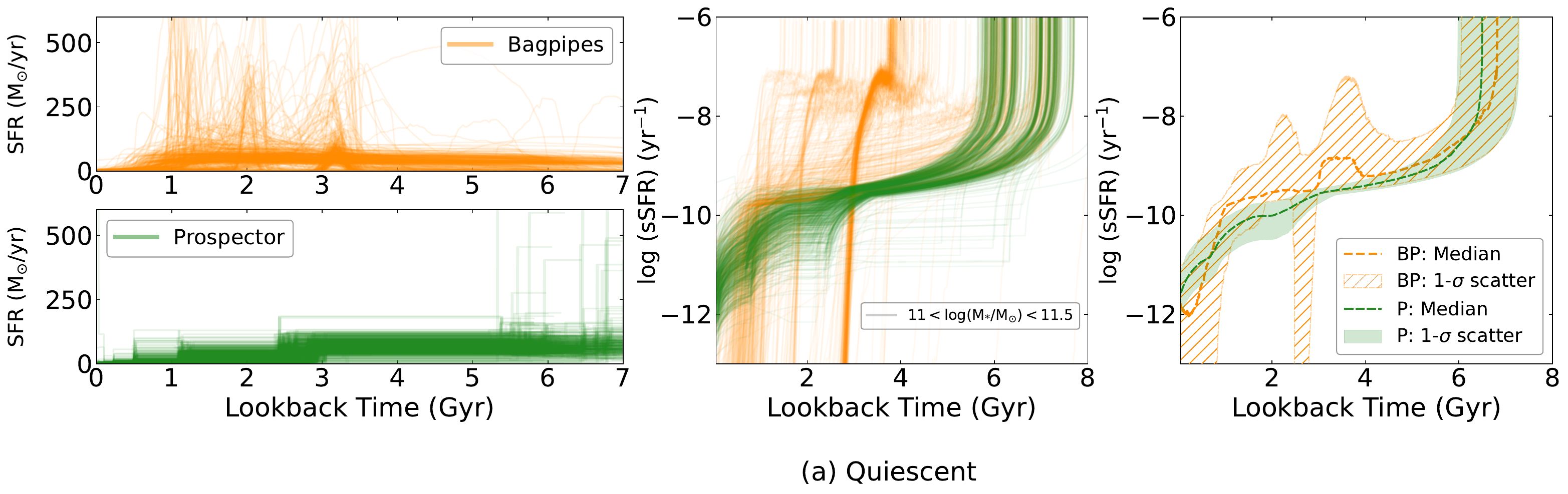}
    \includegraphics[width=\textwidth]{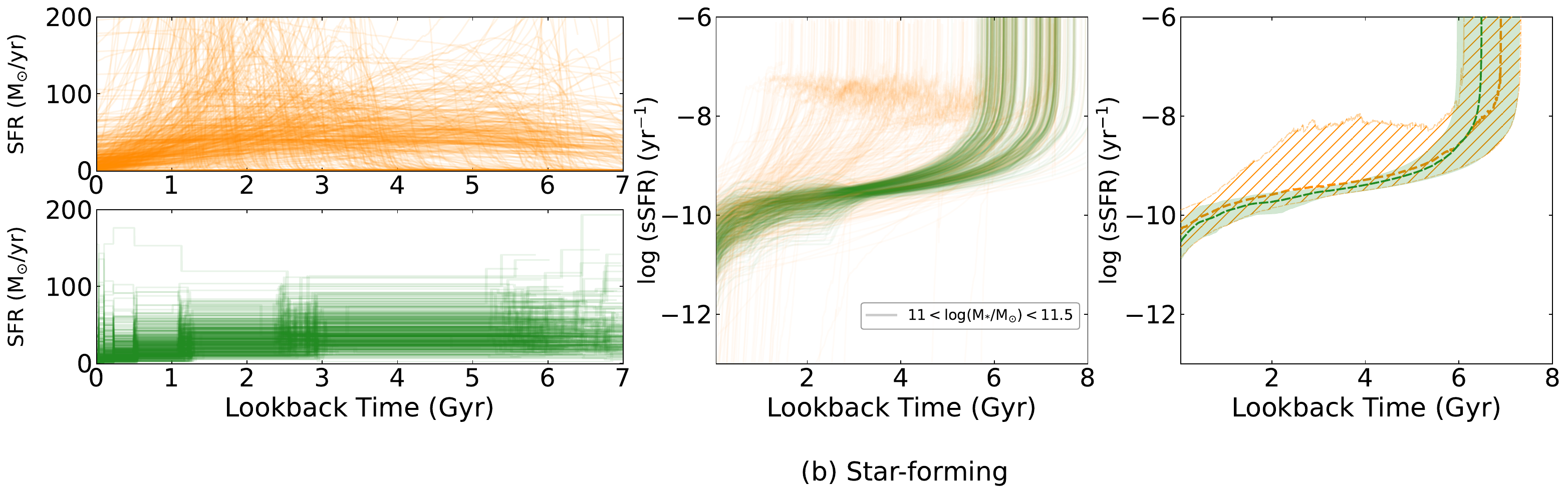}

    \caption{Extended version of Figure \ref{fig:ssfr-example} showing recovered SFHs of lower stellar mass galaxies for \texttt{Bagpipes} and \texttt{Prospector} in SFR ($M_{\odot}/yr$) and log(sSFR) ($yr^{-1}$) plane. Top two rows show galaxies in lowest stellar mass bin 10.5 < log($\mathrm{M_*/M_{\odot}}$) < 11 for quiescent (a) and star-forming (b) sample. Similarly, bottom two row shows the same for galaxies in 11 < log($\mathrm{M_*/M_{\odot}}$) < 11.5 stellar mass bin. Note that the vertical axis limits of SFHs are different for the two populations to zoom into the features that are prominent in sSFR plane.}
    \label{fig:ssfr1}
  \end{minipage}
  \hfill
\end{figure}



Our primary spectro-photometric analysis was conducted on deep LEGA-C spectra with BvrizYJ broadband photometric data. To investigate whether this modeling choice impacts our final results, we fit the full LEGA-C primary sample with 2 more flavors in Bagpipes - (1) spectra + vriz bands and (2) spectra + 20 photometric bands (from B to 24 microns as described in Section \S\ref{data-and-sample}). We compare the median $t_{10}$, $t_{10}$ and $t_{90}$ formation times from these flavors with those presented in this paper from Bagpipes and Prospector using BvrizYJ bands only in Figure \ref{fig:appendix1}. Error bars represent the population scatter within the samples of that mass bin. It is evident that the derived formation times do not depend strongly on the photometric information provided to the modeling.

The population scatter is higher for earlier formation times compared to later ones with Prospector $t_{10}$ dominated by the priors on the very first time bin. For quiescent population, we do see highest variation in lowest mass bin 10.5 < $\mathrm{log(M/M_{\odot}})$ < 11, primarily driven by low number of samples ($\sim$290) in that mass range. \\

Figure \ref{fig:ssfr1} is an extended version of Figure \ref{fig:ssfr-example} and shows SFHs of quiescent and star-forming population combined in the other two low stellar mass bins e.g. 10.5 < log($\mathrm{M_*/M_{\odot}}$) < 11 and 11 < log($\mathrm{M_*/M_{\odot}}$) < 11.5. All symbols and labels are same as Figure \ref{fig:ssfr-example}. Top two rows show lowest mass bin for quiescent and star-forming population whereas bottom two rows show middle mass bins. Note that the vertical axis limits of SFHs in $M_{\odot}/yr$ unit are different for the two populations to zoom into the features that are prominent in sSFR plane. 

\end{document}